\documentclass[useAMS,usenatbib]{mn2e}
\usepackage{graphicx,txfonts,amsfonts,pslatex}
\usepackage{pdfscreen}
\def\idm#1{\mbox{\scriptsize #1}}
\def\cM{{\cal M}}
\def\mA{{\mathbb A}}
\def\cH{{\cal H}}
\def\mH{{\mathbb H}}

\def\mI{{\mathbb I}}

\def\cJ{{\cal J}}

\def\cE{{\cal E}}
\def\mE{{\mathbb E}}
\def\mR{{\mathbb R}}

\def\cP{{\cal P}}
\def\cS{\Sigma}
\def\vx{{\bf x}}
\def\vy{{\bf y}}
\def\RP{{$\cal S$}}
\def\Imut{{{\cal I}_{\mbox{\scriptsize mut}}}}
\def\Izero{i_0}
\def\Hsec{{\cal H}_{\mbox{\scriptsize sec}}}

\def\AMD{\mathrm{AMD}}
\def\nAMD{{\mathcal{\cal A}}}
\def\dt#1{\frac{\mbox{d}#1}{\mbox{d}t}}
\newcommand\Chi{(\chi^2_\nu)^{1/2}}
\def\hide#1{{}}
\def\corr#1{{ #1}}
\title[Equilibria in non-coplanar secular problem]{
Equilibria in the secular, non-coplanar two-planet problem
}
\author[C. Migaszewski and K. Go\'zdziewski]{Cezary Migaszewski$^{1}$\thanks{E-mail:
c.migaszewski@astri.uni.torun.pl} and Krzysztof Go\'zdziewski$^{1}$\footnotemark[1]\thanks{E-mail:
k.gozdziewski@astri.uni.torun.pl}\\
$^{1}$Toru\'n Centre for Astronomy of the Nicolaus Copernicus University, 
  Gagarin Str. 11, 87-100 Toru\'n, Poland}

\begin{document}

\date{Accepted 2009 January 15.  Received 2009 January 12; in original form 2008
May 23}

\maketitle

\label{firstpage}

\begin{abstract}
We investigate the secular dynamics of a planetary system composed of the parent
star and two massive planets in mutually inclined orbits.  The dynamics are
investigated in wide ranges of semi-major axes ratios  (0.1--0.667), and
planetary masses ratios (0.25--2) as well as in the whole permitted ranges of
the energy and total angular momentum. The secular model is  constructed by
semi-analytic averaging of the three-body system. We focus on  equilibria of the
secular Hamiltonian (periodic solutions of the full system), and we analyze
their stability. We attempt to classify  families of these solutions in terms of
the angular momentum integral. We identified new equilibria, yet unknown in the
literature. Our results are general and may be applied to a wide class of
three-body systems, including configurations with a star and brown dwarfs and
sub-stellar objects. We also describe some technical aspects of the
semi-numerical averaging. The HD~12661 planetary system is investigated as an
example configuration.
\end{abstract}
%

\section{Introduction}
Nowadays, about thirty extrasolar multi-planet systems have been
detected\footnote{See Jean Schneider's Extrasolar Planets Encyclopedia 
http://exoplanet.eu for frequent updates on the discoveries and orbital
parameters}. Many of them seem either locked in or close to low-order mean
motion resonances (MMRs). Moreover, there is a class of the so called
hierarchical systems \citep{Lee2003} which can be  characterized by relatively
small ratio of semi-major axes.  Their planetary orbits are well separated and
far from collision zones, hence the long-term, qualitative  dynamics of such
systems may be  effectively investigated with secular theories. The
Hamiltonian of a hierarchical system can be averaged out over mean longitudes
which play the role of fast angles.  In the regime of small eccentricities and
inclinations, this approach leads to the well known, classical Laplace-Lagrange
(L-L) secular theory \citep{Murray2000}. It relies on the expansions of the
disturbing function in power series with respect to eccentricities and
inclinations which are small parameters of the problem.  However, many
multi-planet hierarchical systems do not satisfy the assumption of small
eccentricities, and the L-L theory may fail. 

Still, to deal with the observed diversity of orbital configurations, the
secular theories relying on high-order expansions of the perturbations are used,
e.g., the series in eccentricity
\citep[e.g.,][]{Murray2000,Rodriguez2005,Libert2006,Libert2007a,Libert2007b,Veras2007}
or expansion to the third order in the ratio of semi-major axes $\alpha$, known
as the octuple theory \citep{Ford2000,Lee2003}. This theory can be also
generalized to higher orders \citep[][and references therein]{Migaszewski2008a}.
The analytical expansions are particularly suitable for studies of hierarchical
systems.  Moreover, they are usually valid only in limited ranges of the orbital
parameters, and special cases (like resonant configurations) must be treated
individually. The alternative, recently developed quasi-analytical theory relies
on averaging the perturbing Hamiltonian  numerically
\citep{Michtchenko2004,Michtchenko2006}. In this work, we are heavily  inspired
by these papers and their idea of the semi-analytical technique. Because the
method does not require any expansion of the perturbing Hamiltonian, basically, 
it has no limitations inherent in the analytical theory. For instance, with 
a help of this technique, \cite{Michtchenko2004} found new, non-classic feature
of the secular dynamics of  coplanar system of two planets (the so called 
non-linear secular resonance in the regime of large eccentricity). In the later
work, \cite{Michtchenko2006} consider more general three-dimensional (3-D)
secular model of two-planet system, and present a systematic approach helpful to
investigate the global dynamics of such configurations.  

\begin{figure*}
\centerline{\includegraphics[width=5.6in]{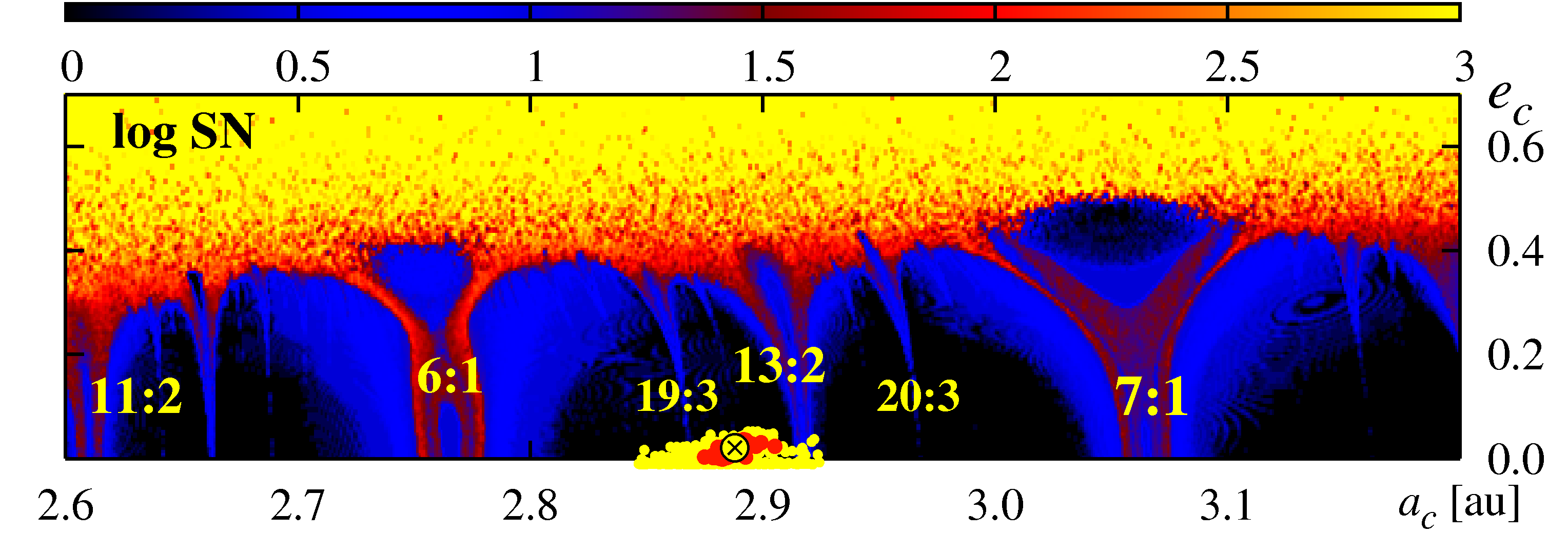}}
\caption{
The dynamical map of the edge-on, coplanar HD~12661 system in the
$(a_{\idm{c}},e_{\idm{c}})$-plane, for the best-fit solution to the RV
data published in \citep{Butler2006}. Large values of the Spectral Number (SN)
marked in yellow
indicate continuous spectrum  of the fundamental frequencies of the system and 
strongly chaotic motions, small SN (black) means discrete frequencies and ordered
motions. Positions of low-order MMRs are labeled.  The best-fit
solution, yielding $\Chi \sim 1.08$ and an rms $\sim 7.46$~ms$^{-1}$, in terms
of parameter tuples ($m$ [$\mbox{m}_{\idm{J}}$], $a$~[au], e, $\omega$~[deg], ${\cal
M}$~[deg]) including planetary mass, semi-major axis, eccentricity, the argument of pericenter
and the mean anomaly at the epoch of the first observation,
$t_0$=JD2,450,831.608380, is the following (2.34, 0.831, 0.361, 296.24, 126.86)
for planet~b, and (1.84, 2.888, 0.021, 52.66, 66.18) for planet~c, respectively.
\hide{Offsets are $0.073$~ms$^{-1}$ and  $-0.012$~ms$^{-1}$, for the Lick and Keck
data, respectively.} The original errors are rescaled by adding stellar jitter of
$3.5$~m/s in quadrature. The mass of the parent star is 1.11~$M_{\sun}$.
Solutions within $1\sigma$ level of the best fit are marked with yellow circles,
fits with marginally worse $\Chi$ are marked in red. 
}
\label{fig:fig1}
\end{figure*}

As an example to study, we choose the HD~12661 planetary system
\citep{Fischer2001,Fischer2003,Butler2006}.  The discovery paper 
\citep{Fischer2001} announces two Jovian planets 
on well separated orbits with semi-major axes of $\sim0.8$~au and
$\sim2.8$~au, respectively, and of moderate eccentricities. We analyzed the
most recent, publicly available data from the catalogue of \cite{Butler2006} and
\citep{Wright2008}, using the $N$-body model of the radial velocities
(RV)  and the so called hybrid
minimization  \citep{Gozdziewski2006}. The results of our analysis  of the RV
data published in \citep{Butler2006} are illustrated in Fig.~\ref{fig:fig1}. The
best fit solution yielding $\Chi\sim 1.08$ and an rms $\sim 7.5$~m/s is marked
with a crossed circle in the dynamical map in terms of  the Spectral Number (SN)
\citep{Michtchenko2001}. The SN is the fast indicator making it possible to
distinguish between chaotic and regular planetary configurations. The osculating
elements of the best fit solution  at the epoch of the first observation are
given in caption to Fig.~\ref{fig:fig1}. In this figure, we mark the semi-major
axis and  eccentricity of the outer planet  derived from an ensemble of fits
within $1\sigma$ of the best fit solution. Clearly, the available data already
constrain  orbital elements of the outer planet very well.  The dynamical maps
reveal orbits well separated from the low-order MMRs. Two most prominent MMRs
within the vicinity of the best fit are 19:3 and 13:2~MMRs, respectively.
Moreover, the best-fits within $1\sigma$ confidence level span the region of
small eccentricities in which the resonances are very narrow.

Hence, the HD~12661 system fits well assumptions of the secular theory.  This system has been studied already in a few papers: with 
the direct numerical integrations \citep{Ji2003b}, with the  analytical octupole
theory of hierarchical systems  \citep{Lee2003}, with mapping of the phase space
by  fast indicators \citep{Gozdziewski2003a}, and with the classic analytical
theory that relies on expansions of the perturbation in eccentricity
\citep{Rodriguez2005,Libert2006}. All the cited works assume that the
HD~12661  system is coplanar and oriented edge-on.  However, we should keep in
mind that a major limitation of the Doppler technique lies in the ambiguity of
orbital inclinations, which cannot be well determined by far. The observational
windows are still relatively narrow, and to remove the inclination degeneracy,
several orbital periods of the outermost orbit are required. Moreover, the
recent formation theories do not fully predict 
mutual inclinations in multi-planet systems. We cannot be certain yet 
whether the common assumption of coplanar orbits really holds true. Likely, 
many different forming scenarios are possible. For instance, the migration in
low-order MMRs may end up with systems characterized by large mutual
inclinations \citep{Thommes2003}.  The dynamical relaxation of initially dense
planetary systems of giant planets \citep{Adams2003} may lead to scattering
events  which produce wide distribution of the mutual inclinations. Indeed,
recent simulations of  \cite{Veras2008}  revealed that the outer planet of the
HD~12661  system undergoes large oscillations for nearly all of the allowed
two-planet orbital solutions. These authors conclude that it might be the effect
of  a perturbation of planet~c, perhaps due to strong scattering of an
additional planet that was subsequently accreted onto the star. Moreover, we
stress that the inclination of the  HD~12661 system is still unknown, hence
the understanding of basic features of its 3-D  dynamics  seems also important. 
This intriguing system is an excellent candidate for tests and numerical
experiments regarding non-coplanar configurations. Moreover,  because we attempt
to study the secular 3-D dynamics globally, our results are  general and valid
for much wider class of three-body systems. The secular theory  which we consider
here, covers planetary systems with different masses and semi-major axes ratios,
and the full range of mutual inclinations. 

The plan of this work is the following. In Sect.~\ref{model}, we recall the
general  mathematical model of the 3-D two-planet system.
Subsection~\ref{sec:average} is devoted to a short technical overview of the
averaging approach and some computational details that may be useful in a
practical implementation of the method. To make the paper self-contained, we
also recall the notion of  representative planes, and energy levels calculated
for fixed values of the total angular momentum integral
(Sect.~\ref{sec:equilibria}). To illustrate the precision of the semi-analytic
approach, we compute the Poincar\'e cross sections, and demonstrate chaotic
behavior of the secular system \citep{Michtchenko2006}. The main results are
described in Sect.~\ref{sec:families}  which is devoted to the analysis  of the
existence and bifurcations of equilibria in the secular, spatial problem of two
mutually interacting planets. In particular, we detect and investigate closely a
few families of these equilibria in a  wide range of planetary mass ratio, $\mu
\in\{0.25,0.5,1,2\}$, and the semi-major axes ratio, $\alpha \in
\{0.1,0.2,0.333,0.667\}$. The results are general and valid as long as the
partition of the Hamiltonian onto the Keplerian, integrable term, and the small
perturbation is reasonable. In that part, our work extends the paper of
\cite{Libert2007b}. After introducing non-singular canonical variables, they
investigate the existence, stability and bifurcations of stationary solutions
emerging from the equilibrium at zero-eccentricities, the so called  {\em
Lidov--Kozai resonance} \citep[][to mention a few papers in an endless list of
references]{Lidov1961,Kozai1962,Lidov1976,Michel1996,Innanen1997,Kinoshita2007}
which was found  and intensively investigated in the restricted  three body
problem. The full three-body problem in the Hill case  (hierarchical
configurations) was also intensively studied by many authors
\citep[e.g.,][]{Krasinsky1972,Krasinsky1974,Lidov1976,Ferrer1994,Miller2002,
Fabrycky2007}. These works rely mostly on the second order expansion of
the secular Hamiltonian in the semi-major axes ratio (the quadrupole
approximation).  In the present work, we focus on two aspects of the problem:
\begin{itemize}
\item we consider the unrestricted problem in wide ranges of semi-major axes
ratio $\alpha$, up to 0.667, and mass ratio $\mu$, 
\item we study equilibria of
the full secular Hamiltonian; the semi-analytic averaging helps us
to compute the secular perturbations beyond convergence limits 
of the usual power series expansions.
\end{itemize}
Thanks to the quasi-analytic averaging, we found new families  of equilibria of
the secular 3D planetary problem which unlikely may be detected with the help of
perturbation techniques. We also study the Lyapunov stability of these solutions
in detail (or to an extent permitted by technical limits of the semi-analytic
algorithm).

\section{The 3-D dynamics of two-planet system}
\label{model} 
The Hamiltonian of the three-body planetary system, expressed  with respect to
canonical Poincar\'e variables \citep[see][]{Laskar1995,Michtchenko2006} 
has the form of:
\begin{equation}
\mathcal{H}=
\underbrace{\sum_{i=1}^{2} {\bigg(\frac{1}{2 \beta^*_i} \mathbf{p}_i^2 - 
\frac{\mu^*_i \beta^*_i}{r_i} \bigg)}}_{\textrm{\small Keplerian part}} - 
\underbrace{\frac{k^2 m_1 m_2}{\Delta}}_{\textrm{\small direct part}} +
\underbrace{\frac{\mathbf{p}_1 \cdot \mathbf{p}_2}{m_0}}_{\textrm{\small indirect
part}},
\label{hamiltonian}
\end{equation}
where  ${\mathbf{r}_i}$ denote the position vectors {\em relative to the star}, 
${\mathbf{p}_i}$ -- their conjugate momenta {\em relative to the barycenter} of
the full three-body system,  ${\Delta=|\mathbf{r}_1-\mathbf{r}_2|}$ is for the
distance between planets,  ${m_0}$ -- is the mass of the parent star;  ${m_1}$,
${m_2}$ -- are for the masses of the planets  (also index $i=1$ is for the inner
planet, and $i=2$ for the outer planet). We denote  also the mass parameters
${\mu^*_i=k^2~(m_0+m_i)}$ and the reduced masses  through
${\beta^*_i=(1/m_i+1/m_0)^{-1}}$. Under the assumption of $m_i \ll m_0$ (or, more
generally, for small enough perturbations of Keplerian orbits), the Hamiltonian
of the system, ${\cal H}$, is a sum of the {\em Keplerian}  term (which would be
integrable in the absence of mutual interactions between planets) and the
interaction term with the so called  {\em direct} and {\em indirect} terms.

Alternatively, the dynamical state of the system,
$({\mathbf{r}_i},{\mathbf{p}_i})$ may be represented through the mass-weighted
canonical angle--action variables of Delaunay \citep{Murray2000}: 
\begin{equation}
\begin{array}{ll}
{\cal M}_i - \textrm{the mean anomaly}, & {L_i=\beta^*_i~\sqrt{\mu^*_i~a_i}},\\
{\omega_i} - \textrm{the argument of pericenter}, & {G_i=L_i~\sqrt{1-e_i^2}},\\
{\Omega_i} - \textrm{the longitude of node}, &{H_i=G_i \cos~I_i,}
\end{array}
\label{delaunay}
\end{equation}
where $a_i$ denote semi-major axes,  $e_i$ -- eccentricities, and $I_i$  stand
for inclinations; $(L_i,G_i,H_i)$ are the conjugate momenta.  The transformation
between $({\mathbf{r}_i},{\mathbf{p}_i})$ and the set of Delaunay elements
$(a_i,e_i,I_i,\Omega_i,\omega_i,{\cal M}_i)$ may be found in
\citep{FerrazMello2005} or \citep{Morbidelli2002}. If orbits are far from strong
MMRs and collision zones then Hamiltonian in Eq.~\ref{hamiltonian} can be
averaged out with respect to the mean anomalies playing the role of  fast
angles, and then we obtain the secular Hamiltonian ${\cal
H_{\idm{sec}}}$ which approximates the  long-term, slow variations of the mean
elements.

To make the paper self-contained, we recall basic facts on the secular 3-D
problem of two planets. We follow \cite{Michtchenko2006} and \cite{Libert2007b}.
The averaged $\mathcal{H}_{sec}$ does not depend on $\cM_1, \cM_2$,
therefore the conjugate actions $(L_1, L_2)$ are constant. After the partial
reduction of nodes, $\Hsec$ depends on $\Delta{\Omega}$ only, and not on
$\Omega_1$ and $\Omega_2$ separately. After the canonical transformation
\citep{Michtchenko2006}:
\begin{equation}
\begin{array}{l}
(\omega_1,G_1)\\
(\omega_2,G_2)\\
(\Omega_1,H_1)\\
(\Omega_2,H_2)
\end{array}
\quad \Rightarrow \quad
\begin{array}{l}
(\omega_1,G_1)\\
(\omega_2,G_2)\\
\left(\theta_1=\frac{1}{2}\left(\Omega_1+\Omega_2\right), \ J_1=H_1+H_2 \right)\\
\left(\theta_2=\frac{1}{2}\left(\Omega_1-\Omega_2\right), \ J_2=H_1-H_2
\right).
\end{array}
\label{trans}
\end{equation}
The secular Hamiltonian does not depend on $\theta_1$, therefore
$J_1=|\mathbf{C}|=\mbox{const}$, where $\mathbf{C}$ is the total angular
momentum of the system.  Moreover, $\theta_2=\pi/2=\mbox{const}$  (in the
Laplace frame, $2\theta_2 \equiv \Delta\Omega = \pm \pi$, after Jacobi's
elimination of the nodes), and:
\[
J_2 = (G_1^2 - G_2^2)/J_1.
\] 
For fixed angular momentum $J_1$ and secular energy $\cH_{\idm{sec}}$, the averaged
system can be reduced to two degrees of freedom.  Instead of $J_1$, 
we may use the so called Angular Momentum Deficit,
$ 
\AMD = L_1 + L_2 - J_1.
$ 
The $\AMD$ is a measure of the system non-linearity
\citep{Laskar2000}.   Coplanar and circular orbits have the minimum of $\AMD=0$.
In configurations with large $\AMD$, crossing orbits are possible and they are
very unstable.

Because the secular Hamiltonian still depends on many parameters, the global
analysis of the long-term dynamics are complex. To simplify the study  of 
their 
basic properties, we fix particular  values of integrals and/or orbital
elements.  We choose the semi-major axes and masses as the primary parameters of
the secular model. Then $L_1$ and $L_2$ are their (scaled) representation. The
maximum of $\AMD$ is equal to $L_1+L_2$,  hence we introduce {\em  the
normalized Angular Momentum Deficit}:
\[
 \nAMD = \frac{\AMD}{L_1+L_2}, \quad \nAMD \in [0,1],
\]
which is an uniform and non-dimensional representation of $\AMD$. 
 Relative to the Laplace
plane, $C_x=C_y=0$, $C_z\equiv C$, hence we have:
\begin{eqnarray*}
&& L_1 \sqrt{1-e_1^2} \cos I_1 + L_2 \sqrt{1-e_2^2} \cos I_2 = J_1, \\
&& L_1 \sqrt{1-e_1^2} \sin I_1 - L_2 \sqrt{1-e_2^2} \sin I_2 = 0.
\end{eqnarray*}
Also the mutual inclination of orbits $\Imut=I_1 + I_2$. Thus, $\cos
\Imut = \cos I_1 \cos I_2 + \sin I_1 \sin I_2 \cos\Delta\Omega$,
or, alternatively,
\begin{eqnarray}
 \cos \Imut (e_1,e_2) &=& \frac{J_1^2-G_1^2-G_2^2}{2 G_1 G_2}, \\
 \cos I_1(e_1,e_2) &=& \frac{J_1^2+G_1^2-G_2^2}{2 J_1 G_1}, \\
 \cos I_2(e_1,e_2) &=& \frac{J_1^2+G_2^2-G_1^2}{2 J_1 G_2}.
\end{eqnarray}
Because $C \equiv C_z > 0$, the above relations are singular for
$I_1=I_2=\pi/2$ or $e_1=e_2=1$ (when $G_1=G_2=0$). A boundary of the manifold
of permitted motions for a given $J_1 \equiv C$ (or $\AMD$), semi-major axes
and planetary masses ratio, can be defined through $\Imut=0,\pi$.  It can be
also shown that when $\AMD$ is fixed and $I_{1,2}<\pi/2$ then  the mutual
inclination at the origin $(e_1=0,e_2=0)$ is maximal. We will denote such value
by $i_0$ from hereafter.

The dynamics of the secular system are expressed through solutions to the
following canonical equations of motion:
\begin{equation}
 \dt{\omega_i} = \frac{\partial{\Hsec}}{\partial{G_i}}, \quad
   \dt{G_i}      = -\frac{\partial{\Hsec}}{\partial{\omega_i}}, \quad
   i=1,2,
\label{eq:odes}
\end{equation}
where $(\omega_1,\omega_2)$ are canonical angles and $(G_1,G_2)$ are  canonical
momenta. Having only the numerical approximation of $\Hsec$ (see below), we must
solve Eqs.~\ref{eq:odes} numerically. For that purpose, we may use a suitable
integrator, for instance, the 4-th order Runge-Kutta scheme \citep{Press1992}.
The partial derivatives appearing in the right-hand side of the equations of
motion, are calculated with the mid-point rule \citep{Press1992}. Moreover, to
calculate precisely the Hessian of $\Hsec$  which is required to determine the
stability (see Sect.~\ref{sec:stability}), we are forced to use higher
order approximations of the second order partial derivatives. 

\subsection{The semi-analytical averaging}
\label{sec:average}
The problem is now to average out the Hamiltonian, Eq.~\ref{hamiltonian}.
We calculate:
\begin{equation}
\mathcal{H}_{sec}=\frac{1}{(2 \pi)^2}~\int_0^{2 \pi} 
\int_0^{2\pi}{\mathcal{H}~d{\cal M}_1~d{\cal M}_2} \equiv
\left<{\cal H}\right>,
\label{secular}
\end{equation}
where $\cH$ is the Hamiltonian of the problem expressed through the canonical
Delaunay elements. For small enough perturbations, the Keplerian part
of $\left<{\cal H}\right>$ depends on constant $L_i$ only and does
not affect the secular evolution of the system. It can be shown that the average
of the indirect part of Hamiltonian equals to a constant \citep{Brouwer1961}. 
In the non-resonant case, we have to average out the direct part of
disturbing Hamiltonian only. 

The analytical calculation of apparently trivial integral, Eq.~\ref{secular},
is  in fact a difficult problem. Usually, the Hamiltonian is expanded in  power
series with respect to appropriate small parameter (eccentricity, inclination
or semi-major axes ratio). Then with the help of a suitable  canonical
transformation, we can ``remove'' particular terms of the Hamiltonian. However,
the secular series converge for relatively  small  values of 
parameters. Instead, as we
mentioned already, the secular Hamiltonian, Eq.~\ref{secular}, can be computed
numerically, without  troublesome power series expansions. This bright idea of
\cite{Michtchenko2004} is quite simple to apply.

Apparently, to compute integral in Eq.~\ref{secular}, we must evaluate
$\mathcal{H}$ in a discrete grid of the mean anomalies. That would imply multiple
(and in fact unnecessary) solution of the Kepler equation.  To get rid of this
problem, we can change the variables under the double integral using the well
known expressions relating the mean $({\cal M}_i)$, true ($f_i$) and eccentric ($\cE_i$)
anomalies, respectively.
\begin{figure*}
 \centerline{
 \hbox{\includegraphics[height=6.6cm]{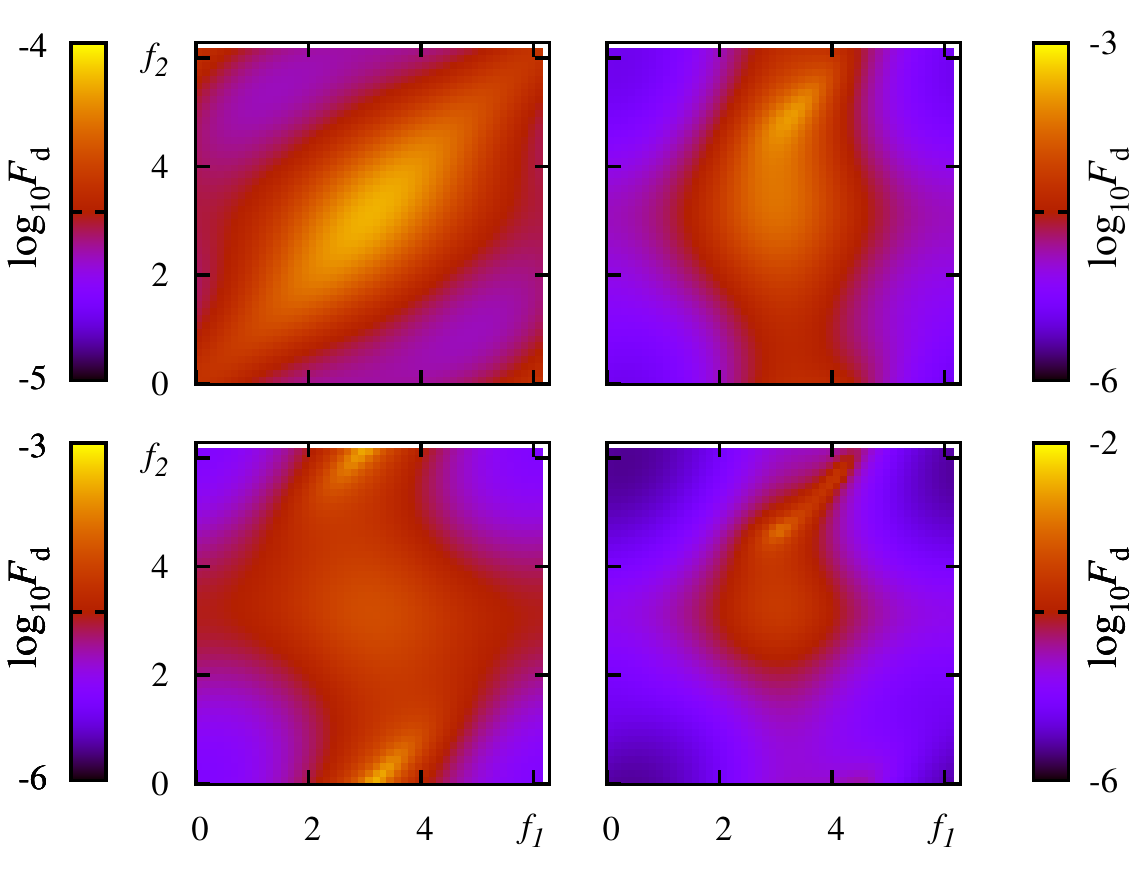}
       \hspace*{0.6cm}
       \includegraphics[height=6.6cm]{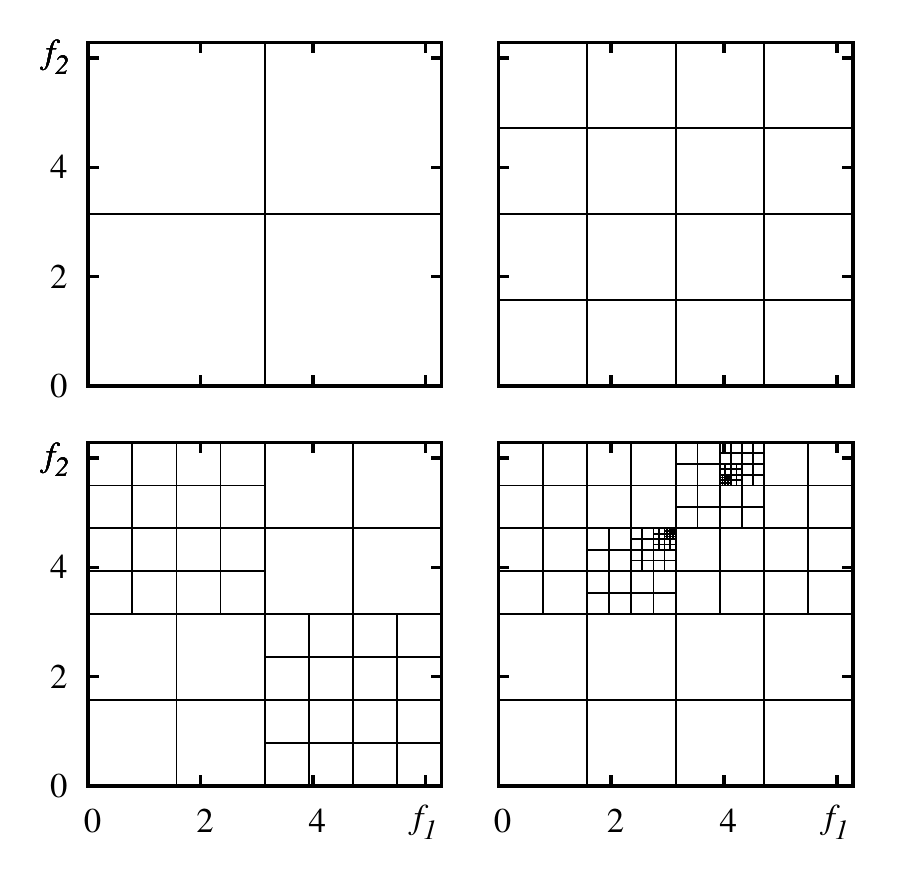}}}
 \caption{
 The left-hand panels are for contour levels of function 
 $(\mathcal{F}_{\idm{d}}$ in the $(f_1,f_2)$--plane, computed for  the coplanar
 two-planet system and orbital parameters: $m_0=1~M_{\odot}$,
 $m_1=1~\mbox{m}_{\idm{J}}$, $m_2=3~\mbox{m}_{\idm{J}}$, $a_1=1$~au, $a_2=3$~au.
 Eccentricities and $\Delta{\varpi}$ are different at each panel: the top
 left-hand panel is for $e_1=0.1$, $e_2=0.2$, $\Delta{\varpi}=0$, the top
 right-hand panel is for $e_1=0.6$, $e_2=0.5$, $\Delta{\varpi}=\pi/2$, the
 bottom left-hand panel is for $e_1=0.4$, $e_2=0.5$, $\Delta{\varpi}=\pi$, the
 bottom right-hand panel is for $e_1=0.6$, $e_2=0.7$, $\Delta{\varpi}=\pi/2$.  
 Panels in the right-hand column  illustrate the AMR-like division of the 
 integration domain, as  depending on the variability of the integrand
 function. 
 }
 \label{fig:fig2}
 \end{figure*}
To express the double integral through the true anomalies, we differentiate the
Kepler equation ${\cal M}_i = \cE_i - e_i \sin{\cE_i}$, with respect to  ${\cal
M}_i, {\cal E}_i$, and then we find that  $d\,{\cal M}_i = {\cal
J}_i~\mbox{d}\,f_i$, where:
\begin{equation} 
{\cal J}_i \equiv {\cal J}_i (e_i,f_i)=\left(1-e_i^2\right)^{3/2}~\left(1+e_i
\cos{f_i}\right)^{-2},  \quad i=1,2.
\end{equation}
The secular Hamiltonian has the following form:
\begin{equation}
\mathcal{H}_{sec}=\frac{1}{(2 \pi)^2}~\int_0^{2 \pi} 
\int_0^{2\pi}{{\cal F}~df_1~df_2}, 
\quad {\cal F} = {\cal H} \cJ_1 \cJ_2.
\label{secularJ}
\end{equation} 
We may also express the double integral  through eccentric anomalies that leads
to even simpler expressions for functions $\cJ_i$. Next, to calculate the
integral in Eq.~\ref{secularJ}, we apply an adaptive-grid integration algorithm
that relies on the Gauss-Legendr\'e quadrature of the 64-th order. The
adaptive algorithm is forced by large variability of the integrand
function. To illustrate that issue, we analyse a few typical examples shown in
Fig.~\ref{fig:fig2}. The left-hand contour plots in this figure are for the
shape of {\em direct} term of ${\cal H}$ (Eq.~\ref{hamiltonian})
multiplied by ${\cal J}_1 {\cal J}_2$,
${\cal F}_{\idm{d}}$, in the $(f_1, f_2)$-plane.  These plots are computed for
different values of eccentricities and $\Delta{\varpi} = \varpi_1 - \varpi_2$,
where $\varpi_{1,2}$ are the longitudes of periastron. In this experiment, the
system is coplanar.   In the top-left panel of Fig.~\ref{fig:fig2} (see its
left-hand part), which corresponds to relatively small eccentricities, ${\cal
F}_{\idm{d}}$ is  weakly varying function  of $(f_1,f_2)$. But for large eccentricities,
it may have narrow extrema in some parts of the $(f_1,f_2)$-plane (see the
bottom-right contour plot in Fig.~\ref{fig:fig2}).  In these areas, to reach a
desired accuracy, the integral must be computed on a dense grid of the
$(f_1,f_2)$-plane. However, in other parts of the grid, such a large number of
quadrature nodes is not necessary and, under the requirement of fixed accuracy, 
it would cause significant CPU overhead. Thus, the optimal computation  of the
double integral is possible with the non-uniform grid in the $(f_1,f_2)$-plane,
following an idea of adaptive quadratures [see, for instance,
\cite{Press1992}]. In the right-hand part of Fig.~\ref{fig:fig2}, we illustrate
the steps of our simple adaptive mesh integration by appropriate divisions of
the integration subintervals. Typically, the number of divisions is small but 
in some parts of the $(f_1,f_2)$-plane, it may be as large as 8--9, in order to
obtain the relative error of $\epsilon \sim 10^{-12}$ in two subsequent
steps of the integration.  
%
\section{Equilibria in the 3-D secular problem}
%
\label{sec:equilibria}
According to the classic methodology of Poincar\'e, to understand the dynamics,
one should investigate whole families of solutions. Isolated orbits in the phase
space tell us little on the global properties of the system. The most simple
class of solutions that can be investigated efficiently in any two degree of
freedom Hamiltonian system are equilibria defined through  {\em
algebraic} equations:
\begin{equation}
\dt{\omega_i} = 0, \quad
  \dt{G_i} = 0, \quad i=1,2.
\label{eq:eq}
\end{equation}
Typically, one tries to find the phase-space coordinates of these equilibria, 
their number and bifurcations as well as to determine their  Lyapunov stability
(at least, the linear stability).
The analysis of the existence and bifurcations of equilibria in the secular 3D
system are quite complex because they depend on many parameters ($\AMD$, the total energy, particular orbital elements, masses of planetary
companions). Hence, to investigate such solutions globally, we have to choose a
proper representation of the phase space regarding these  parameters. Moreover,
to avoid limitations of the analytical approach, the whole analysis should be
done numerically, by the semi-analytical averaging. Hence, a
reduction of the dimension of the phase space is critically important.
%
\subsection{The representative planes of the energy}
%
To simplify the search for equilibria of $\Hsec$, we choose a specific
two-dimensional plane of initial conditions that makes it possible to represent
the stationary  solutions in the 4-D phase-space of the secular system. We
follow \cite{Michtchenko2006} and \cite{Libert2007b}.  The {\em representative
plane} of initial conditions (the \RP{}-plane from hereafter) should have common
points with each phase trajectory of the secular system. In
\cite{Michtchenko2006}, the \RP{}-plane is defined through:  
\[
\cP_M = \{ e_1 \cos{\Delta{\varpi}} \times e_2 \cos{2 \omega_1} \},
\]
where $e_{1,2} \in [0,1]$ and angles $(\Delta{\varpi},2\omega_1)$ are fixed to
pairs of angles $(0,0)$, $(\pi,0)$, $(0,\pi)$, and $(\pi,\pi)$, respectively. 
In that notion, the \RP{}-plane  comprises of four subsets of points which
coordinates span the range of $e_1 \cos\Delta\varpi \in [-1,1]$ and $e_2 \cos
2\omega_1 \in [-1,1]$.

\begin{figure*}
\centerline{
\hbox{
\hbox{
\includegraphics[width=2.2in]{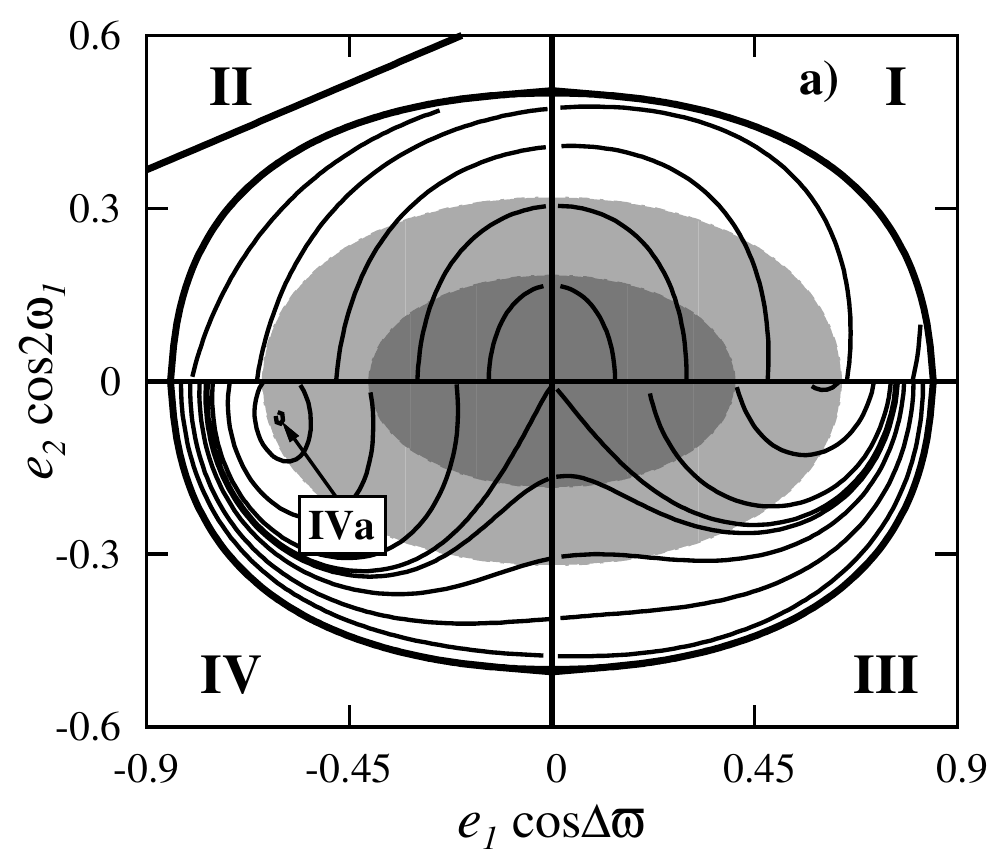}
\includegraphics[width=2.2in]{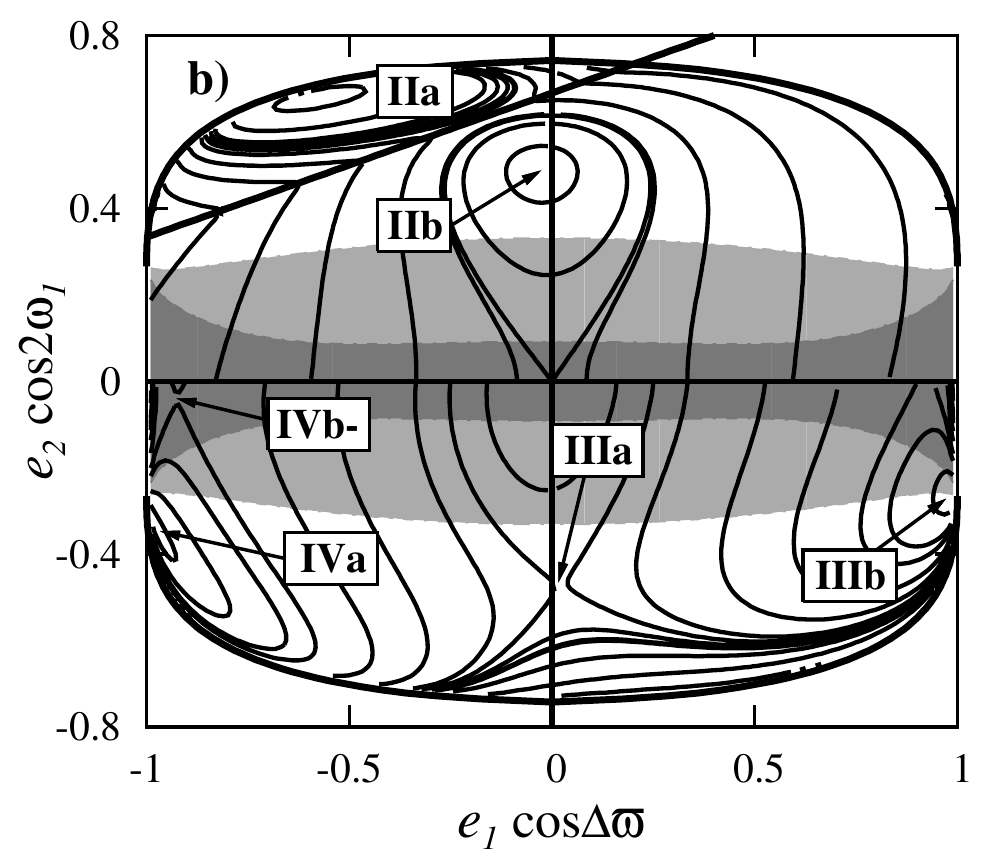}
\includegraphics[width=2.2in]{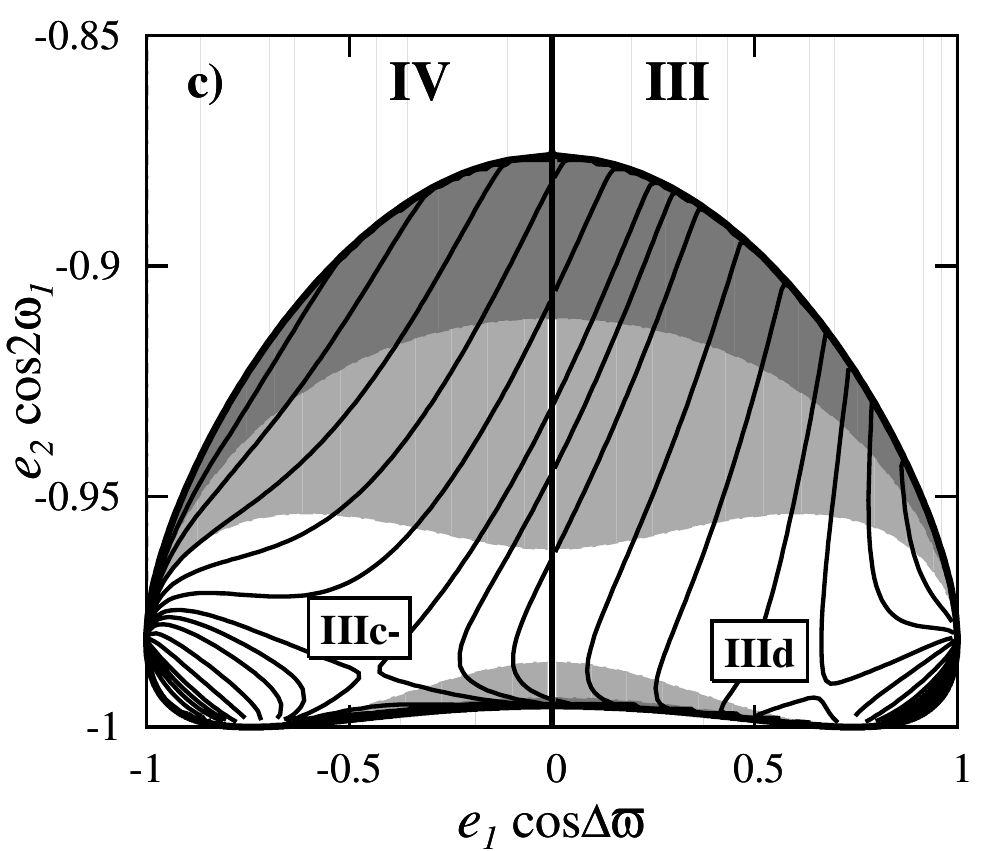}
}
}
}
\caption{
Generic plots of $\Hsec$ in the $\cP_M$-plane of 
$(e_1 \cos{\Delta\varpi},~e_2 \cos{2\omega_1})$ obtained for
$\alpha=0.333$, $\mu=0.5$ and $\nAMD=0.10$ ({\bf a}) and  $\nAMD=0.25$ ({\bf b})
and   $\nAMD=0.85$ ({\bf c}), respectively.  Small Roman numbers
label families  of
stationary solutions identified in this work (see the text for more details).
Solid thick line marks the collision line of orbits defined through $a_1 (1 \mp
e_1)=a_2 (1 - e_2)$.  Shaded regions mark mutual inclinations. Panel ({\bf a})
$[0^{\circ},50^{\circ}]$ ({\em white}), $[50^{\circ},60^{\circ}]$ ({\em light
gray}), and $\ge 60^{\circ}$ ({\em dark gray}), respectively. The limit values
of inclinations for panel ({\bf b}) are, $0^{\circ}$, $95^{\circ}$ and
$105^{\circ}$, respectively; for panel ({\bf c}): $0^{\circ}$, $140^{\circ}$ and
$155^{\circ}$.
}
\label{fig:fig3}
\end{figure*}
Subsequent panels of Fig.~\ref{fig:fig3} show generic views of the 
$\cP_M$-plane derived for different values of  $\nAMD$ integral and the same
primary parameters, ($\alpha,\mu$). In particular, these plots
are drawn for $\Hsec$ levels which are found  numerically as solutions to
$\cH_{sec}-E_0=0$, where $E_0$ is a fixed value,  for the following values of
the semi-major axes and masses ratios: $\alpha=a_1/a_2=0.333$,
$\mu=m_1/m_2=0.5$, and  $\nAMD=0.10$  (the left-hand panel),  $\nAMD=0.25$  (the
middle panel), and $\nAMD=0.85$ (the right-hand panel), respectively. According
with the general construction of the \RP{}-plane,  it is divided by four
quadrants and, for a reference, labeled with Roman numbers in the left-hand panel
of Fig.~\ref{fig:fig3}: quadrant~I  ($\Delta\varpi=0$, $\omega_1=0$),  
quadrant~II  ($\Delta\varpi=\pi$, $\omega_1=0$),   quadrant~III
($\Delta\varpi=0$, $\omega_1=\pi/2$),   and quadrant~IV  ($\Delta\varpi=\pi$,
$\omega_1=\pi/2$). 

Alternatively, we also use another definition of the \RP{}-plane:
\begin{equation}
\cP_S = \{ e_1 \sin{\omega_1} \times e_2 \sin{\omega_2}:
\omega_1,\omega_2=\pm\pi/2; e_{1,2} \in [0,1)\},
\label{plane2}
\end{equation}
\citep[see][]{Libert2007b}. The $\cP_S$-plane helps to avoid a discontinuity of the levels
of $\Hsec$ at the $x$-axis. In fact, that plane carries out  the same
information as the negative ($y<0$) part of the $\cP_M$-plane.  Obviously, pairs of
angles of the $\cP_S$ representation:
($\omega_1=+\pi/2,\omega_2=-\pi/2$),
($\omega_1=+\pi/2,\omega_2=+\pi/2$),
($\omega_1=-\pi/2,\omega_2=-\pi/2$),
($\omega_1=-\pi/2,\omega_2=+\pi/2$),
correspond to the following pairs of angles in the $\cP_M$ representation:
($\Delta\varpi=0,2\omega_1=\pi$),
($\Delta\varpi=\pi,2\omega_1=\pi$),
($\Delta\varpi=\pi,2\omega_1=-\pi$ ),
($\Delta\varpi=0,2\omega_1=-\pi$).
Hence, two bottom quadrants of the $\cP_S$-plane are equivalent to quadrants~IV
and III  of the $\cP_M$-plane. Two upper quadrants of the $\cP_S$-plane are 
their central reflections with respect to the origin. It follows from the
definition of coordinate axes through $e_i \sin() $ and $e_j \cos()$ functions
(where $i,j=1,2$).  Apparently, the $\cP_S$-plane contains redundant
information. However, the energy levels are continuous in this plane and
their interpretation is easier than in the $\cP_M$-plane
[see also \citep{Libert2007b}]. The  central projections of quadrants III and IV
can be obtained by reversing signs of $\omega_1$ and $\omega_2$ (or measuring
angles in opposite direction).

We define one more \RP{}-plane, which makes it possible to obtain
a smooth representation of quadrants II and I of the $\cP_M$-plane:
\begin{equation}
\cP_{C} = \{ e_1 \cos{\omega_1} \times e_2 \cos{\omega_2}:
\omega_1,\omega_2=0,\pi; e_{1,2} \in [0,1) \}.
\label{plane3}
\end{equation}
Because we are interested in possibly global  and transparent representation of
the equilibria in the secular problem (see below), we will use not only the
primary notion of the \RP{}-plane by \cite{Michtchenko2006} but also the two
other definitions.

An important observation which is very helpful to justify the choice of the
\RP{}-planes for the search for equilibria, is the symmetry of 
$\Hsec$ with respect to \corr{the characteristic plane}. It can 
be shown as follows. For
the defined above pairs ($\omega_1^0, \omega_2^0$) of the \RP{}-plane:
\begin{equation}
\dt{G_i}\big|_{(\omega_1^0, \omega_2^0)}=
-\frac{\partial{\Hsec}}
     {\partial{\omega_i}}\Big|_{(\omega_1^0, \omega_2^0)}=0.
\label{eq:g}
\end{equation}
Indeed, from the general formulae of the secular
Hamiltonian expressed by Fourier series we have:
\[
{\mathcal{H}_{sec}} = \sum_{k,l,m \in (-\infty,\infty)}  
h_{k,l,m}(a_1,a_2,e_1,e_2,I_1,I_2) \cos \Phi_{k,l,m},
\]
where $k,l,m$ are integers,  $h_{k,l,m}$ are coefficients of the expansion, and
$\Phi_{k,l,m} = k \omega_1 + l \omega_2 + m \Delta\Omega$ is the generic angle
argument of the expansion. (Further, we shall assume that the series
converge). According with the analytic properties of the Fourier expansion, indices $k$ and
$l$ must have the same parity \citep{Brumberg1995,Michtchenko2006}. Also, after
the Jacobi's elimination  of nodes, $\Delta\Omega = \pm \pi$. Now, the
derivatives of ${\mathcal{H}_{sec}}$ over $\omega_i$  (Eq.~\ref{eq:g}) are:
\[
 \dt{G_i} =  \sum_{k,l,m \in (-\infty,\infty)} h_{k,l,m}
 (a_1,a_2,e_1,e_2,I_1,I_2)
 \sin \Phi_{k,l,m} \frac{\partial \Phi_{k,l,m}}{\partial\omega_i},
\]
and because coefficients $h_{k,l,m}$ can be considered as  functionally
independent,  the derivatives may vanish only when all $\sin\Phi_{k,l,m} \equiv
0$. This is only possible when $\Phi_{k,l,m} = n\pi$, $n \in {\mathbb Z}$,
hence, when $k \omega_1 + l \omega_2 = \pm (n-m) \pi$, for any integers $k,l$
of the same parity, and when $\omega_i=\pm\pi/2,0,\pi$. That also means,
that $2\omega_1=0,\pi$ and $\Delta\varpi=\varpi_1-\varpi_2=0,\pi$. 

The zeros of the derivatives of the secular Hamiltonian over $\omega_i$ may be
also deduced geometrically,  relying on the symmetry of interacting mean
orbits. The mean orbits may be understood as material elliptic rings 
(the Gauss approximation), which interact gravitationally. The potential of 
interaction has symmetries with respect to the particular angles $\Delta\varpi,
2\omega_1$ or ($\omega_1,\omega_2$) which define the \RP{}-plane. Points
($G_1^0, G_2^0$) in the \RP{}-plane, fulfilling conditions:
\begin{equation}
\frac{\partial{\mathcal{H}_{sec}}}{\partial{G_i}}
\Big|_{(G_1^0, G_2^0, \omega_1^0, \omega_2^0)}=0,
\label{eq:motion}
\end{equation}
may be identified with stationary solutions of the secular problem. We solve the
above equations with respect to unknown $(G_1^0, G_2^0)$ or, $(e_1^0,e_2^0)$ for
pairs of fixed angles $(\omega_1^0,\omega_2^0)$ in the given quadrants of the
\RP{}-plane and for fixed $C$. Hence, the  notion of the \RP{}-plane is
particularly suitable for the analysis of equilibria.

Figure~\ref{fig:fig3} reveals numerous stationary solutions labeled accordingly
with the quadrant of the $\cP_M$-plane and a letter  labeling a specific type (a
family) of solutions.  The equilibria appear as local extrema (or rather as
elliptic or quasi-elliptic points) or saddle points of $\Hsec$ in the
\RP{}-plane. At these critical points, the derivatives with respect to all phase
variables must be equal to zero. After fixing the ($\alpha$, $\mu$)-pair, $e_1$
and $e_2$ may be varied in ranges permitted by constant $C\equiv J_1$ (or $\AMD$). The
thick curve is for the boundary  of the energy level defined for a given value
of $\AMD$. The eccentricities and mutual inclination are coupled 
again through $J_1$ (or $\AMD$). To indicate boundaries of the
mutual  inclination permitted for a given range of $(e_1,e_2)$, the regions in
which the mutual inclination is grater than a prescribed value are shaded. We
mark a few such shaded regions in the \RP{}-plane (lighter shade means smaller
mutual inclination).  The mutual inclinations at their boundaries are quoted in
the caption to Fig.~\ref{fig:fig3} (also in captions to other plots of the
\RP{}-plane).

To avoid the geometric singularity of the equations of motion at the origin of
the \RP{}-plane  and at the coordinate axes ($x\equiv e_1=0$, $y \equiv e_2=0$),
we follow \cite{Libert2007b}, and introduce the following non-singular, canonical
variables: 
\[
p_i = \sqrt{2 (L_i - G_i)}\, \cos \omega_i, \quad
q_i = \sqrt{2 (L_i - G_i)}\, \sin \omega_i, \quad i=1,2.
\]
We denote $\vx \equiv (p_1,q_1,p_2,q_2)$ from hereafter.
These non-singular variables are convenient for a quasi-global continuation 
of stationary solutions in the \RP{}-plane. 

\begin{figure}
\centerline{
\vbox{
\hbox{
\includegraphics[width=4.20cm]{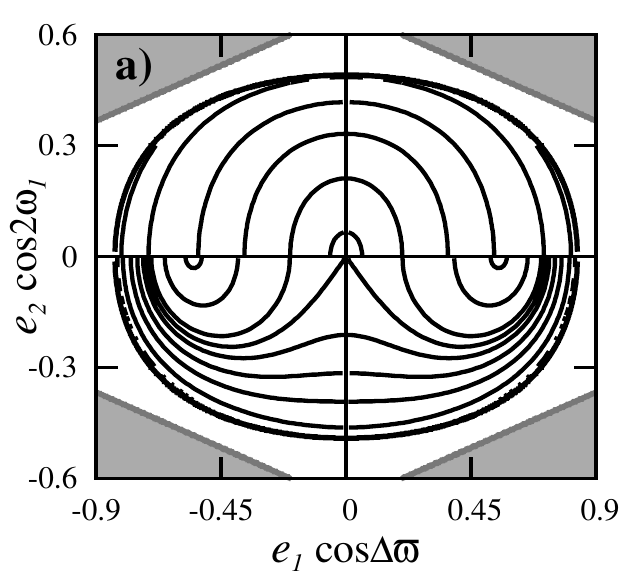}
\includegraphics[width=4.20cm]{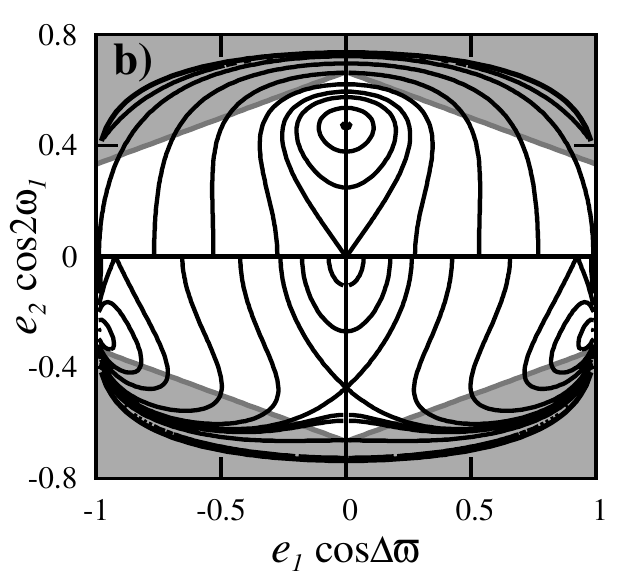}
}
\hbox{
\includegraphics[width=4.20cm]{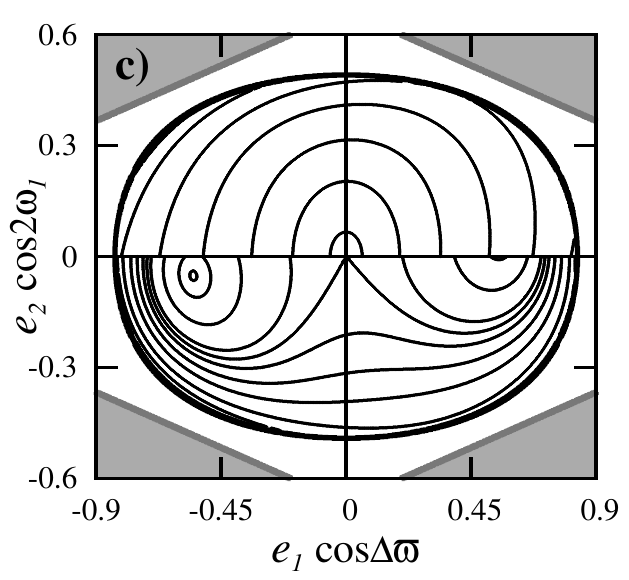}
\includegraphics[width=4.20cm]{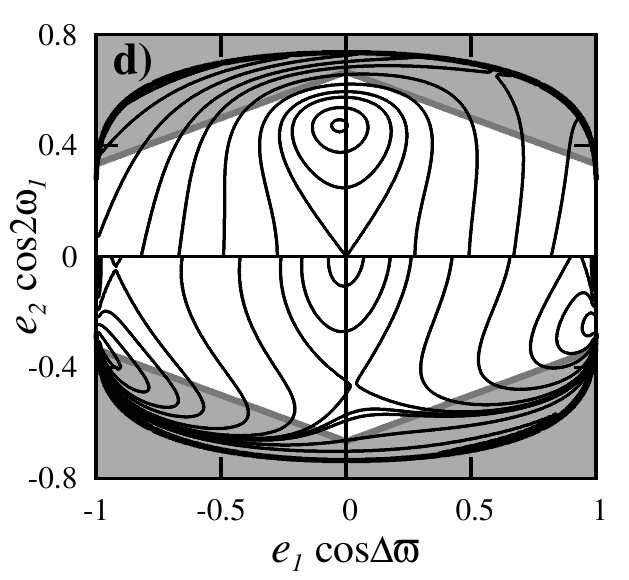}
}
}
}
\caption{
Levels of the secular Hamiltonian in the quadrupole approximation (top panels),
and in the octupole approximation (bottom panels). These plots are obtained for
$\alpha=0.333$, $\mu=0.5$ and $\nAMD=0.1$ (the left-hand panels), and
$\nAMD=0.25$ (the right  hand panels). Compare with the semi-analytic model in
Fig.~\ref{fig:fig3}. The shaded areas mark parameters for which
the expansion of $\Hsec$ in terms of $\alpha$ would diverge. 
}
\label{fig:fig4}
\end{figure}

Finally, to show the relevance of the semi-analytic averaging, we calculated
the energy levels in the \RP{}-plane when only the quadrupole  ($\sim
\alpha^{2}$) and octuple ($\sim \alpha^{3}$) terms of the perturbing
Hamiltonian are accounted for. These terms are averaged analytically. The
results are illustrated in four panels of Fig.~\ref{fig:fig4} which are derived
for the same values of $\alpha=0.333$ and $\mu=0.5$ as in Fig.~\ref{fig:fig3}.
Panels in the top row are for the quadrupole-order secular theory, panels in
the bottom row are for the octupole theory. The left-hand plots are for
$\nAMD=0.1$, the right-hand plots are for $\nAMD=0.25$.  Shaded areas mark
regions of the parameter plane which lie beyond the limit of convergence of the
expansion of $\Hsec$ in $\alpha$, and obviously we cannot obtain there  a
proper representation of equilibria solutions.  The quadrupole term leads to
exactly symmetric view of the \RP{}-plane --- in fact, the quadrupole
Hamiltonian does not depend on $\Delta\varpi$. The octupole approximation fits
much better to the semi-analytic secular model (compare with
Fig.~\ref{fig:fig3}a,b), nevertheless the energy levels are still significantly
distorted and some features are missing at all; for instance, there is no
quasi-elliptic point over the collision line in quadrant~II (see
Fig.~\ref{fig:fig3}b); instead, we may found {\em a false} saddle solution
close to the border in quadrant~I. Although the tested configuration has
relatively small $\alpha=0.333$, in such a case both analytic approximations of
$\Hsec$ introduce artifacts which can be only avoided by an application of the
semi-analytic averaging.
%
\subsection{Lyapunov stability and critical inclinations}
\label{sec:stability}
The stability of equilibria may be investigated with the help of  Lyapunov
theorem \citep[see, e.g.,][]{Markeev1978,Khalil2001}.  If the Hamiltonian is
positive (negative) definite function in a neighborhood of an equilibrium ${\bf
x}_0$, then the equilibrium is Lyapunov stable. At a stable equilibrium, the
parameters of the averaged system are constant, hence the orbital elements do
not change in the secular-time scale, and the  orbital conïfiguration evolves
in the short-time scale only. In the 3-D secular problem, that is equivalent to
conditions for local extrema of $\Hsec$ in the phase space. We recall that 
such extrema  must appear as elliptic points in  2-D plots of the
\RP{}-plane. We should also remember that   these quasi-elliptic points may be
in fact related to saddles in  two remaining and ``hidden'' dimensions of the
phase space. To determine, whether the secular Hamiltonian is sign definite
function of the phase variables in the neighborhood of a critical point, we
compute its Hessian, $\mH_2 = \partial^2\,\Hsec/\partial\,\vx^2$ at the
equilibrium, and then we determine whether it is sign-definite matrix.  

\begin{figure}
\centering
\vbox{
 \hbox{\includegraphics[width=3.32in]{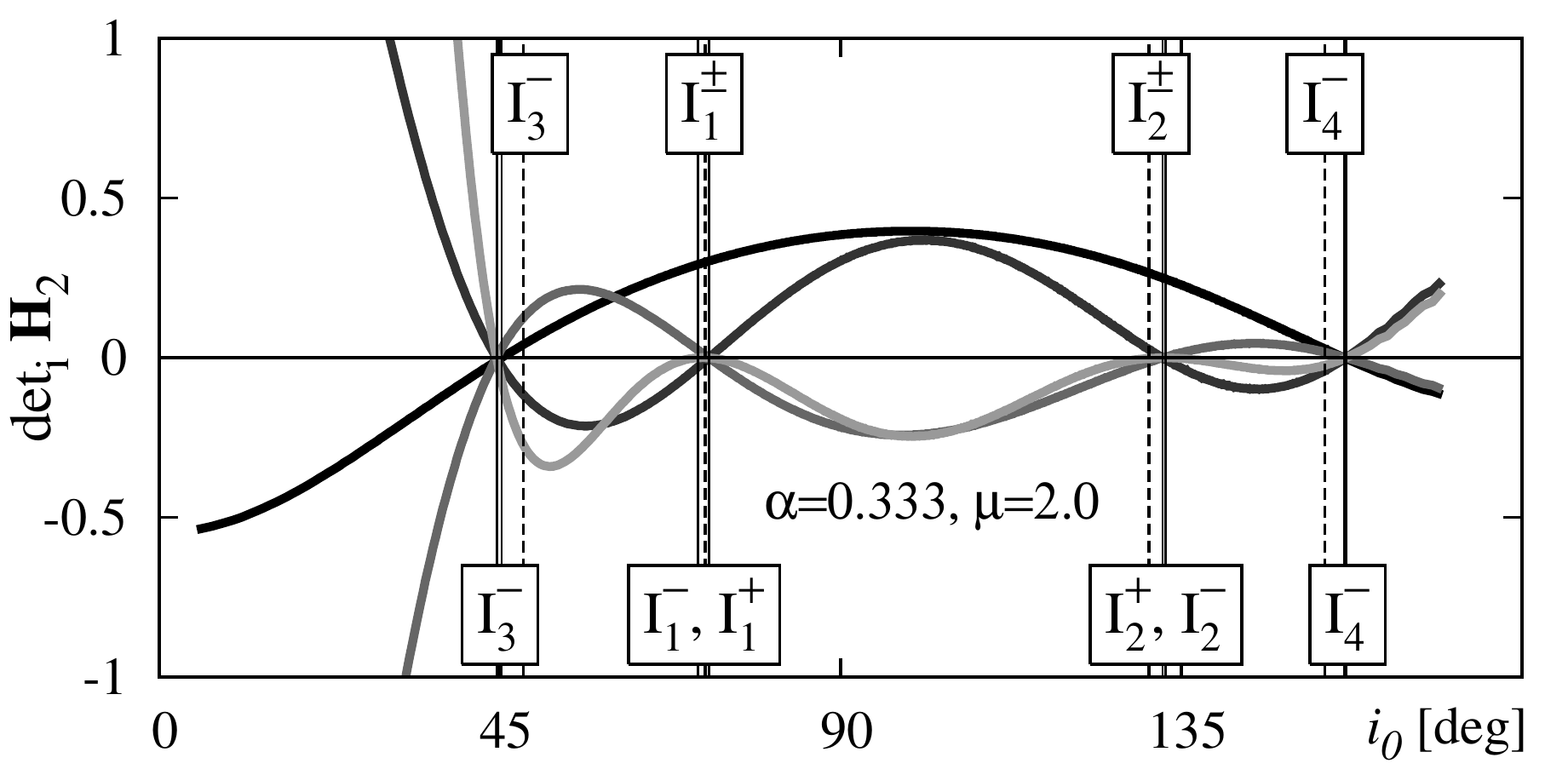}}
 \hbox{\includegraphics[width=3.32in]{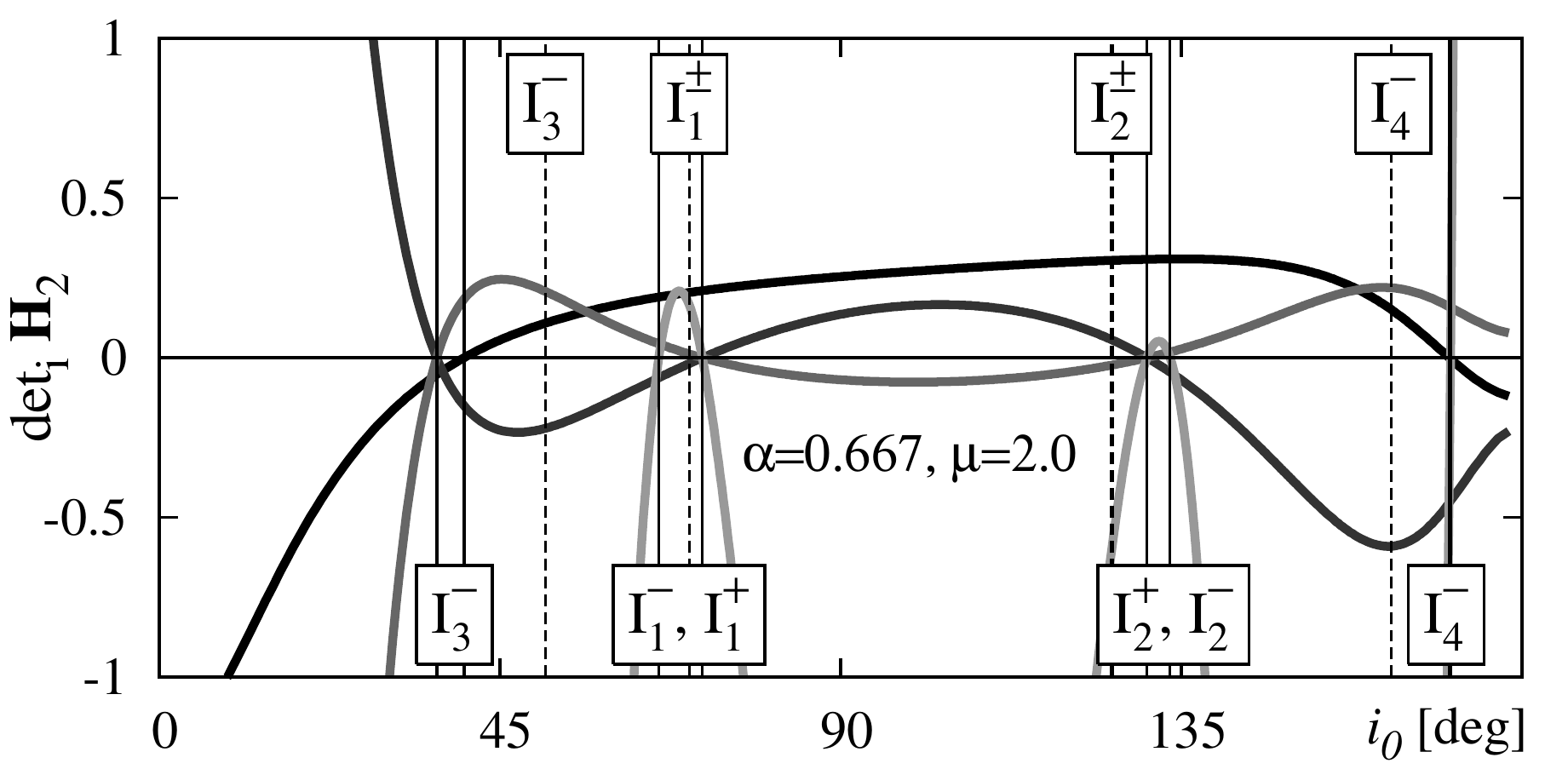}}
}
\caption{
The sub-determinants of the Hessian of the secular Hamiltonian, $\det_1 \mH_2$,
$\det_2 \mH_2$, $\det_3 \mH_2$ and $\det_4 \mH_2$  evaluated at the origin of
the \RP{}-plane. The sub-determinants are plotted with gradually shaded curves
as functions of $\nAMD$ or $i_0$, starting with the black curve for $\det_1
\mH_2$, and lightest gray curve for $\det_4 \mH_2$. The  sub-determinants are
expressed in relative units. The vertical lines mark the bifurcational
mutual inclinations of orbits at the origin ($i_0$)  that  correspond to $\det_i \mH_2 =
0$ for one (or more) $i=1,2,3,4$.  These inclinations are labeled with
$I^{+,-}_{1,2,3,4}$.  Thin, dashed, vertical lines labeled at the top are for
the quadrupole order theory, and the thin solid lines labeled at the bottom are for
the semi-analytic theory. See the text for more details.
}
\label{fig:fig5}
\end{figure}

As an illustration, we show two plots of the sub-determinants of Hessian
$\mH_2$  in Fig.~\ref{fig:fig5} which are computed  at the origin  as functions
of the  mutual inclination of circular orbits, $i_0$. The top-panel is for
$\alpha=0.333$, $\mu=2$ and the bottom panel is for $\alpha=0.667$, $\mu=2$. 
For some particular values of $i_0$, which are related to the given value of
$\nAMD$, the sub-determinants may vanish  and then we cannot determine
sign-definiteness of the Hessian.   For instance, if $\alpha=0.333$  then it
takes place for $i_0$ close to $\sim 45^{\circ}$, $70^{\circ}$, $130^{\circ}$,
and $155^{\circ}$, respectively.  

In fact, these values of $\Izero$  are related to {\em bifurcational
inclinations} of the ``trivial'' equilibrium  at the origin 
\citep{Krasinsky1972,Krasinsky1974} and changes of its stability and global
topology of $\Hsec$. The later work gives explicitly their values in terms of
parameter $\beta \equiv L_1/L_2 \sim \mu \sqrt{\alpha}$ which were calculated
for the second order secular Hamiltonian (the quadrupole term). For a
reference, the vertical lines in Fig.~\ref{fig:fig5} mark the bifurcations
derived with the quasi-analytic theory (thin lines), and with the quadrupole
Hamiltonian (thick, dashed lines). Bifurcational values of $i_0$ are labeled
with $I^{+,-}_{1,2,3,4}$. Following  terminology of \cite{Krasinsky1974}, the
``+'' sign means that the bifurcation of the origin leads to {\em nontrivial
solution of the positive type}, the ``--'' sign means {\em nontrivial solution
of the negative type}. The positive type solutions are characterized by
$\omega_{1,2}=0,\pi$,  hence  bifurcations take place in the $\cP_C$-plane; 
the negative type equilibria ($\omega_{1,2}=\pm \pi/2$) appear after
bifurcations in the $\cP_S$-plane. An inspection of Fig.~\ref{fig:fig5}
reveals, that the bifurcational inclinations may be very different in both
theories, and it may be particularly well seen for $\alpha=0.667$ (bottom panel
of Fig.~\ref{fig:fig5}). In the later case, the bifurcational values of
inclination are clearly splitted, and the bifurcations takes place for {\em
different} $\nAMD$ (or $J_1\equiv C$).  Note that they only depend on $\beta$ in the
quadrupole theory, and on $\mu$ and $\alpha$  separately in the full model.
Actually,  angles $I_{1}^{+} =I_{1}^{-}$  and $I_{2}^{-} =I_{2}^{+}$ are
degenerated in the quadrupole theory (note that the octupole theory breaks the
symmetry).  All that means that the topology of the phase space must be
different in the two secular theories.

When the Hamiltonian  evaluated at a critical point is not a sign definite
function then the analysis of stability become much more difficult than in the
case of an extremum. In general, only the linear stability of the equilibrium
can be determined relatively easy. We accomplish that by solving the
eigenproblem of matrix $\mA$ of the linearized equations of motion. The
variational equations in terms of new canonical variables $\vy$, where $\vx =
\vx_0 + \vy$:
\[
 {\dt{\vy}} = \mA\,\vy, \quad \mA = \mI\,\mH_2(\vx_0),
\]
and $\mI$ is the symplectic unit. In general,
for a conservative Hamiltonian system,
$\mA$ has complex eigenvalues 
\[
\lambda_i = \pm \rho_i \pm \mbox{i}\sigma_i, \quad
\rho_i,\sigma_i \in \mR, \quad \rho_i \ge 0,\quad\sigma_i \ge 0, \quad i=1,2.
\] 
We can find them easily as the roots of symmetric characteristic polynomial
$p(\lambda) = \det(\mA - \lambda \mE) =0$,
where $\mE$ is the unit matrix. It is well known that the necessary
and sufficient condition for the linear stability is fulfilled if $\lambda_i =
\pm \mbox{i}\sigma_i$ are purely imaginary and matrix $\mA$ is diagonalizable
($\sigma_i$ are the characteristic frequencies).

In the case of two-degree of freedom conservative Hamiltonian systems, we can
apply the theorem of Arnold--Moser \citep[e.g.,][]{Meyer1986} to conclude that
equilibria which are linearly stable are generically Lyapunov stable. However,
there is no such implication if the characteristic frequencies are involved in
resonances up to the 4-th order, i.e., when $p\sigma_1+q\sigma_2=0$ for
$0<|p|+|q| \le 4$, with $p, q \in \mathbb{Z}$, or when coefficients of the
Birkhoff's normal form of the Hamiltonian expanded near the equilibrium fulfill
a particular condition  involving $\sigma_i$ [see \citep{Meyer1986} or
\citep{Markeev1978}  for details]. In resonant cases, we should examine each
particular  normal form of the polynomial expansion of the Hamiltonian with
respect to variations $\vy$. This can be done with the help of constructive
theorems by Markeev and Sokolskii [see, e.g., \cite{Markeev1978,Sokolskii1975}
or \cite{Gozdziewski2003c} for an example application of these theorems, and
references therein]. Moreover, because high-order expansions are required, 
such an extensive study is hardly possible because we must average out $\Hsec$
{\em and} calculate its derivatives numerically. A precise enough determination
of the  second order derivatives becomes very difficult.  Hence, we are forced
to limit the stability analysis to the linear, non-resonant case. Nevertheless,
recalling the implications of the Arnold-Moser theorem, a study of the linear
stability provides valuable information on the generic Lyapunov stability. 
%
\subsection{A general view of the \RP{}-plane}
%
While in Sect.~4 we describe the results regarding new families of stationary
solutions found in this paper in some systematic way, here we refer to generic
properties of the \RP{}-plane identified through many numerical experiments.
Fixing $\nAMD$, we can obtain typical views of the \RP{}-plane  which are shown
in three panels of Fig.~\ref{fig:fig3}. The left panel of Fig.~\ref{fig:fig3}
illustrates configurations recently investigated  in \cite{Michtchenko2006} and
\cite{Libert2007b}. We can see a maximum of $\Hsec$ at quadrant~IV 
($\omega_1=\pi/2$, $\omega_2=\pi/2$) of the \RP{}-plane. It corresponds to
equilibrium marked with IVa and  known  as the Lidov-Kozai resonance,   with
the analogy to the restricted problem \citep{Lidov1961,Kozai1962}.   In the
vicinity of equilibrium IVa of the non-restricted problem, angles $\omega_1$
and $\omega_2$ librate around $\pi/2$. Simultaneously, these librations of
$\omega_{1,2}$ are related to   large-amplitude, anti-phase variations of the
eccentricity of the inner orbit and of the mutual inclination.
This mechanism may lead to strong instability. We observed it already in the
case of hierarchical two-planet configurations \citep{Gozdziewski2004}. 

\begin{figure*}
\centerline{
\hbox{
\hskip+1mm\includegraphics[width=4.30cm]{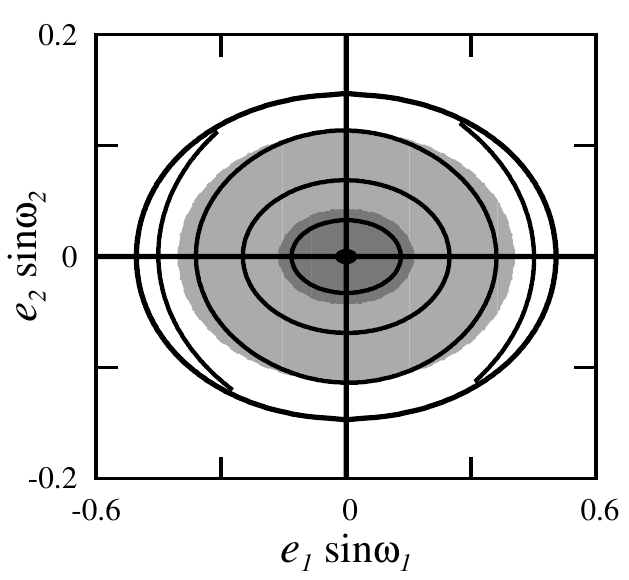}
\hskip+1mm\includegraphics[width=4.30cm]{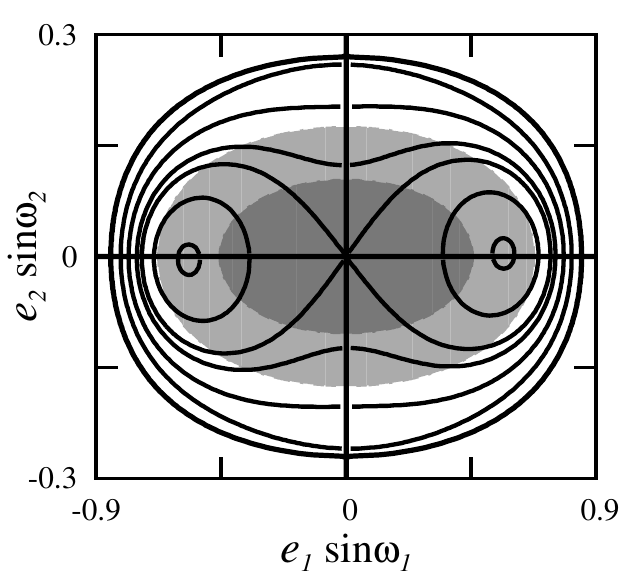}
\hskip+1mm\includegraphics[width=4.30cm]{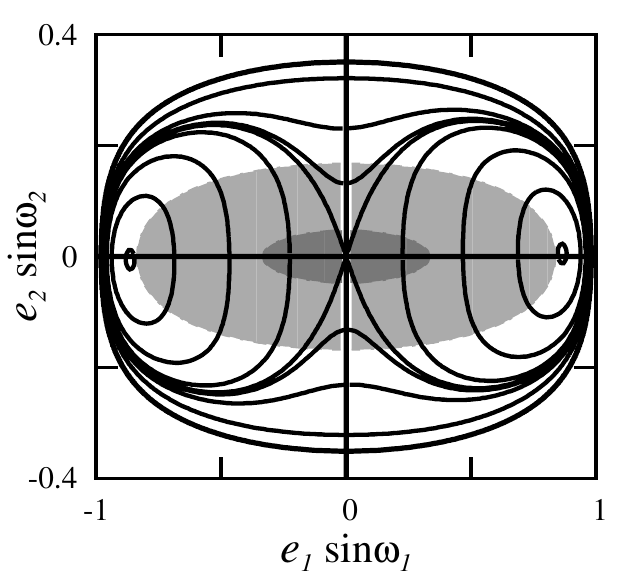}
\hskip+1mm\includegraphics[width=4.30cm]{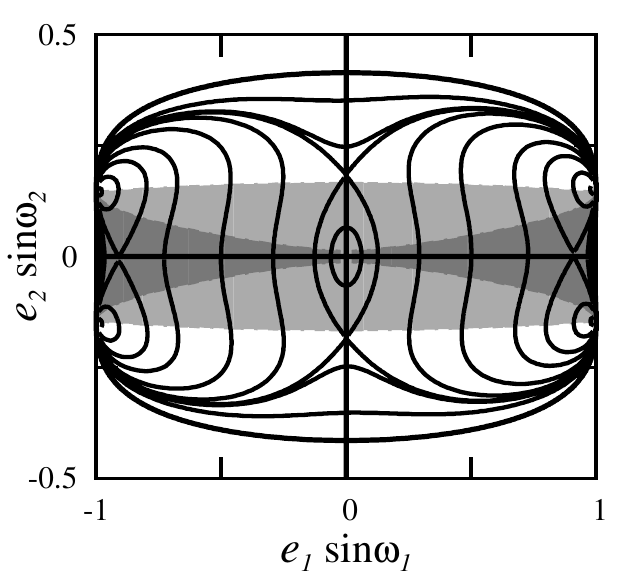}
}
}
\caption{
Levels of the secular energy ($\Hsec$) at the $\cP_S$-plane defined through  
$(e_1 \sin{\omega_1},~e_2 \sin{\omega_2})$ with $\omega_1,\omega_2=\pm\pi/2$,
for $\alpha=0.1$, $\mu=0.25$ and  $\nAMD$ are (from the left-hand panel to
right-hand panel):  $0.01, 0.03, 0.06, 0.08$, respectively.  Shaded areas
indicate ranges of the mutual inclination: 
$(0^{\circ},20^{\circ},30^{\circ})$ for the left-hand panel, 
$(0^{\circ},50^{\circ}, 60^{\circ})$ and $(0^{\circ},70^{\circ}, 80^{\circ})$
for the inner panels, respectively, and $(0^{\circ},95^{\circ}, 105^{\circ})$
for the right-hand panel.
}
\label{fig:fig6}
\end{figure*}

Due to  discontinuity of the $\cP_M$-plane at the $x$-axis,  it is difficult
to follow the evolution of geometric structure of the L-K resonance.
Instead, the $\cP_S$ and $\cP_{C}$ representative planes are more convenient
for that purpose, particularly near the origin.  A sequence of plots shown in
Fig.~\ref{fig:fig6}, reproduces the analytical results of \cite{Libert2007b}
which were obtained for $\alpha=0.1$ and $\mu=0.25$. For $\nAMD=0.01$ (the
left-hand panel of Fig.~\ref{fig:fig6}), the origin is stable, permitting
mutual inclination of circular orbits $i_0 \sim 30^{\circ}$. With increasing
$\nAMD=0.03$,   the inclination grows, and for $i_0 \sim 43^{\circ}$, the
stable stationary point become unstable and bifurcates. Three new solutions
appear: one is unstable and two are stable. This phenomenon may be called the
L-K bifurcation. At the bifurcation point, some sub-determinants of $\mH_2$ are
equal to zero and the stability cannot be determined (see Fig.~\ref{fig:fig5}
and the previous Section for details).  With further increase of $\nAMD$, the
L-K resonance centers move toward large values of $e_1$ (see the third panel in
Fig.~\ref{fig:fig6} plotted for $\nAMD=0.06$) and approach $e_1\sim 1$ for
$\nAMD=0.08$ (see the last, fourth panel in Fig.~\ref{fig:fig6}). While we
refer to the analytic work of \citep{Libert2007b}, these authors did not follow
the L-K equilibrium for this value of $\nAMD$. The semi-analytical algorithm
makes it possible to continue the family of L-K solutions up to such a value, for
which we observe new bifurcations of the equilibria. From each bifurcation of
stable L-K equilibrium emerge three new solutions: one linearly stable (a
saddle point in the \RP{}-plane) and two elliptic points. One of them is
Lyapunov stable, the other one is unstable.  The elliptic points approach $e_1
\sim 1$ and moderate $e_2$. The solution at the origin bifurcates the second
time but it remains unstable (note that it appears as an {\em elliptic} point
in the \RP{}-plane) and two unstable equilibria (saddles of $\Hsec$) located at
$(e_1\sim 0,e_2>0)$ also appear.

In the second  plot of the \RP{}-plane  (see the middle panel  of
Fig.~\ref{fig:fig3}), we consider a configuration with $\alpha=0.333$,
$\mu=0.5$  for $\nAMD=0.25$. We can recognize the L-K resonance after a
bifurcation: the bottom-left quadrant of the \RP{}-plane  reveals a local
extremum labeled with IVa and a saddle point IVb-. In remaining three
quadrants, we can find also other new equilibria labeled with IIa, IIb, IIIa,
IIIb, respectively.  Curiously,  the maximum marked with IIa lies {\em beyond
the  geometrical crossing line} of orbits defined implicitly through $a_1 (1
\pm e_1) = a_2 (1-e_2)$.  Finally, in the right-hand panel of
Fig.~\ref{fig:fig3}, we draw the $\cP_M$-plane for large $\nAMD=0.85$ which
lead to a discontinuity of the energy plane in the regime of large $e_2$. In
this case, the inclination reaches very large values.  

These examples indicate that the 3-D problem is much more complex and rich in
dynamical phenomena than the co-planar problem of two planets.  We recall that
in this case \citep{Michtchenko2004}, the phase space of the secular system is
spanned by librations of $\Delta\varpi$ around $0$ (mode~I), librations of
$\Delta\varpi$ around $\pi$ (mode~II), and circulations of $\Delta\varpi$.
There is also possible the so called non-linear secular resonance (the true
secular resonance) which is present in the regime of moderate and large
eccentricities.

Generally, the equilibria are not isolated in the parameter space   of
($\mu,\alpha)$ and the $\AMD$ integral. Yet, as the example of the L-K
resonance demonstrates, the stationary solution may evolve in the parameter
space,   they can bifurcate, and may change their stability. Hence they form
families of solutions and their behavior depends on a complex way on problem
parameters. To investigate these families, we require a continuation method for
determining  bifurcations of the equilibria and their stability.
%
\section{Parametric survey of equilibria}
%
\label{sec:families}
To resolve the families of equilibria, we apply a simple continuation method
with respect to $\nAMD$ as the primary parameter. For fixed parameters $\alpha$
and $\mu$, $\nAMD \in [0,1]$.  We increase this quantity by small steps, and we
compute the secular energy map in the \RP{}-plane.  An inspection of the
characteristic plane makes it possible to detect the origin and development of
basic dynamical structures. In particular, we can determine critical values of
$\nAMD$ for which new equilibria (represented by elliptic or saddle points in
the \RP{}-plane) appear (see, e.g., Fig.~\ref{fig:fig6}). Having an overall
view of the \RP{}-plane, we may follow a given solution along some path in the
parameters space  with the help of a minimization algorithm (see below).

\begin{figure*}
\centerline{
\vbox{
\hbox{
\hskip+1mm\includegraphics[width=4.00cm]{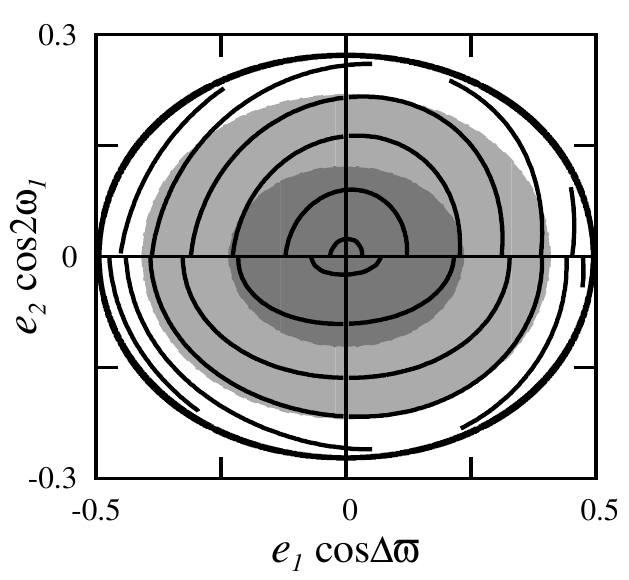}
\hskip+1mm\includegraphics[width=4.00cm]{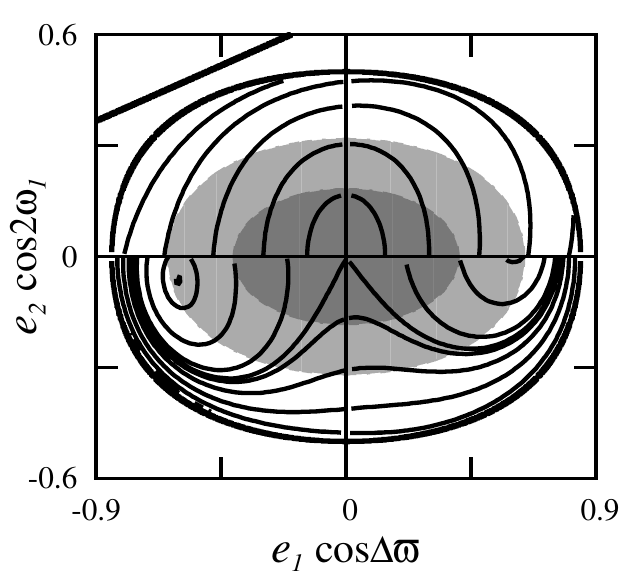}
\hskip+1mm\includegraphics[width=4.00cm]{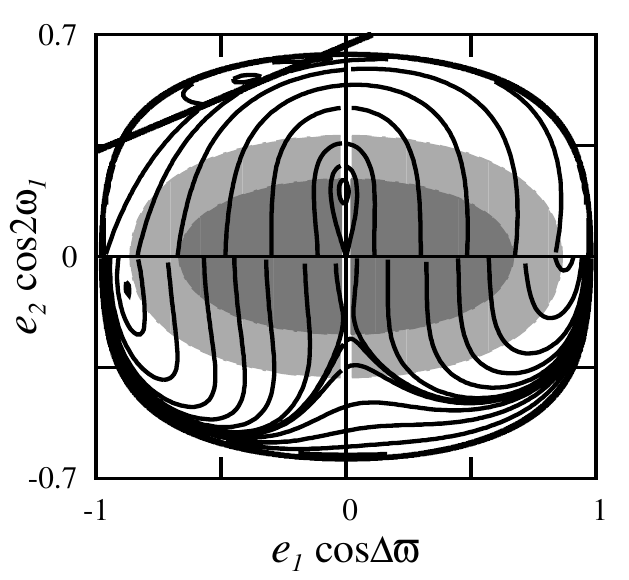}
\hskip+1mm\includegraphics[width=4.00cm]{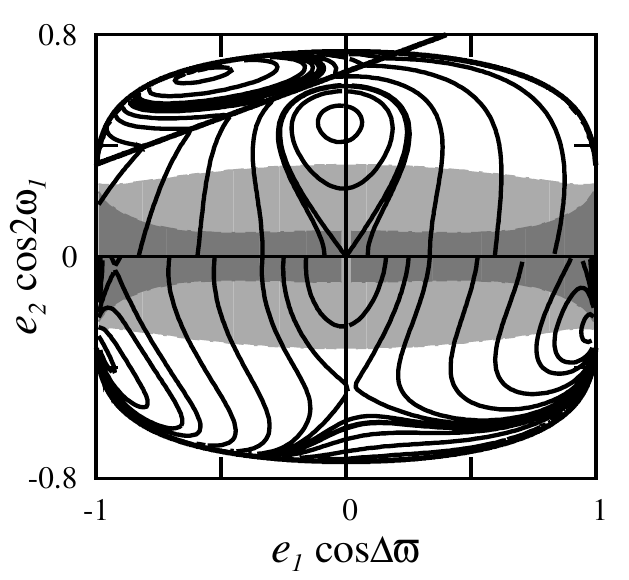}
}
\hbox{
\hskip+1mm\includegraphics[width=4.00cm]{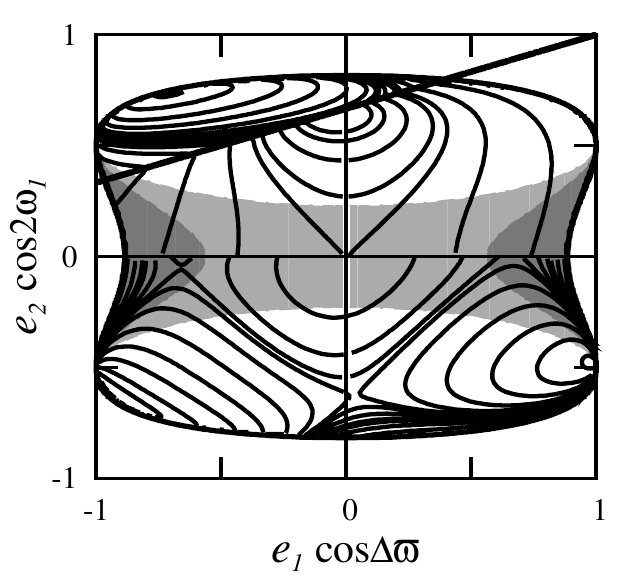}
\hskip+1mm\includegraphics[width=4.00cm]{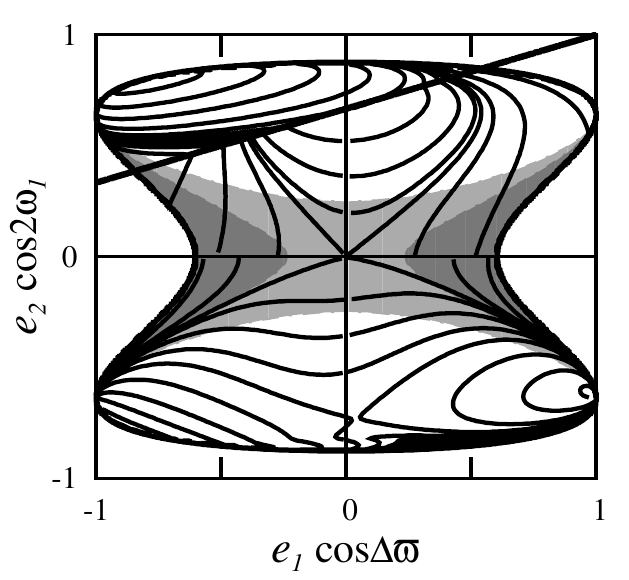}
\hskip+1mm\includegraphics[width=4.00cm]{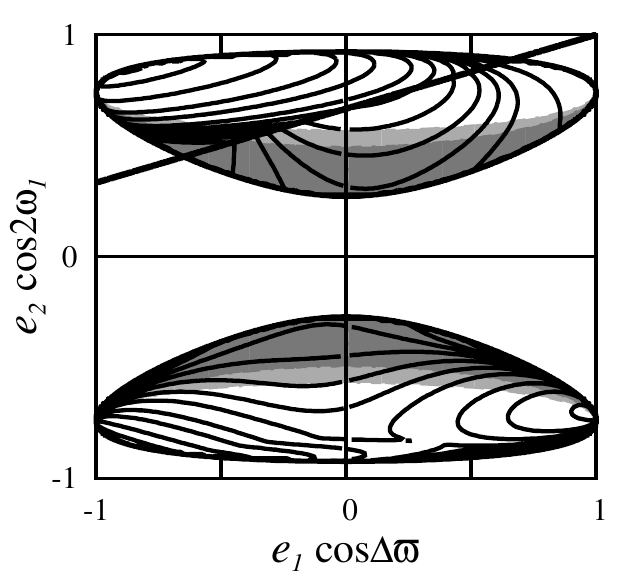}
\hskip+1mm\includegraphics[width=4.00cm]{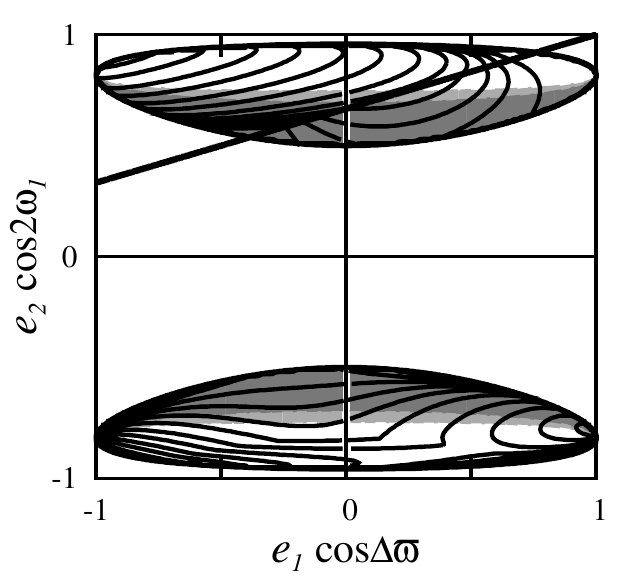}
}
}
}
\caption{
Levels of the secular energy ($\Hsec$) at the $\cP_M$-plane defined through  
$(e_1 \cos{\Delta\varpi},~e_2 \cos{2\omega_1})$ with $\Delta\varpi=0,\pi$,
$2\omega_1=0,\pi$, for $\alpha=0.333$, $\mu=0.5$ and varied $\nAMD$ (or the
mutual inclination of circular orbits, $i_0$). Values of $\nAMD$ are (counting
plots from the left to the right and from the  top to the bottom):  $0.03, 0.1,
0.18, 0.25, 0.33, 0.4, 0.46, 0.50$, respectively. Each panel has two shaded 
regions indicating ranges of the mutual inclinations (corresponding to  the
fixed above of $\nAMD$): $(20^{\circ}, 30^{\circ})$, $(50^{\circ}, 60^{\circ})$,
$(70^{\circ}, 80^{\circ})$,  $(95^{\circ}, 105^{\circ})$, $(120^{\circ},
130^{\circ})$, $(140^{\circ}, 150^{\circ})$, and $(130^{\circ}, 140^{\circ})$,
$(120^{\circ}, 130^{\circ})$, respectively.
}
\label{fig:fig7}
\end{figure*}

\begin{figure*}
\centerline{
\vbox{
\hbox{
\hskip+1mm\includegraphics[width=4.00cm]{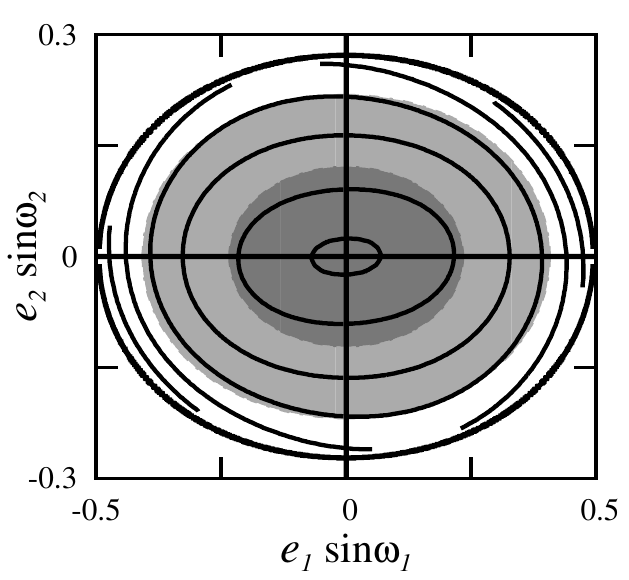}
\hskip+1mm\includegraphics[width=4.00cm]{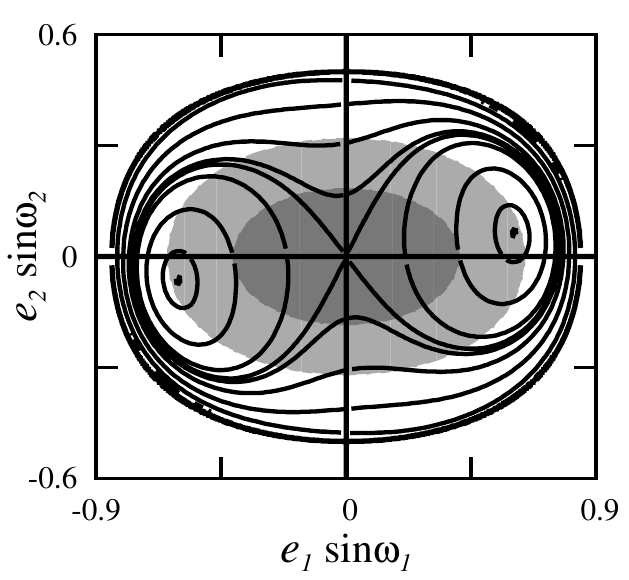}
\hskip+1mm\includegraphics[width=4.00cm]{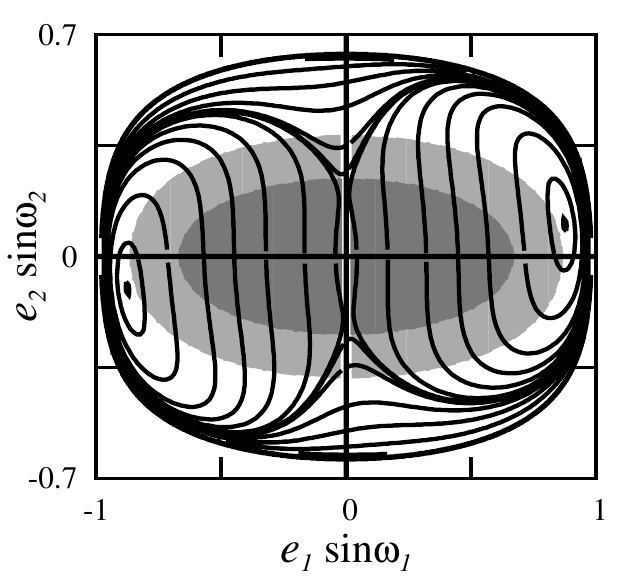}
\hskip+1mm\includegraphics[width=4.00cm]{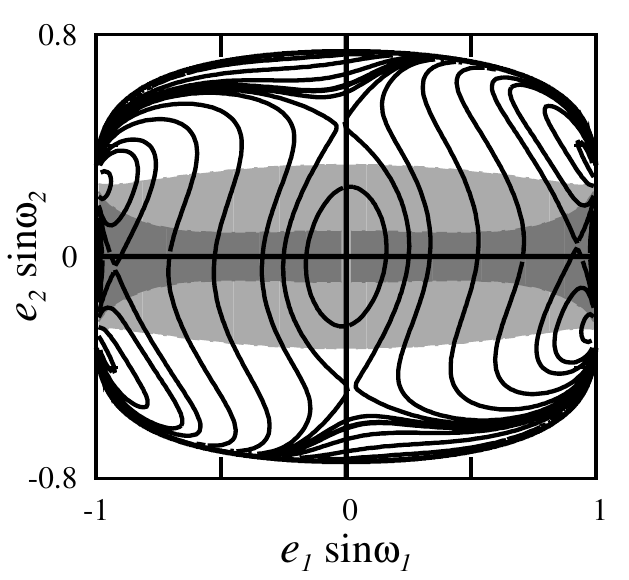}}
\hbox{
\hskip+1mm\includegraphics[width=4.00cm]{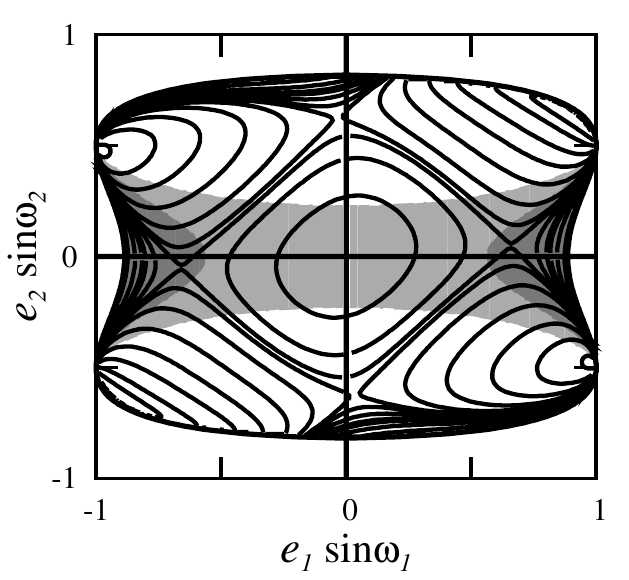}
\hskip+1mm\includegraphics[width=4.00cm]{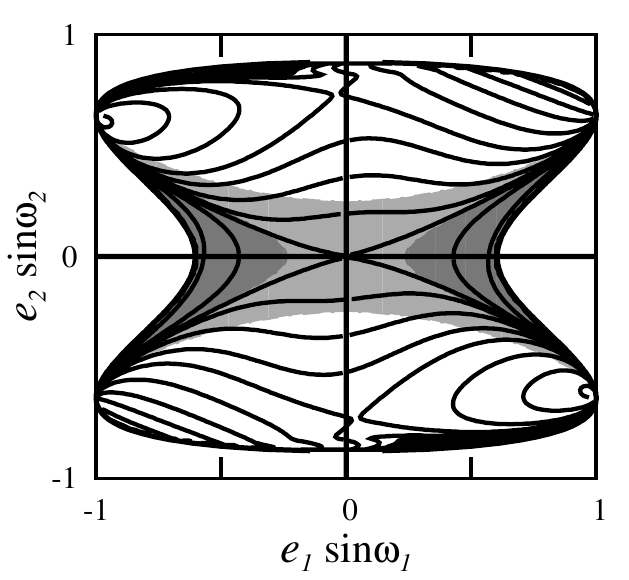}
\hskip+1mm\includegraphics[width=4.00cm]{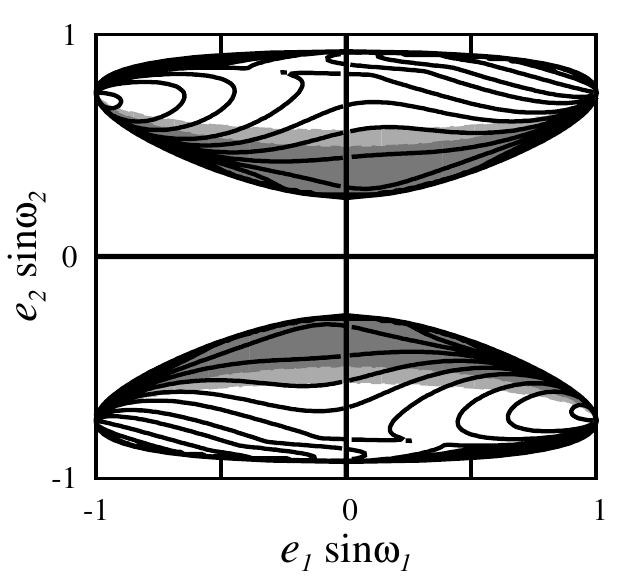}
\hskip+1mm\includegraphics[width=4.00cm]{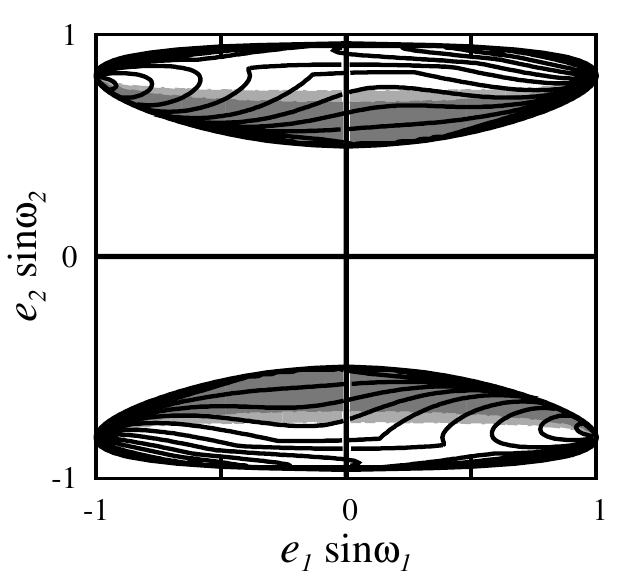}
}
}
}
\caption{
The energy levels plotted in the $\cP_{S}$ representative plane of $(e_1
\sin{\omega_1},~e_2 \sin{\omega_2})$ with $\omega_1,\omega_2=\pm\pi/2$.
Parameters are $\alpha=0.333$, $\mu=0.5$, the values of  $\nAMD$ and inclination
ranges are the same as in Fig.~\ref{fig:fig7}.
}
\label{fig:fig8}
\end{figure*}

\begin{figure*}
\centerline{
\vbox{
\hbox{
\hskip+1mm\includegraphics[width=4.00cm]{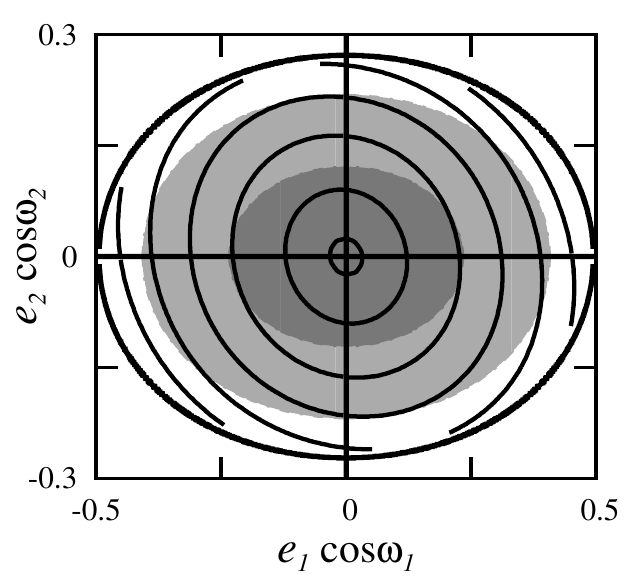}
\hskip+1mm\includegraphics[width=4.00cm]{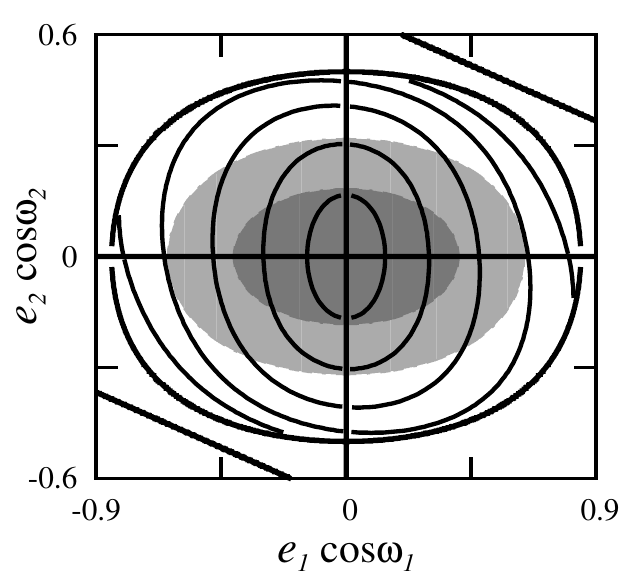}
\hskip+1mm\includegraphics[width=4.00cm]{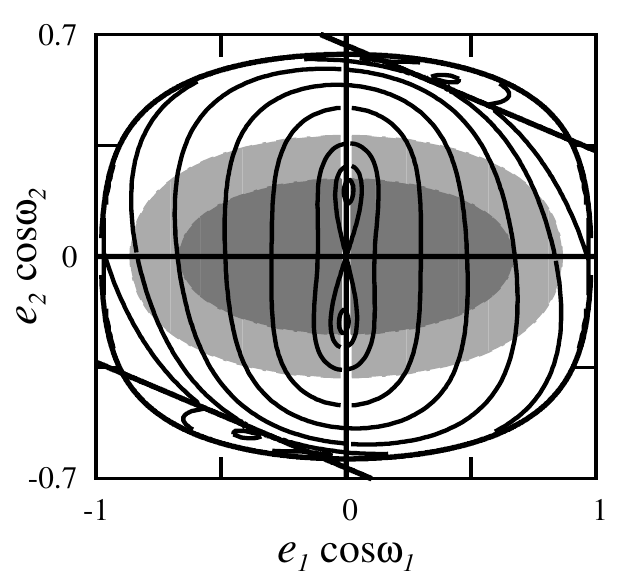}
\hskip+1mm\includegraphics[width=4.00cm]{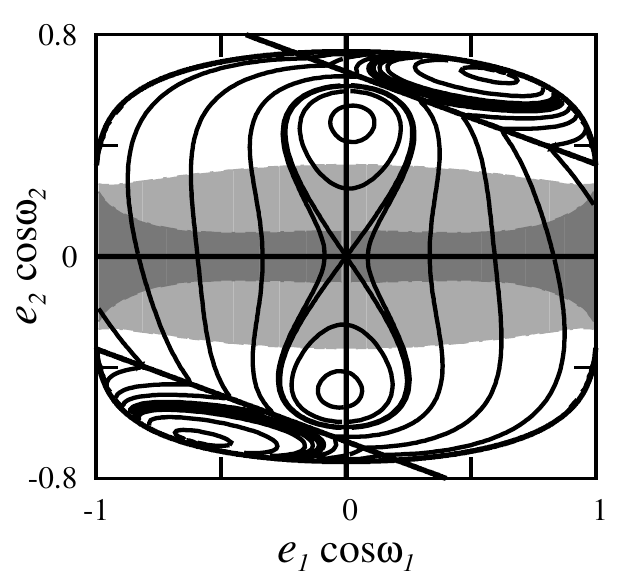}
}
\hbox{
\hskip+1mm\includegraphics[width=4.00cm]{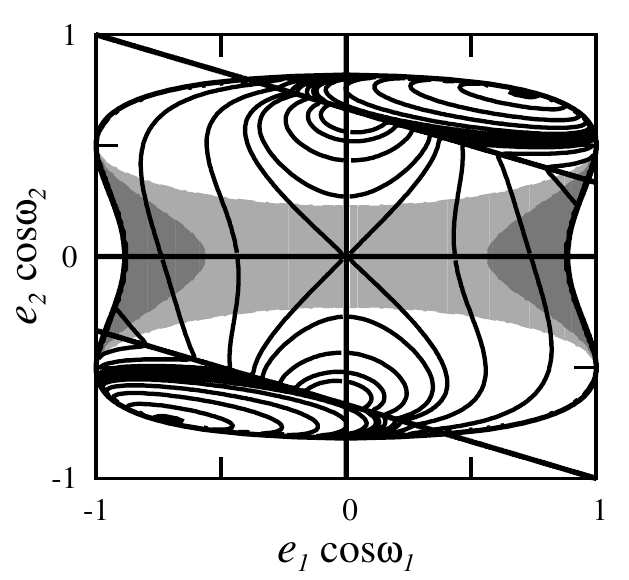}
\hskip+1mm\includegraphics[width=4.00cm]{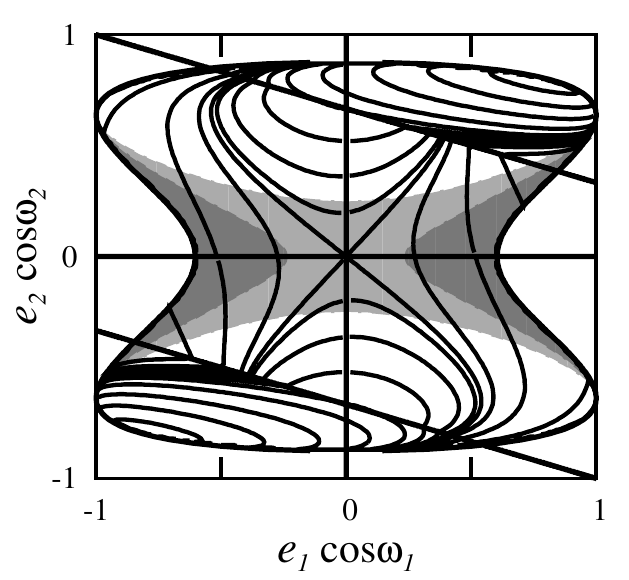}
\hskip+1mm\includegraphics[width=4.00cm]{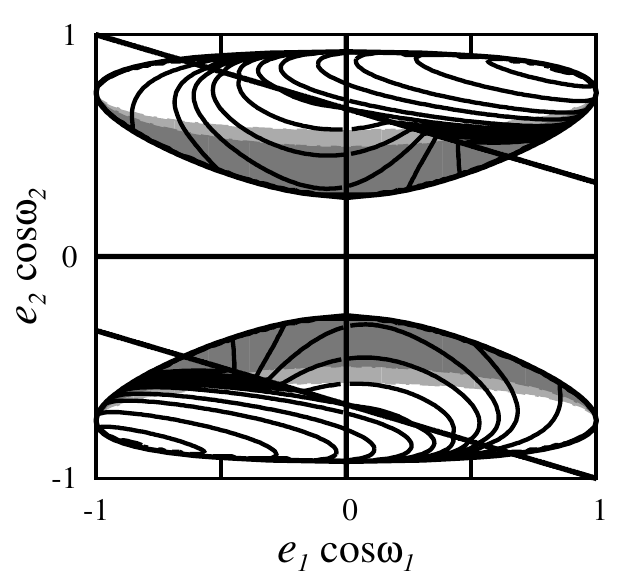}
\hskip+1mm\includegraphics[width=4.00cm]{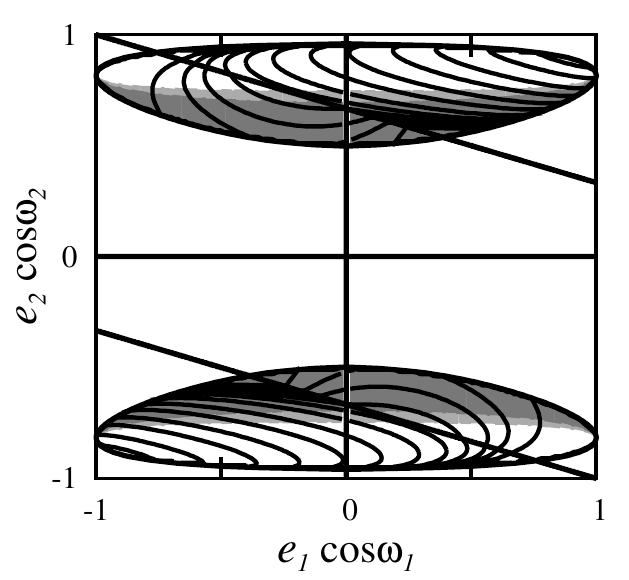}
}
}
}
\caption{
The energy levels plotted in the $\cP_{C}$ representative plane of  $(e_1
\cos{\omega_1},~e_2 \cos{\omega_2})$ with  $\omega_1,\omega_2=0,\pi$. Parameters
are $\alpha=0.333$, $\mu=0.5$, the values of $\nAMD$ and inclination ranges are
the same as in Fig.~\ref{fig:fig7}.
}
\label{fig:fig9}
\end{figure*}

\begin{figure*}
\centerline{
\vbox{
\hbox{
\hskip+1mm\includegraphics[width=4.00cm]{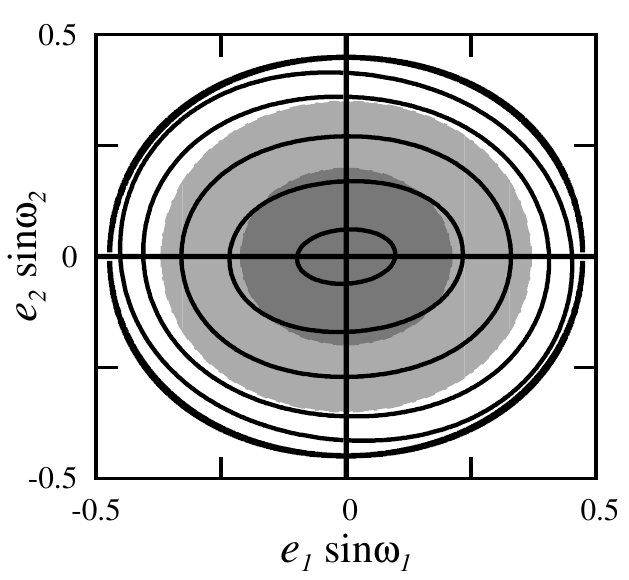}
\hskip+1mm\includegraphics[width=4.00cm]{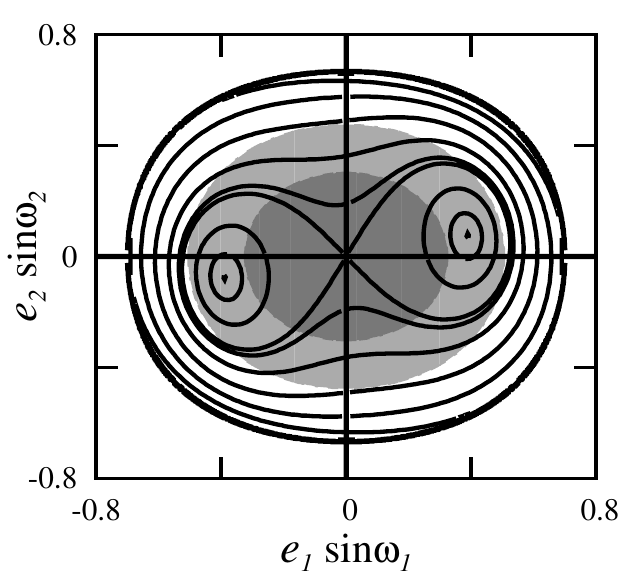}
\hskip+1mm\includegraphics[width=4.00cm]{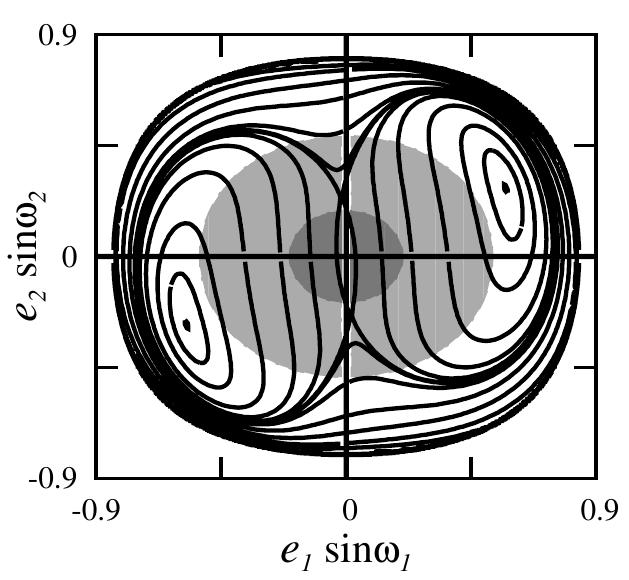}
\hskip+1mm\includegraphics[width=4.00cm]{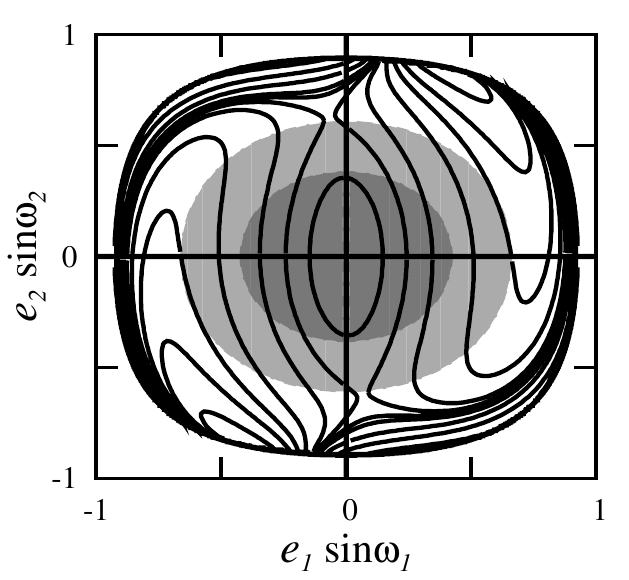}
}
\hbox{
\hskip+1mm\includegraphics[width=4.00cm]{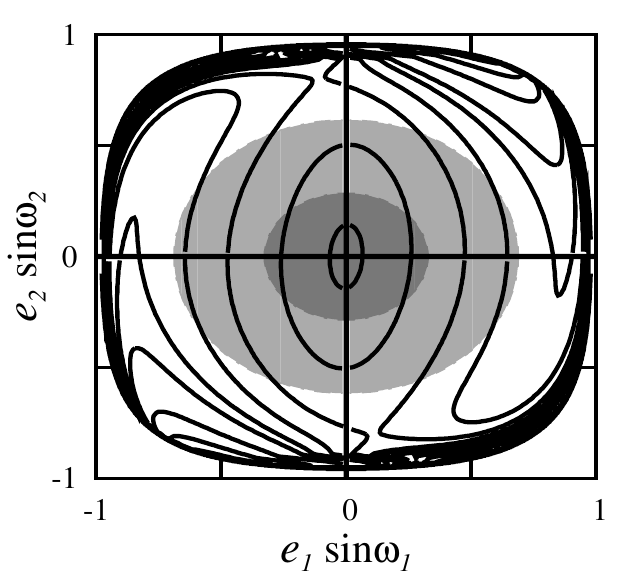}
\hskip+1mm\includegraphics[width=4.00cm]{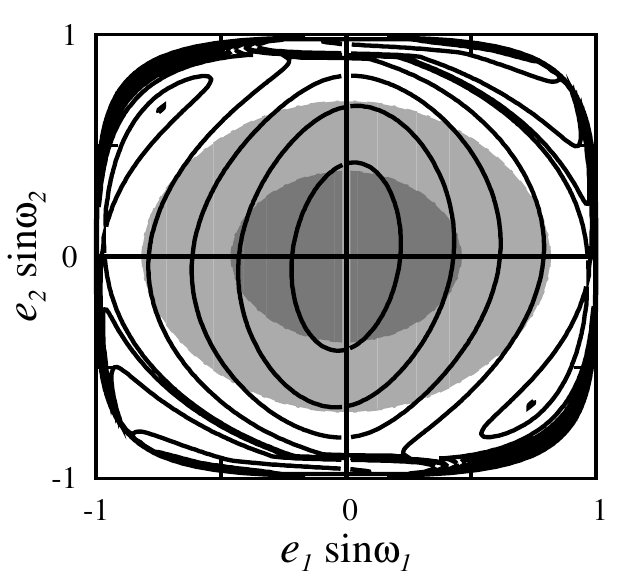}
\hskip+1mm\includegraphics[width=4.00cm]{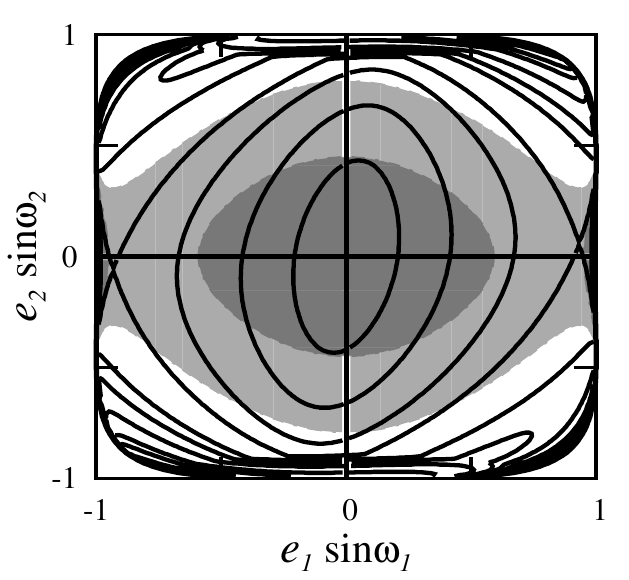}
\hskip+1mm\includegraphics[width=4.00cm]{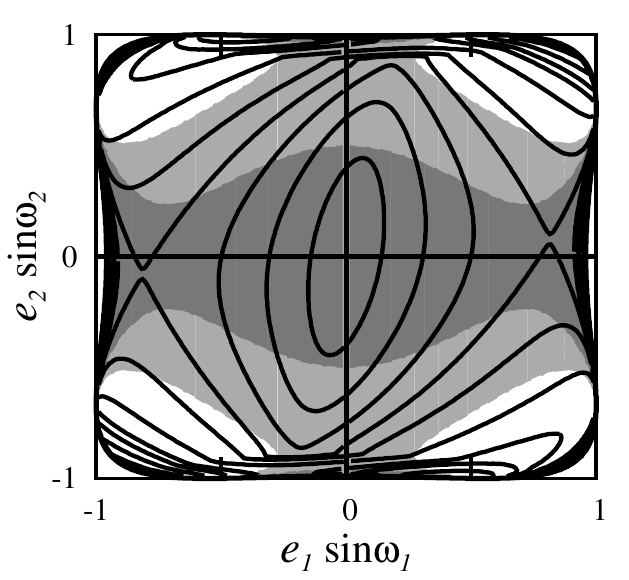}
}
\hbox{
\hskip+1mm\includegraphics[width=4.00cm]{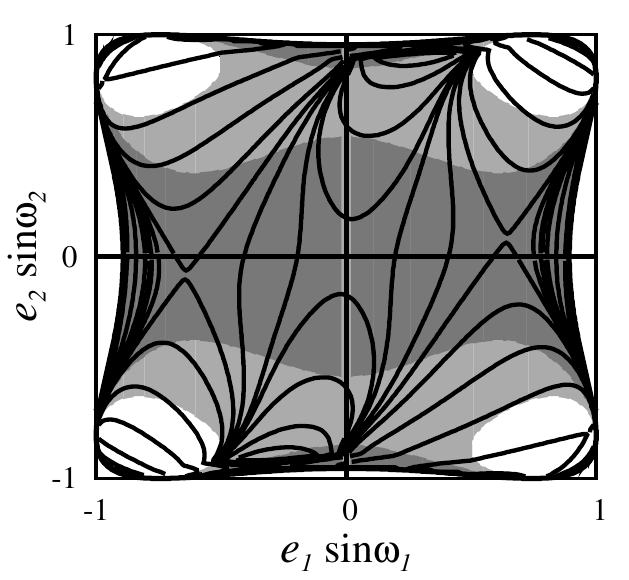}
\hskip+1mm\includegraphics[width=4.00cm]{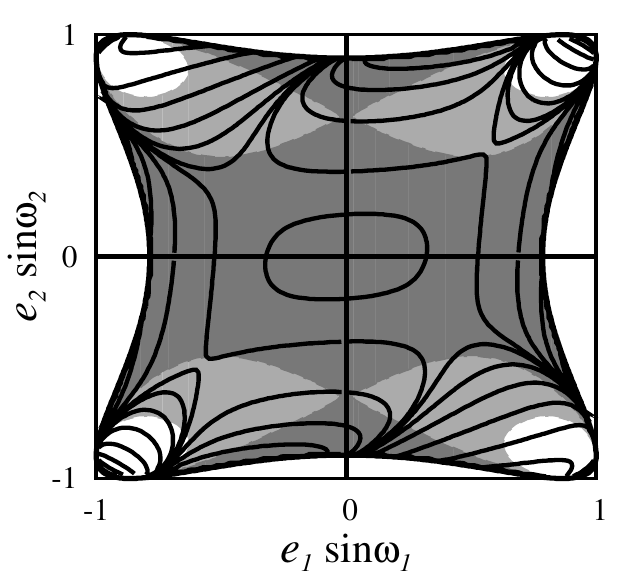}
\hskip+1mm\includegraphics[width=4.00cm]{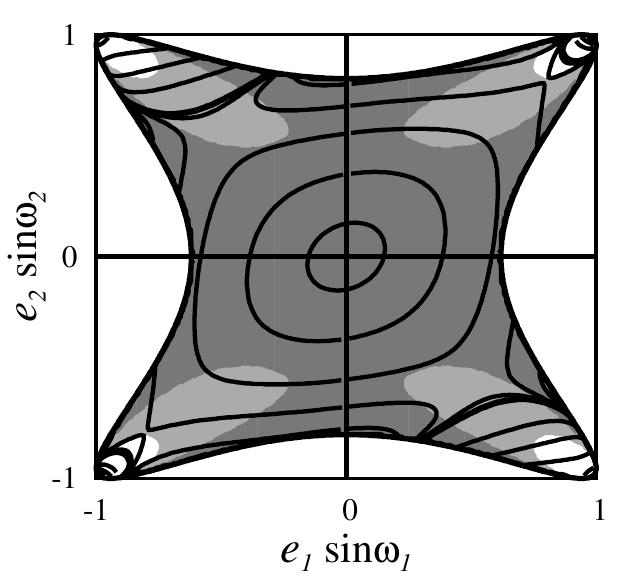}
\hskip+1mm\includegraphics[width=4.00cm]{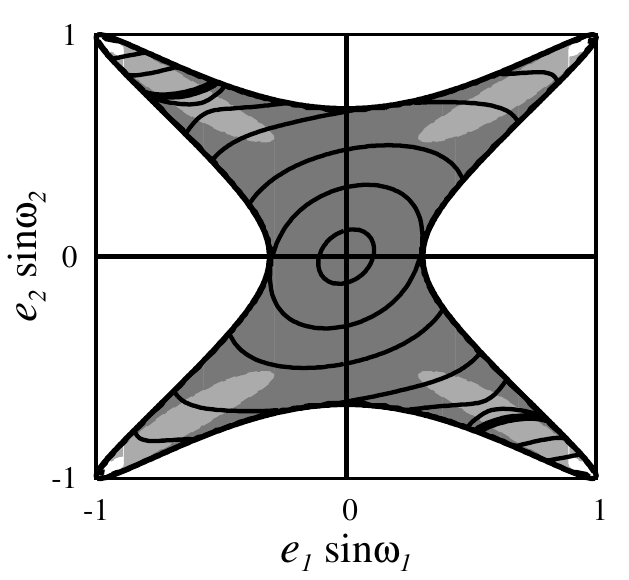}
}
}
}
\caption{
Levels of the secular energy ($\Hsec$) at the $\cP_S$-plane defined through  
$(e_1 \sin{\omega_1},~e_2 \sin{\omega_2})$ with $\omega_1,\omega_2=\pm\pi/2$,
for $\alpha=0.2$ and $\mu=2.0$ and for varied $\nAMD$ values. From {\em the
left} to  {\em the right} and  from {\em the top} to {\em the bottom}: 
$0.06, 0.13, 0.21, 0.29$ (the top row), 
$0.37, 0.45, 0.53, 0.61$ (the middle row), and
$0.69, 0.76, 0.84, 0.92$ (the bottom row), respectively.  
Shaded regions illustrate mutual inclinations in prescribed ranges (lower
inclination --- lighter shade, larger inclination --- darker shade). For each 
fixed value of $\nAMD$, there are two levels of $I_{mut}$ which are marked in 
subsequent panels:
$(25^{\circ}, 35^{\circ})$, $(45^{\circ}, 55^{\circ})$  
$(65^{\circ}, 75^{\circ})$,  $(75^{\circ}, 85^{\circ})$ for the upper row; 
$(90^{\circ}, 100^{\circ})$, $(100^{\circ}, 110^{\circ})$,  
$(110^{\circ}, 120^{\circ})$, $(120^{\circ}, 130^{\circ})$ for the middle row;
$(130^{\circ}, 140^{\circ})$,  $(140^{\circ}, 150^{\circ})$,
$(150^{\circ}, 160^{\circ})$, $(160^{\circ}, 170^{\circ})$
for the bottom row, respectively.
}
\label{fig:fig10}
\end{figure*}

\begin{figure*}
\centerline{
\vbox{
\hbox{
\hskip+1mm\includegraphics[width=4.00cm]{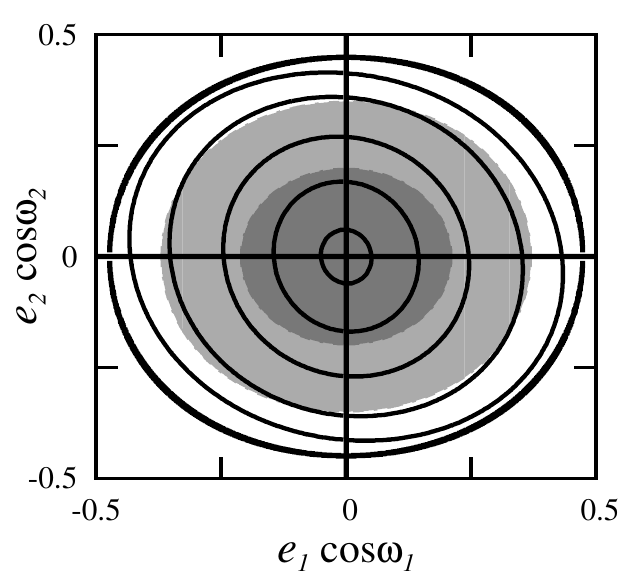}
\hskip+1mm\includegraphics[width=4.00cm]{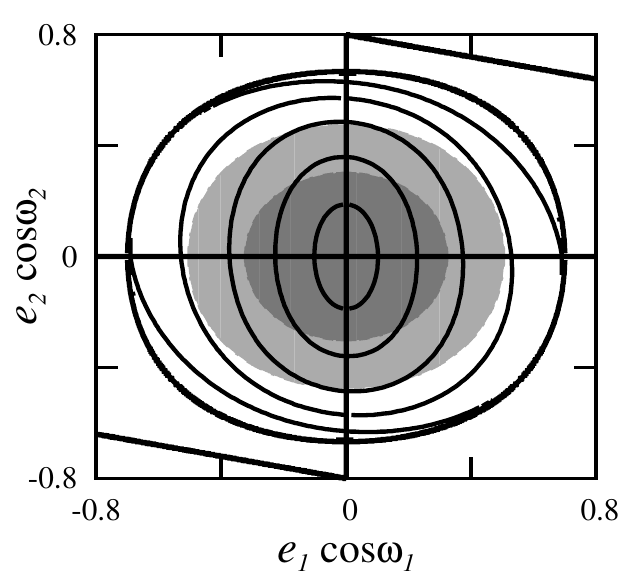}
\hskip+1mm\includegraphics[width=4.00cm]{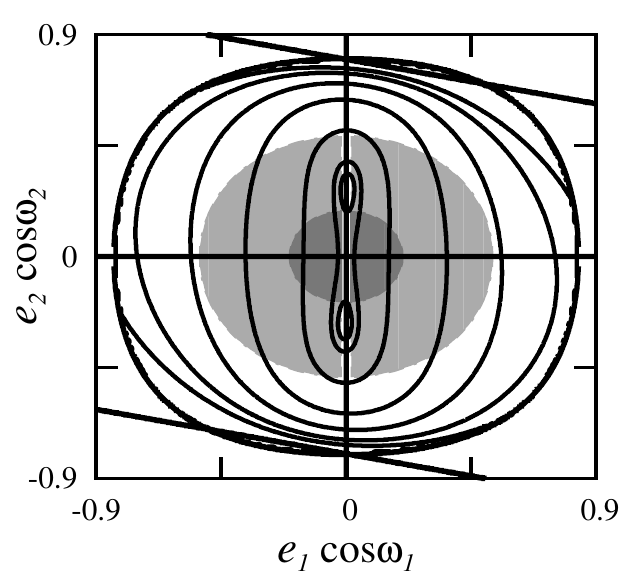}
\hskip+1mm\includegraphics[width=4.00cm]{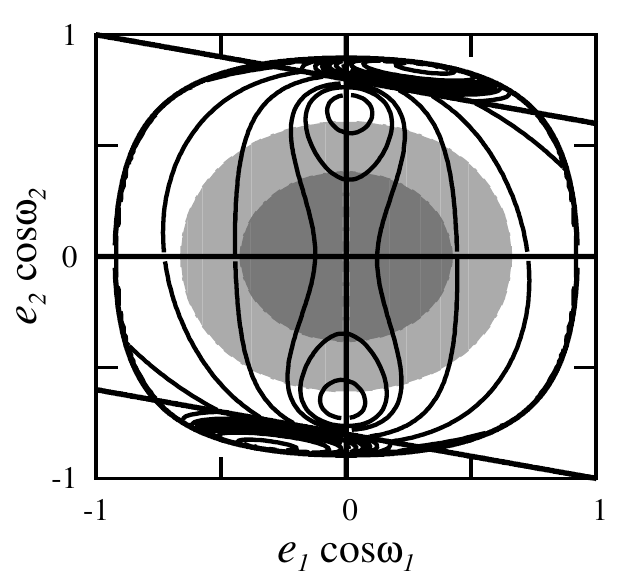}}
\hbox{
\hskip+1mm\includegraphics[width=4.00cm]{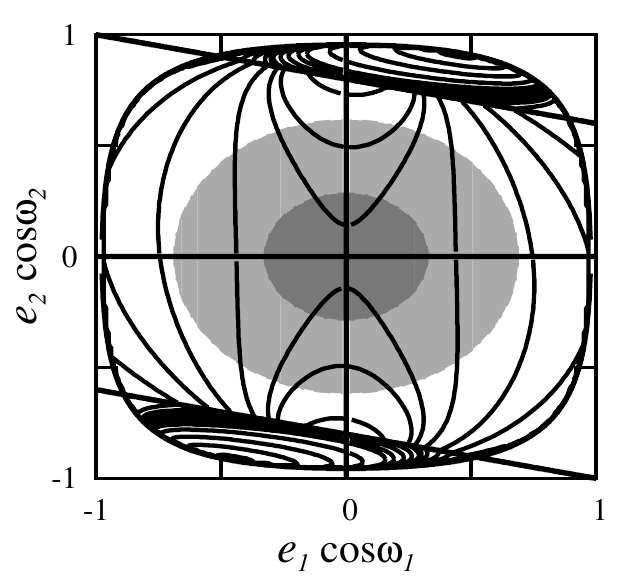}
\hskip+1mm\includegraphics[width=4.00cm]{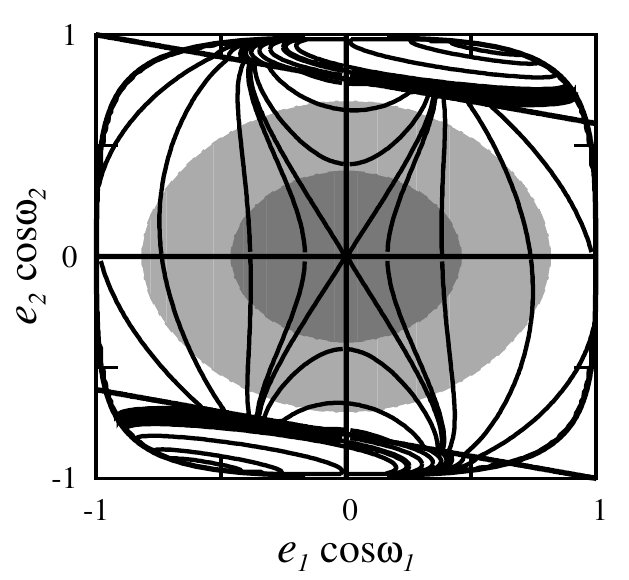}
\hskip+1mm\includegraphics[width=4.00cm]{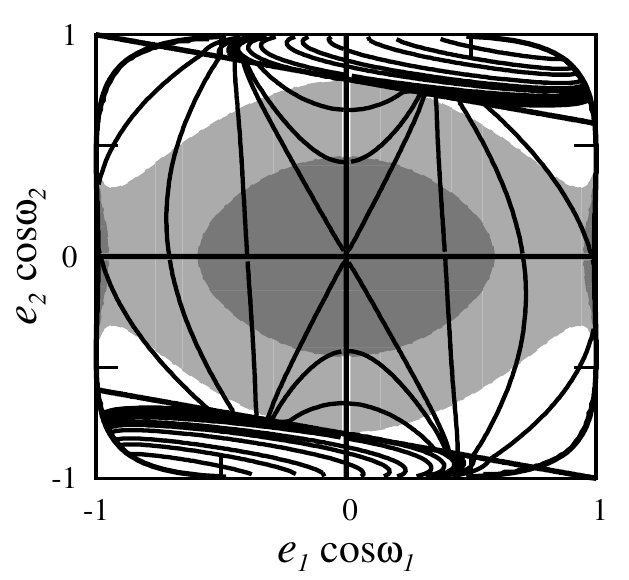}
\hskip+1mm\includegraphics[width=4.00cm]{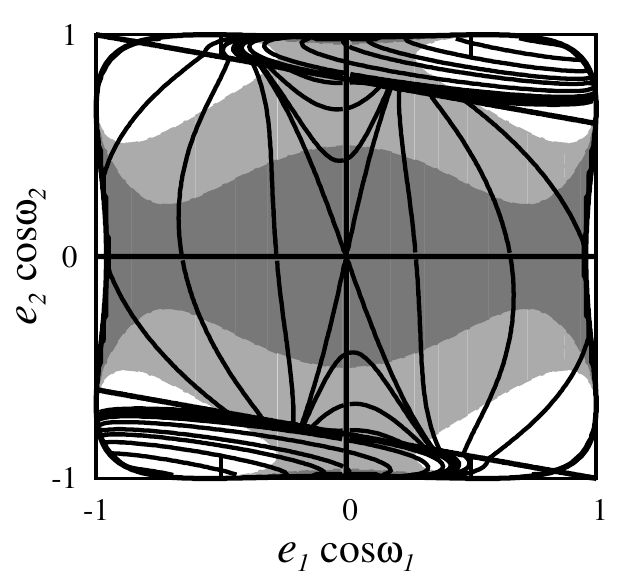}}
\hbox{
\hskip+1mm\includegraphics[width=4.00cm]{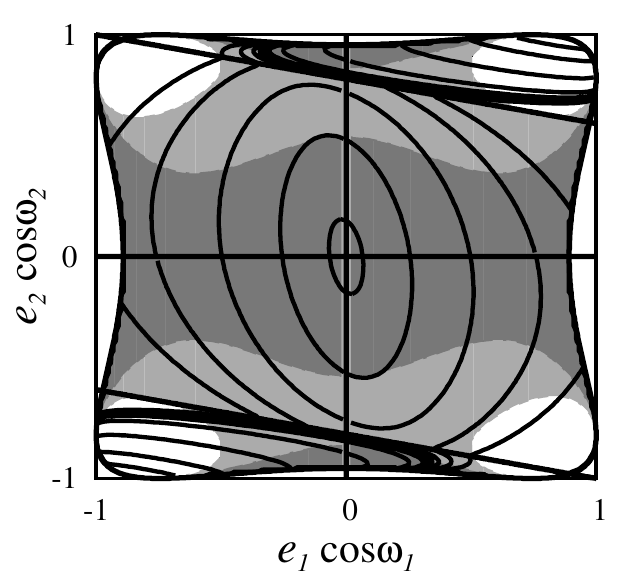}
\hskip+1mm\includegraphics[width=4.00cm]{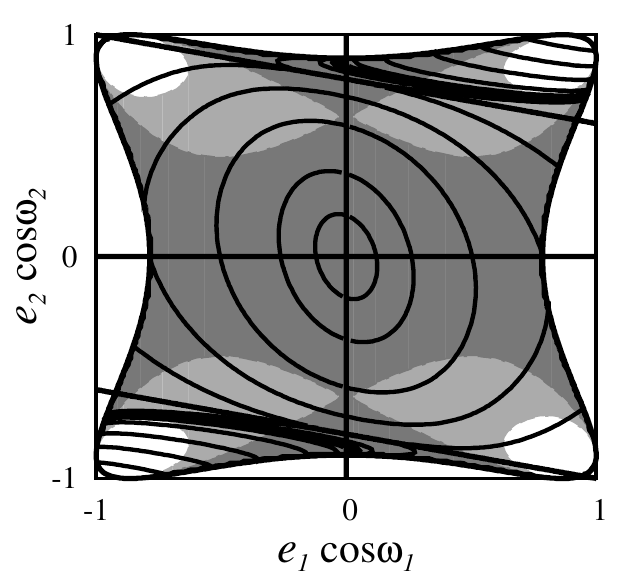}
\hskip+1mm\includegraphics[width=4.00cm]{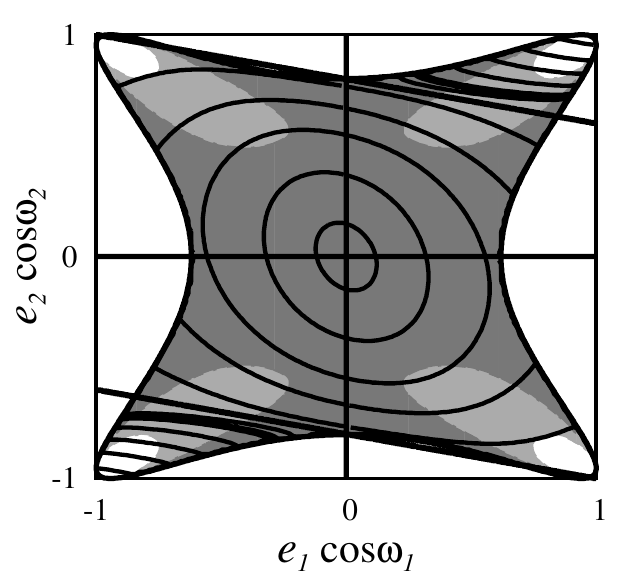}
\hskip+1mm\includegraphics[width=4.00cm]{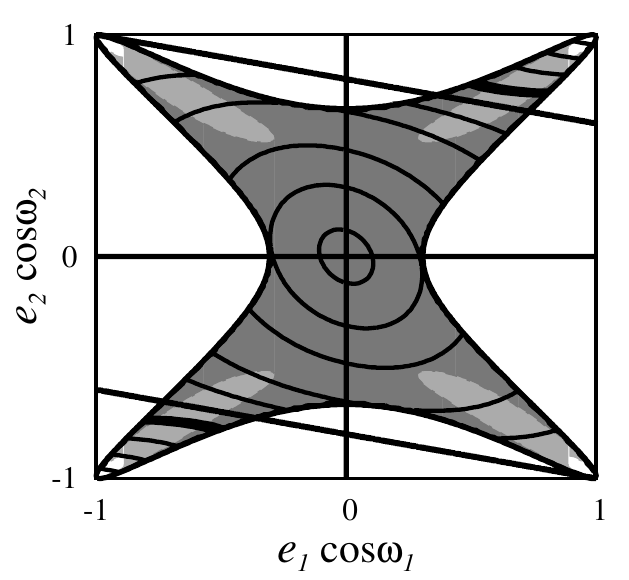}
}
}
}
\caption{
Levels of $\Hsec$ for orbital parameters specified in caption to
Fig.~\ref{fig:fig10} and plotted in the $\cP_C$-plane of  $(e_1
\cos{\omega_1},~e_2 \cos{\omega_2})$, where $\omega_1,\omega_2=0,\pi$. Values of
$\nAMD$ and ranges of mutual inclinations are the same as in that figure.
}
\label{fig:fig11}
\end{figure*}

Here, we show two example sets of the \RP{}-plane derived for two pairs of
$(\alpha,\mu)$. Figures~\ref{fig:fig7}--\ref{fig:fig9} are for $(\alpha,\mu)
\equiv (0.333,0.5)$, while figures \ref{fig:fig10}--\ref{fig:fig11} are for
$(\alpha,\mu) \equiv (0.2,2)$. We start to look more closely at the first set
of the energy diagrams. Figure \ref{fig:fig7} comprises of a number of panels
derived for varied values of $\nAMD$. Each value of $\nAMD$ is related to  the
mutual inclination at the origin, $i_0$. Shaded ares in these plots mark ranges
of the mutual inclination permitted by the given and fixed $\nAMD$. Lighter
shadings encode smaller mutual inclinations. Thin black curves encompass the
region of permitted motion according with $\Imut=0,\pi$. We show the
\RP{}-plane defined as $\cP_M$ (see Fig.~\ref{fig:fig7}) as well as  the
$\cP_{S}$-plane (Fig.~\ref{fig:fig8}) and $\cP_{C}$-plane
(Fig.~\ref{fig:fig9}).  We can see very clearly that the $\cP_{S,C}$-planes
provide a {\em continuous} representation of energy levels.

A sequence of energy diagrams shown in Figs.~\ref{fig:fig7}--\ref{fig:fig9}
helps us to understand the development of a few families of equilibria found in
this work. We start with $\nAMD=0.03$ (the top left-hand panel). In this case,
the origin is the global extremum (the maximum) of the secular energy. This
corresponds to the well known classic zero-eccentricity equilibria investigated
in detail in \citep{Krasinsky1972,Krasinsky1974,Libert2007b}. When $\nAMD=0.1$ (the next
panel in the top row), we see a saddle at the origin and a maximum in
quadrant~IV of $\cP_M$-plane, which can be better seen in the $\cP_{S}$-plane.
The extremum can be identified with the Lidov-Kozai resonance. Clearly, in that
case, the neighboring trajectories characterized by librations of angle
$\omega_1$ around $\pi/2$. We can also notice {\em a non-classic} feature,
regarding the non-restricted model of the L-K resonance: close to the
quasi-elliptic point, also  angle $\Delta\varpi$ may librate around $\pi$ (it
means that $\omega_2$ librates around $\pi/2$).  This effect is possible for
compact systems.

When  $\nAMD=0.18$ (the next panel in the top row of
Figs.~\ref{fig:fig7}--\ref{fig:fig9}), new structures appear: a saddle at the
origin of the $\cP_{C}$-plane (see appropriate panel in Fig.~\ref{fig:fig9})
with two elliptic points close to the $e_1=0$ axis, as well as an elliptic
point {\em above} the collision line (marked with thin lines).   The change of
topology of $\Hsec$ is also visible in the respective panels of the
$\cP_S$-plane. As we already noticed, it is related to the second bifurcation
of the origin.  Simultaneously, the center of the L-K resonance moves toward
large $e_1$. When the $\nAMD=0.25$ (the top right-hand panel of
Fig.~\ref{fig:fig8}) we observe a further development of the structure around
the origin and {\em a bifurcation} of the extremum identified with the L-K
resonance onto a saddle point and {\em two} new  elliptic points appearing in
the regime of large $e_1$. These structures are particularly well seen in
$\cP_{S,C}$-planes, respectively, as illustrated in Fig.~\ref{fig:fig8} and
Fig.~\ref{fig:fig9}.

A similar analysis might be carried out for the second set of parameters
$(\mu,\alpha)$. We do not present it in detail, however an inspection of the
energy diagrams shown in Fig.~\ref{fig:fig10}, and Fig.~\ref{fig:fig11} reveals
that the sequence of plots ends in a different dynamical situation: even for
large $\nAMD$, the region of permitted motion remains closed. Obviously, the
evolution of equilibria appearing for the second pair of parameters
($\alpha,\mu$) is different from the first case. 

\begin{figure*}
\centerline{
\hbox{
\includegraphics[width=14.8cm]{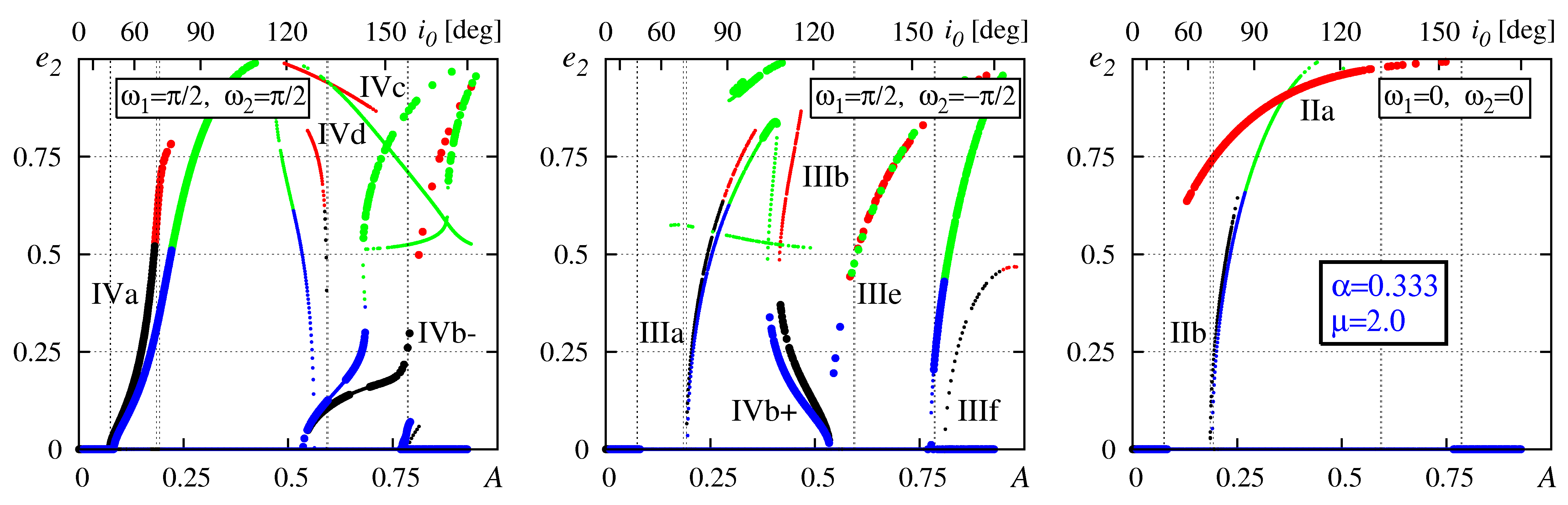}
}
}
\caption{
A comparison of families of stationary solutions found with the octupole
expansion of the secular Hamiltonian (green and blue filled circles) and with
the help of quasi-analytic method (black and red filled circles). Parameters are
for $\alpha=0.333$ and $\mu=2$. Large filled circles are for stable solutions,
and smaller circles are for unstable equilibria. The red and green filled
circles are for equilibria beyond the formal convergence limit of $\Hsec$
expanded in $\alpha$.
}
\label{fig:fig12}
\end{figure*}

Finally, we consider one more experiment devoted to a comparison of the results
derived with the help of the octupole theory and the quasi-analytic averaging
algorithm. Figure~\ref{fig:fig12} illustrates a few families of equilibria
derived for $\alpha=0.333$ and $\mu=2$. Black and red filled circles are for
the semi-analytic theory, blue and green filled circles are for the octupole
Hamiltonian approximation.  The red and green points indicate equilibria found
beyond the formal limit of convergence of the expansion of $\Hsec$ in $\alpha$.
We may notice significant differences between some branches of equilibria
already in the regime of moderate $e_2$. There are also solutions permitted by
the octuple theory, represented by green horizontal branches, which are absent
in the quasi-analytic model. This test confirms that the study of equilibria in
compact systems benefit from the application of the semi-analytic (basically
exact) averaging.

Hence, we should follow a more systematic procedure. Once we identify a solution
of a given family for fixed pair of ($\alpha,\mu$),  we may continue  that
family by searching for the zeros of the right hand sides of the  equations of
motion, Eqs.~\ref{eq:motion}. This task may be accomplished by minimization of
the norm of the partial derivatives of the secular Hamiltonian.  To speed up the
minimization, we apply the fast Levenberg-Marquardt algorithm
\citep{Press1992}. Simultaneously, we examine the stability of solutions which
are found after the L-M algorithm converged. Perhaps a more elaborate
algorithm of the continuation of the equilibria might be applied, nevertheless,
even with such a simple approach, we are able to identify a few families of
solutions that are, to the best of our knowledge, unknown in the literature.
Finally, Figs.~\ref{fig:fig13}--\ref{fig:fig15} illustrate the results of the
continuation globally. Each set of panels is derived for fixed $(\mu,\alpha)$
chosen as combinations of parameters $\alpha=0.2$, $0.333$, $0.667$ and  
$\mu=0.25$, $0.5$, $1$, $2$, respectively. The continuation of the families of
equilibria is done in the whole possible range of $\nAMD$. We plot $\Imut$,
$e_1$ and $e_2$ of the found stationary solutions, as functions of $\nAMD$
(and, if the value of $\nAMD$ permit circular orbits, as a function of $i_0$).
Columns in each group of diagrams are for particular quadrants of the
\RP{}-plane. Note, that we skipped panels for quadrant~I because in that
quadrant we found only  one family~Ia of unstable solutions for a limited range
of $\mu \geq 1$ (see Sect.~\ref{sec:Ia} for details). Simultaneously, Lyapunov
stable (or linearly stable) equilibria are marked with large filled circles,
and unstable solutions are marked with small filled circles. Families of
equilibria are classified accordingly with the quadrant of \RP{}-plane in which
they appear, and they are labeled with corresponding Roman numbers. Let us note
that the red filled circles indicate equilibria found beyond the formal limit
of convergence of the expansion of $\Hsec$ in $\alpha$.

In this way, we can obtain quite a deep insight into the secular equilibria and
their stability in wide ranges of the primary parameters.  Below, we describe
the identified families of stationary solutions in more detail, and we try to
characterize the associated dynamical behaviors of the secular system. A likely
position of the HD~12661 system in the diagrams
Figs.~\ref{fig:fig13}--\ref{fig:fig15} may be deduced from its currently known
orbital elements, $\alpha \sim 0.3$ and $\mu \sim 1$. For relatively small
$\nAMD \sim 0.1$, the system might be found in the top, right-hand panels of 
Fig.~\ref{fig:fig14}, in the regime of the L-K bifurcation, still with moderate
$e_1 \sim 0.3$ and $e_2 \sim 0.1$.

%
\subsection{Family~0 at ${\mathbf{(e_1,e_2)=(0,0)}}$}
\begin{figure*}
\centerline{
\vbox{
\hbox{
     \includegraphics[width=89mm]{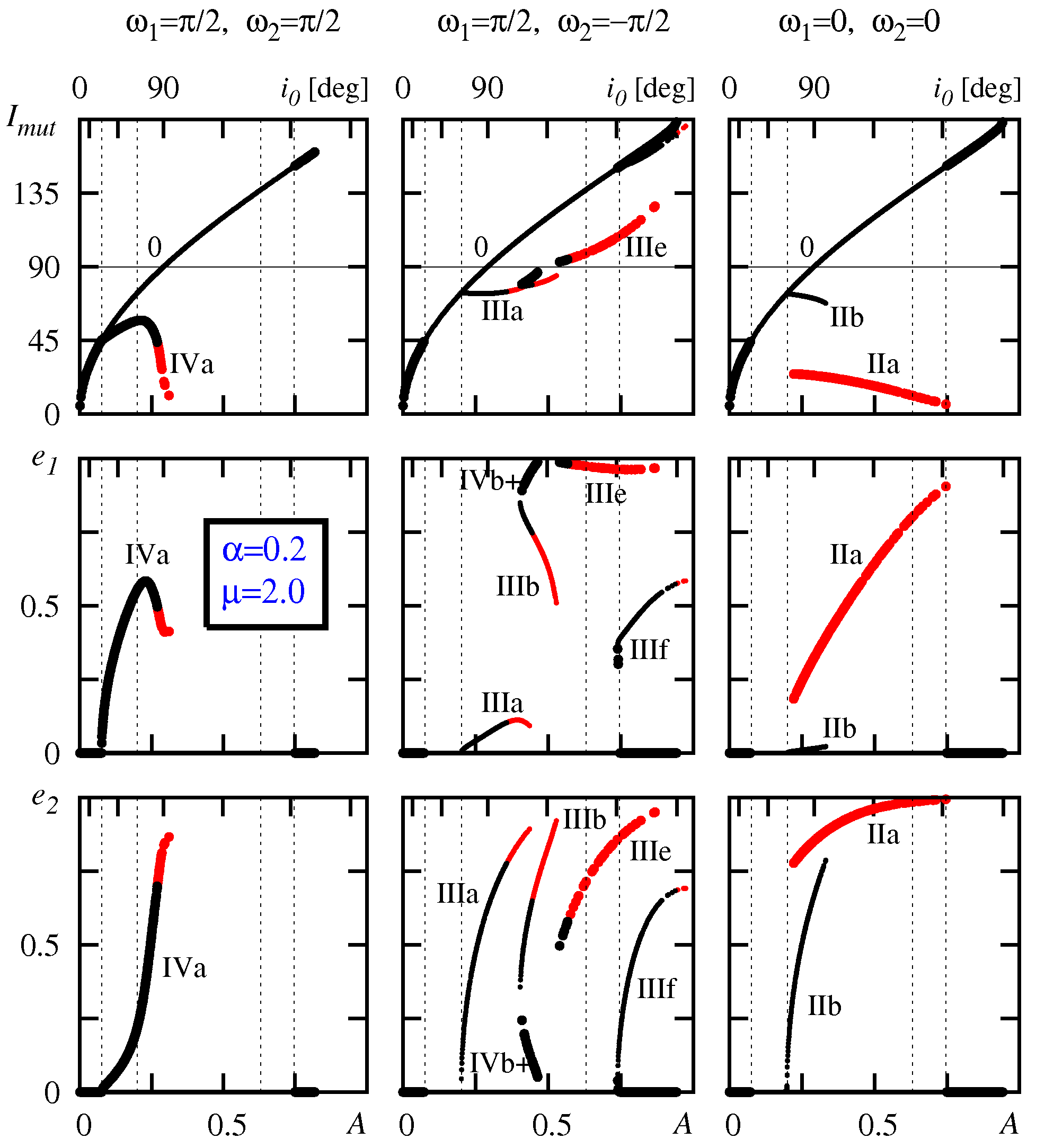}
     \includegraphics[width=89mm]{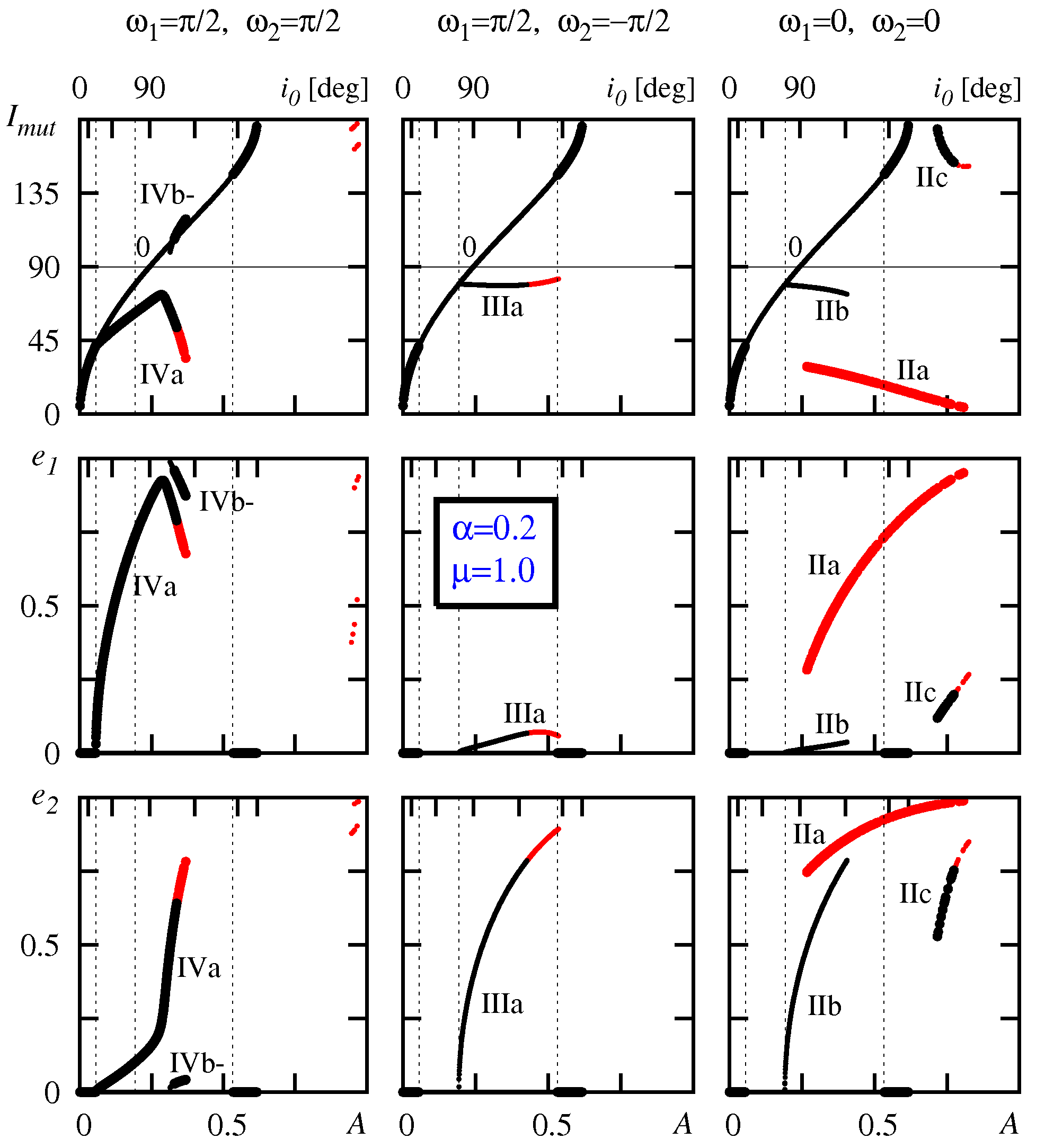}
    }
\vskip 0.25cm
\hbox{
     \includegraphics[width=89mm]{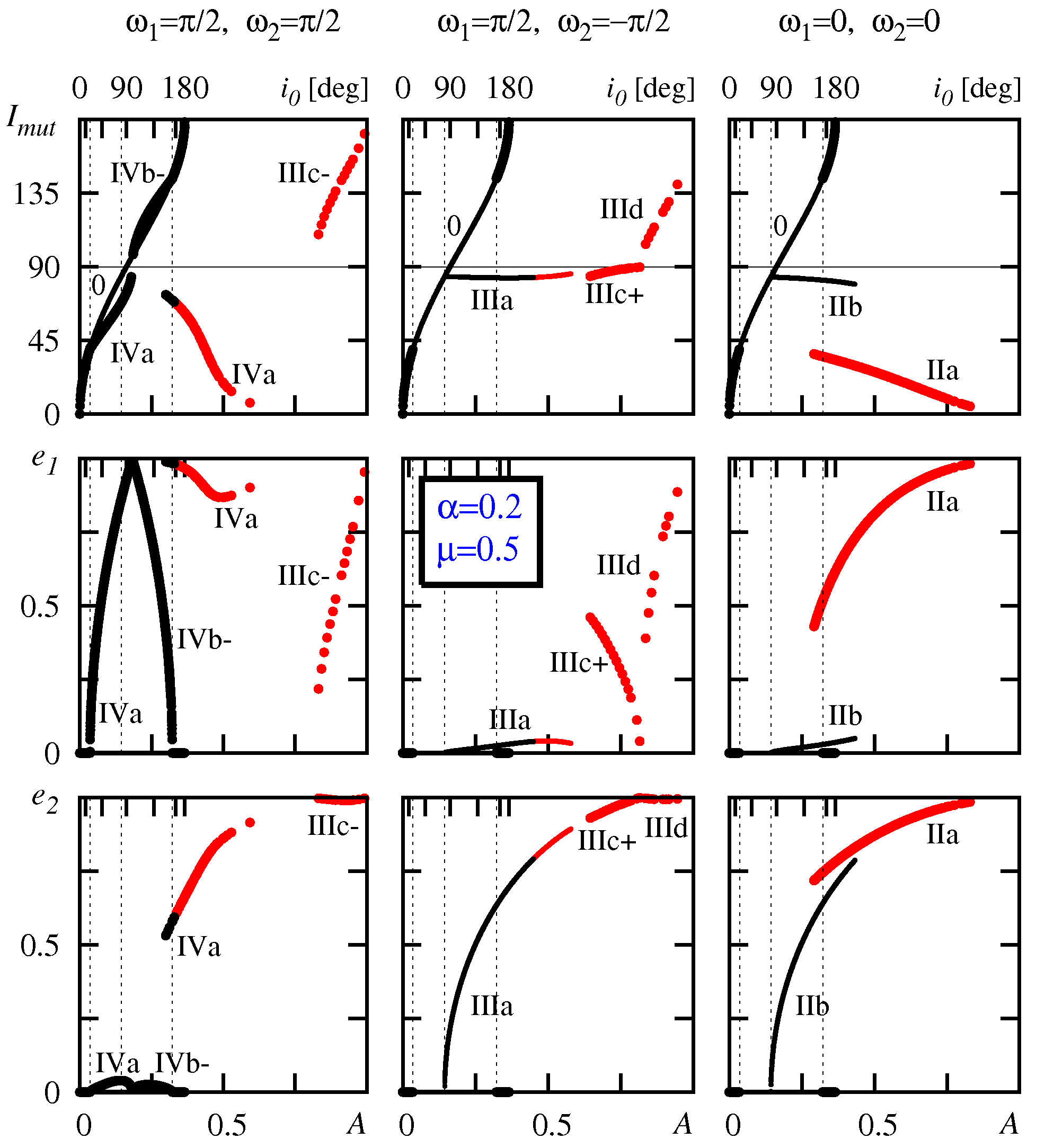}
     \includegraphics[width=89mm]{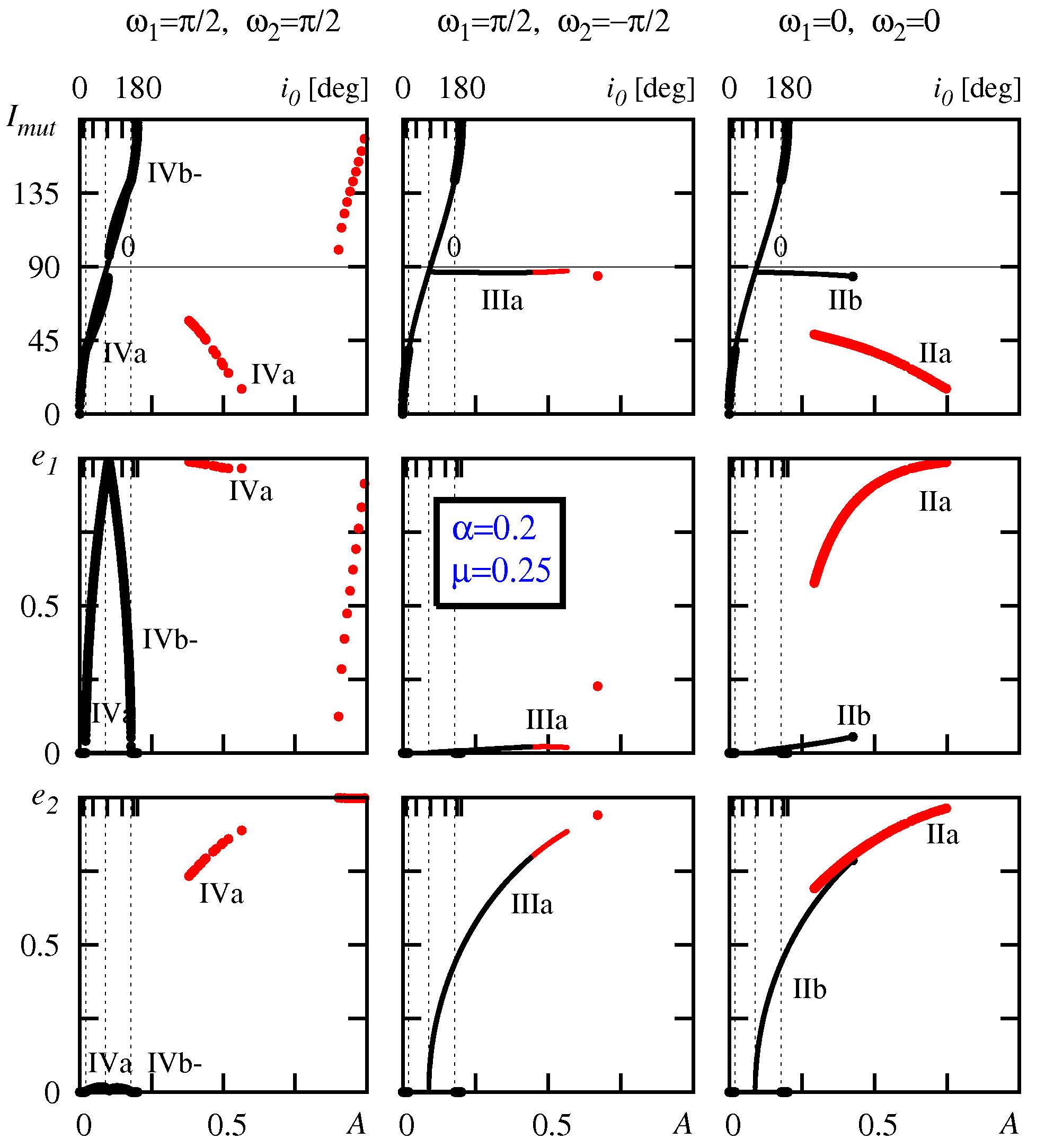}
    }
}
}
\caption{
Families of stationary solutions obtained for $\alpha=0.2$ and the  following
mass ratios: $\mu=2.0$ -- the top left-hand plots,  $\mu=1.0$ -- the top
right-hand plot, $\mu=0.5$ -- the bottom left-hand, and $\mu=0.25$ --  the
bottom right-hand plots. Large filled circle are for Lyapunov stable (or
linearly stable) equilibria, smaller filled circles are for unstable
equilibria, red filled circles are for solutions found in regions, where the
power series of $\Hsec$ in $\alpha$ would diverge. The stationary solutions are
classified according with the quadrant of the \RP{}-plane, in which they
appear, hence the columns in each sub-group of diagrams [for fixed
$(\alpha,\mu$) written in the  legend] are for the   following
$(\omega_1,\omega_2)$-pairs:  $(\pi/2,\pi/2)$ -- the left column, 
$(\pi/2,-\pi/2)$ -- the middle column, and  $(0,0)$ --  the right column. Each
sub-group of stability diagrams has panels for the mutual inclination (the top
row), and for the eccentricities (the middle and the bottom rows,
respectively). The $x$-axis of each diagram is labeled by $\nAMD$ and  $i_0$.
Particular families of solutions which are identified in this work are labeled
with Roman numbers and appropriate Latin letters.  See the text for more
details.
}
\label{fig:fig13}
\end{figure*}

\begin{figure*}
\centerline{
\vbox{
\hbox{
     \includegraphics[width=89mm]{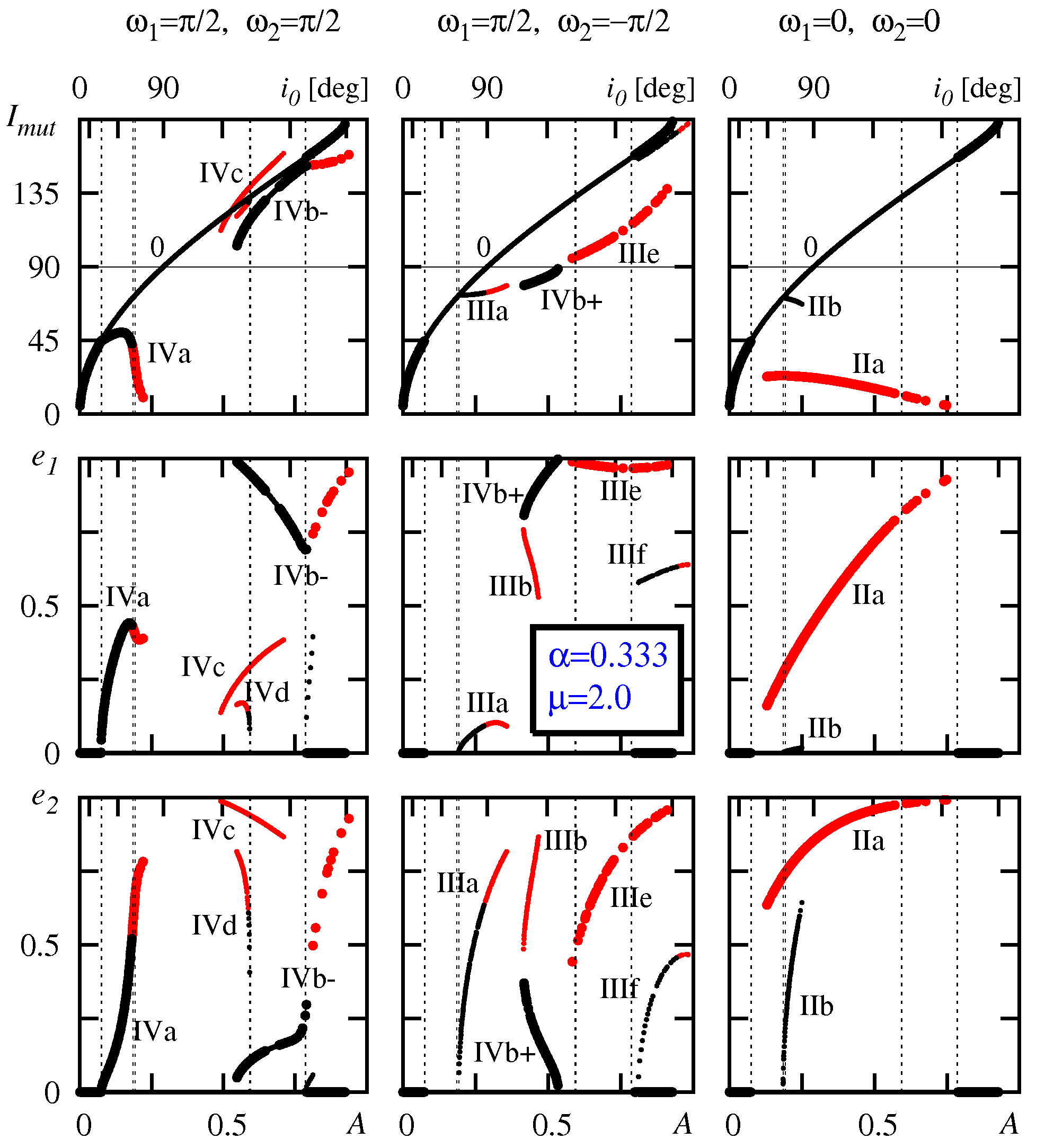}
     \includegraphics[width=89mm]{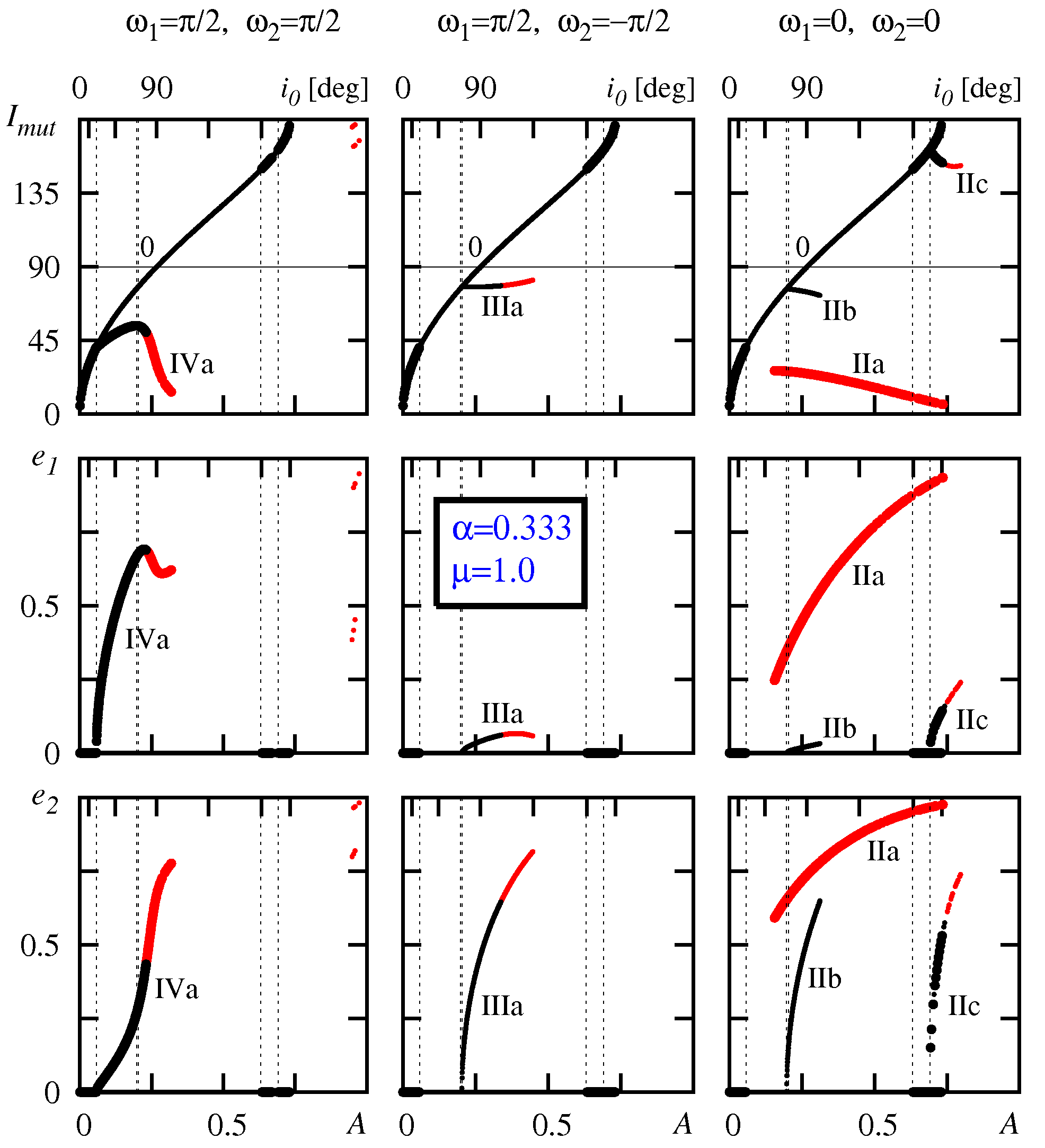}
    }
\vskip 0.25cm
\hbox{
     \includegraphics[width=89mm]{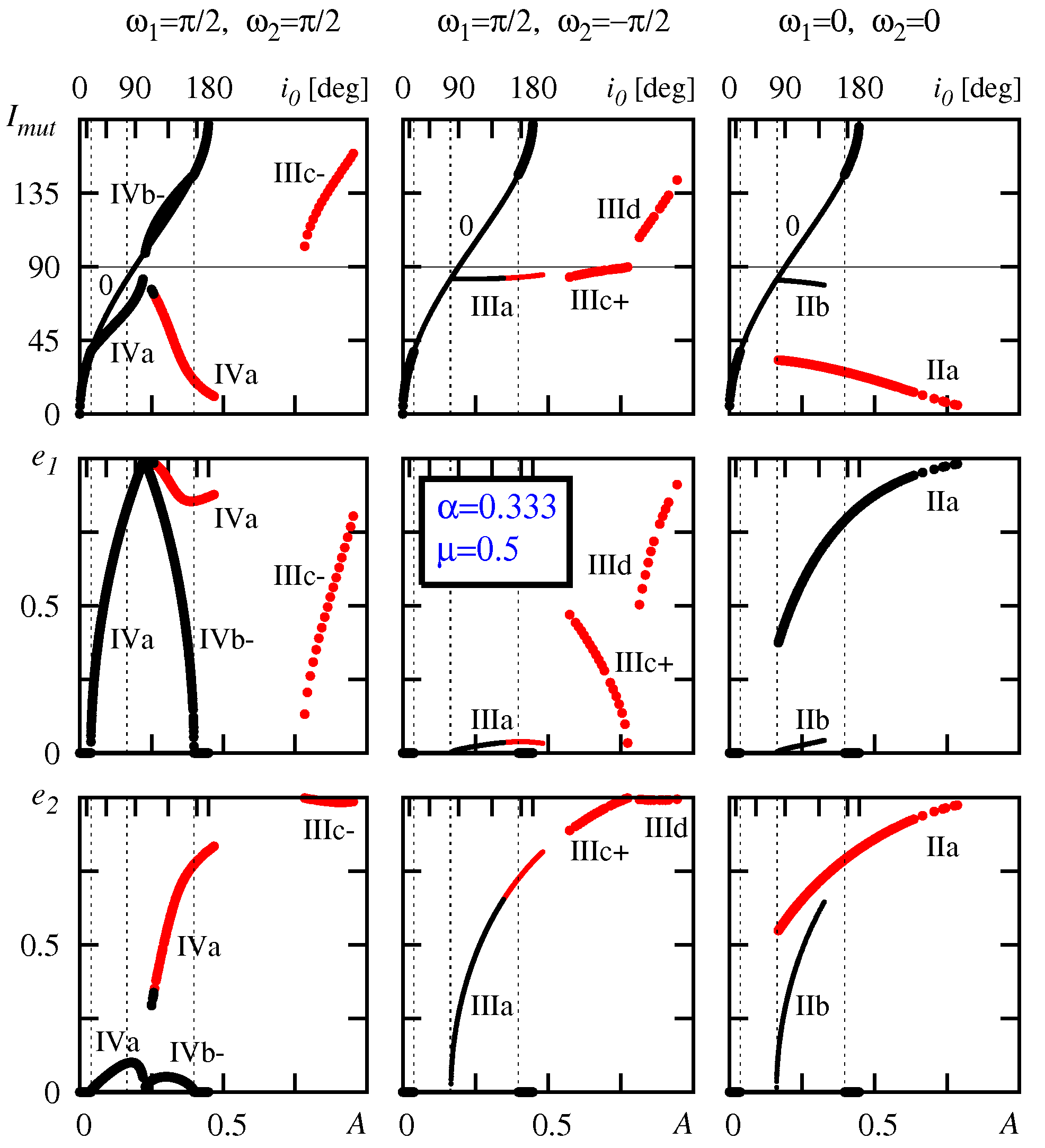}
     \includegraphics[width=89mm]{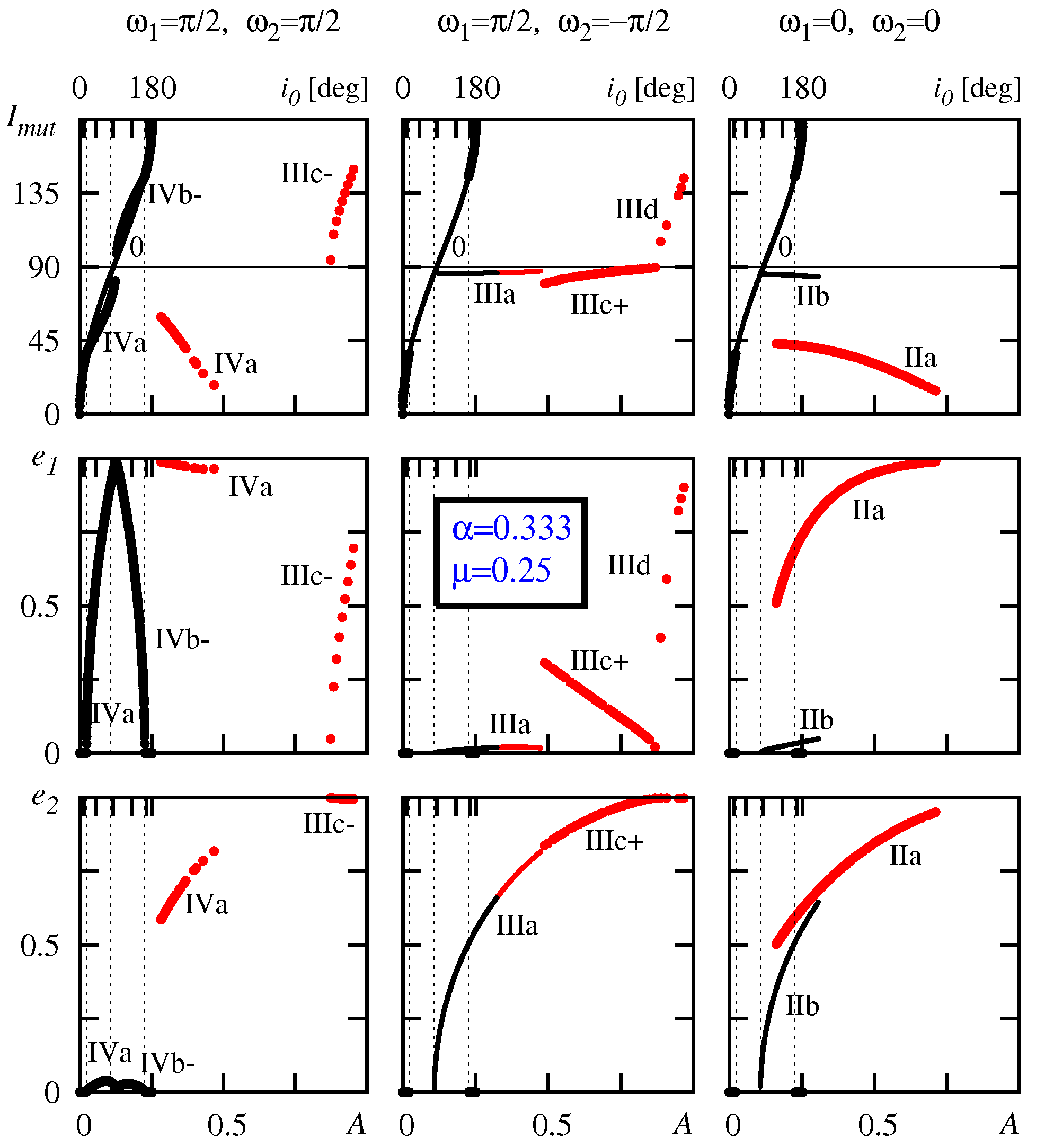}
    }
}
}
\caption{
Families of stationary solutions obtained for $\alpha=0.333$ and the  following
mass ratios: $\mu=2.0$ -- the top left-hand plots,  $\mu=1.0$ -- the top
right-hand plot, $\mu=0.5$ -- the bottom left-hand, and $\mu=0.25$ --  the
bottom right-hand plots. Large filled circle are for Lyapunov stable (or
linearly stable) equilibria, smaller filled circles are for unstable
equilibria, and red filled circles are for solutions found in regions, where
the power series of $\Hsec$ in $\alpha$ would diverge. The stationary solutions
are classified according with the quadrant of the \RP{}-plane, in which they
appear, hence the columns in each sub-group of diagrams [for fixed
$(\alpha,\mu$) written in the  legend] are for the following
$(\omega_1,\omega_2)$-pairs:  $(\pi/2,\pi/2)$ -- the left column, 
$(\pi/2,-\pi/2)$ -- the middle column, and  $(0,0)$ --  the right column. Each
sub-group of stability diagrams has panels for the mutual inclination (the top
row), and for the eccentricities (the middle and the bottom rows,
respectively). The $x$-axis of each diagram is labeled by $\nAMD$ and $i_0$.
Particular families of solutions which are identified in this work are labeled
with Roman numbers and appropriate Latin letters. See the text for more
details.
}
\label{fig:fig14}
\end{figure*}

\begin{figure*}
\centerline{
\vbox{
\hbox{
     \includegraphics[width=89mm]{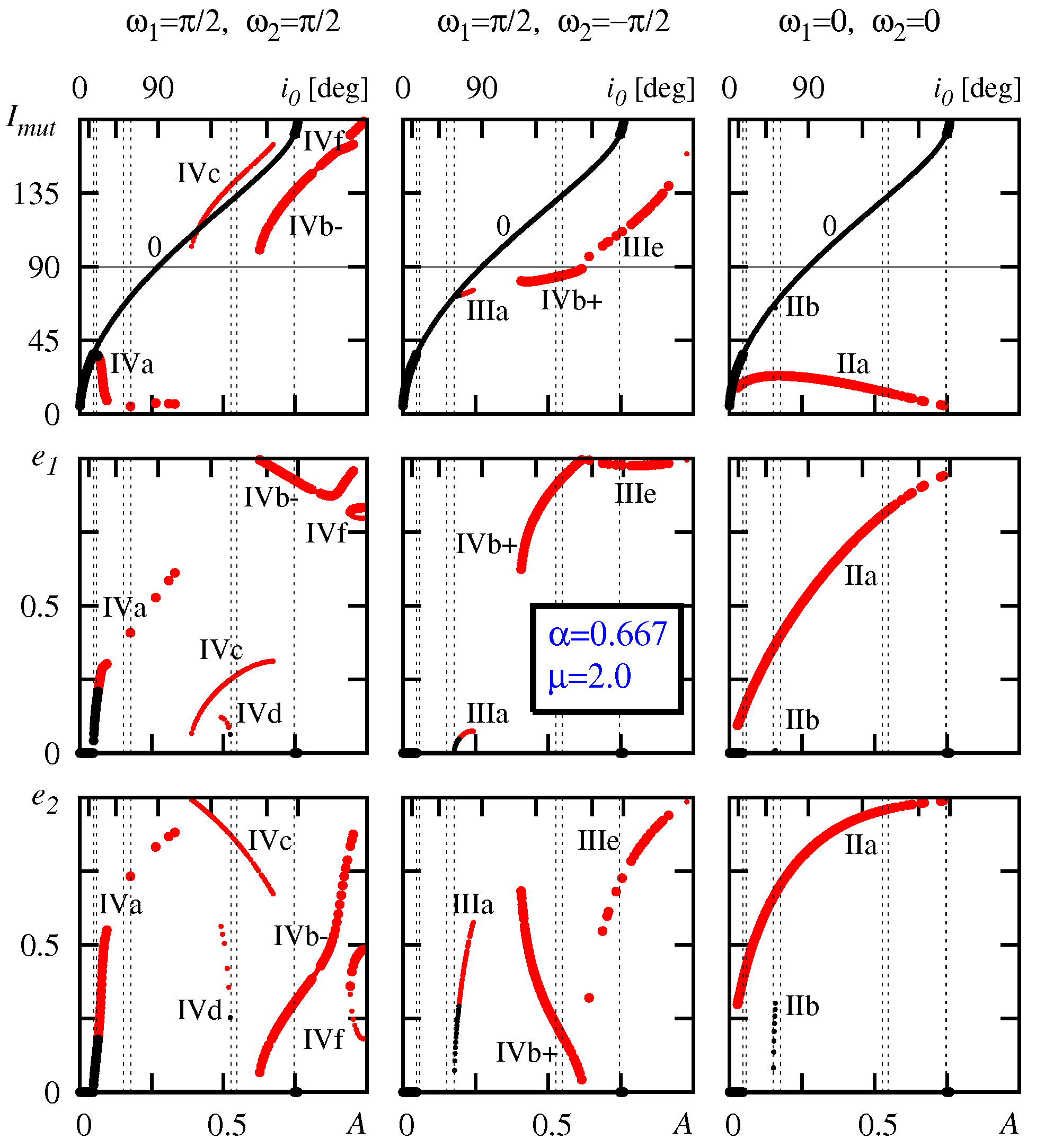}
     \includegraphics[width=89mm]{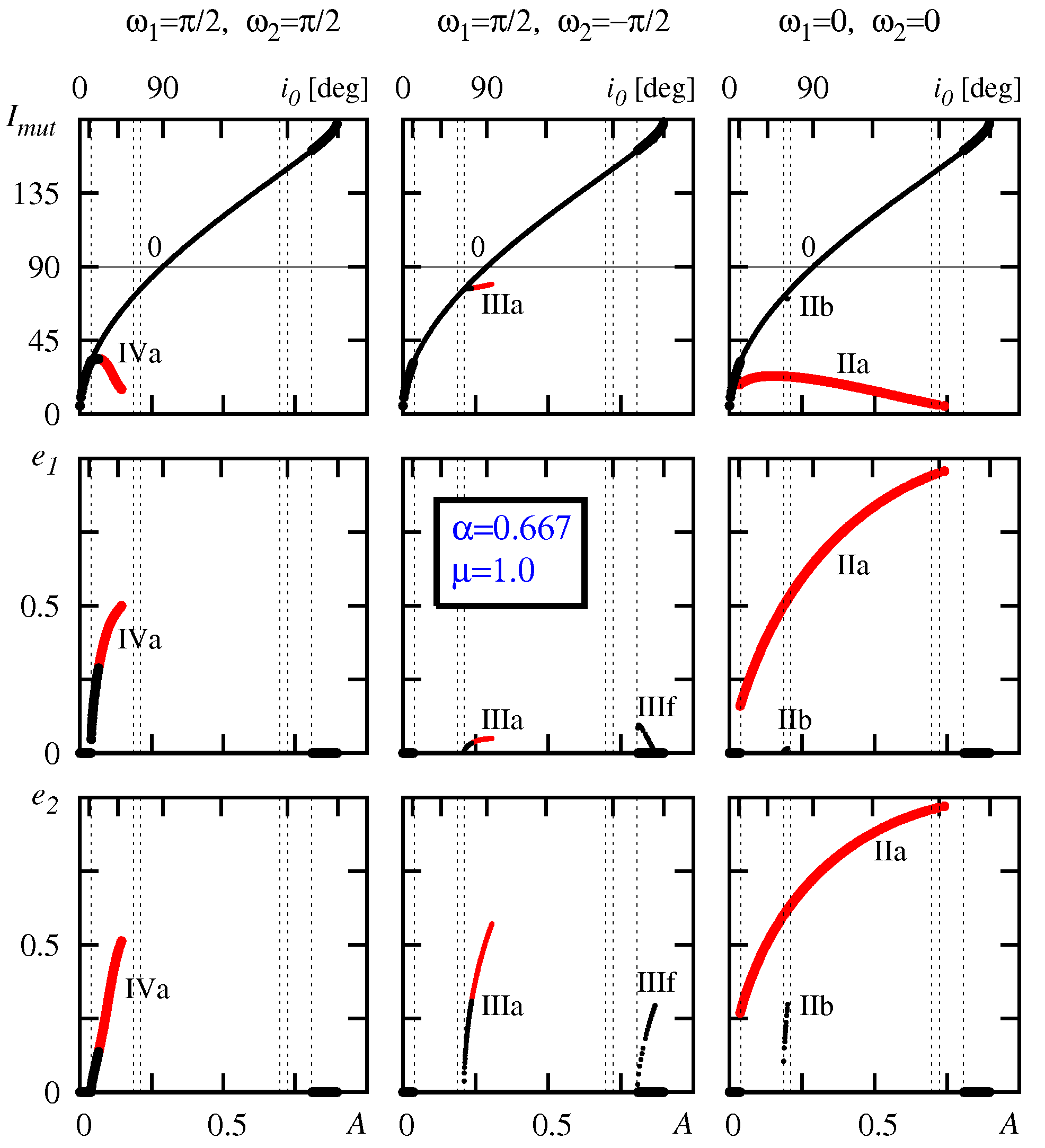}
    }
\vskip 0.25cm
\hbox{
     \includegraphics[width=89mm]{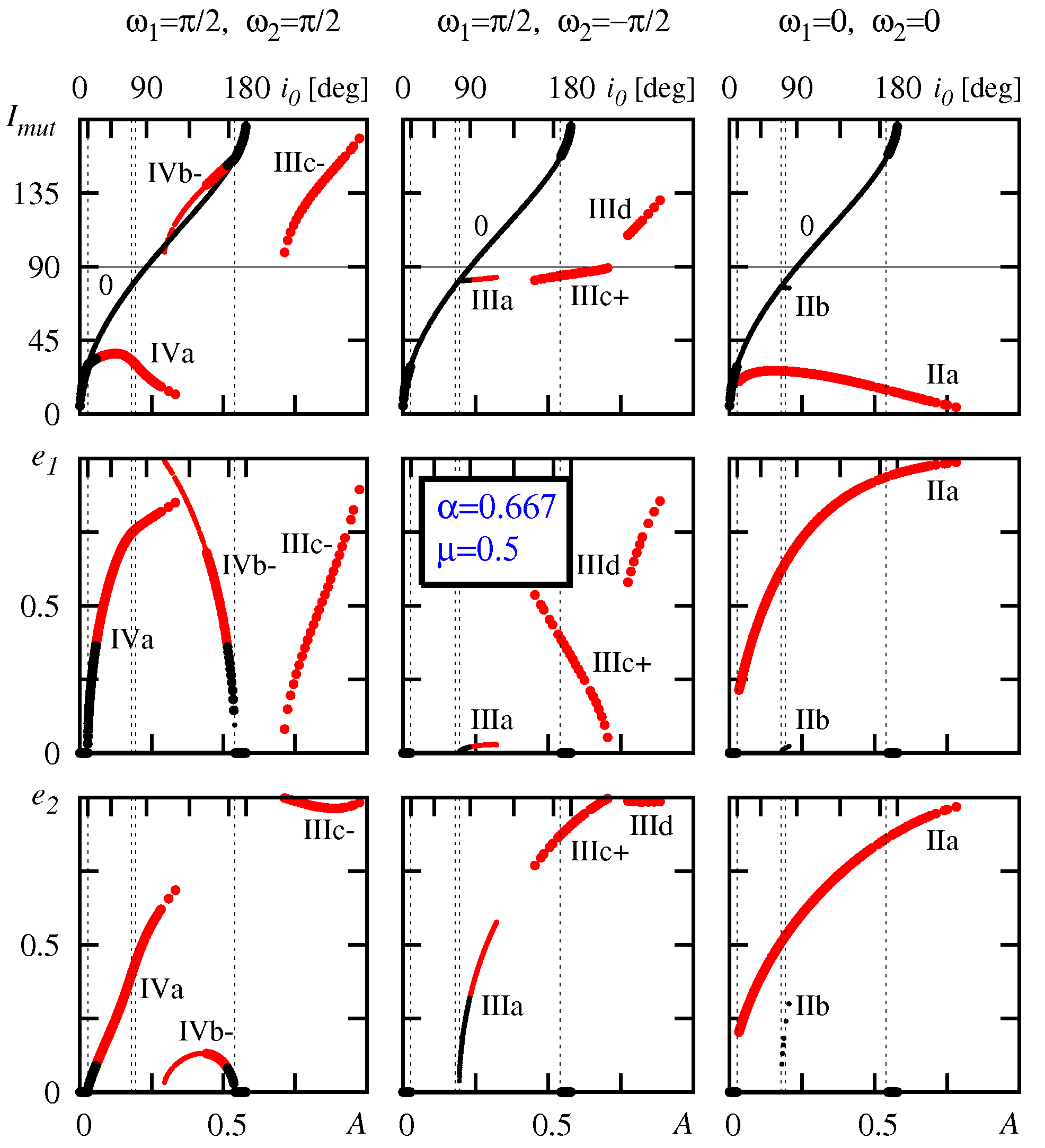}
     \includegraphics[width=89mm]{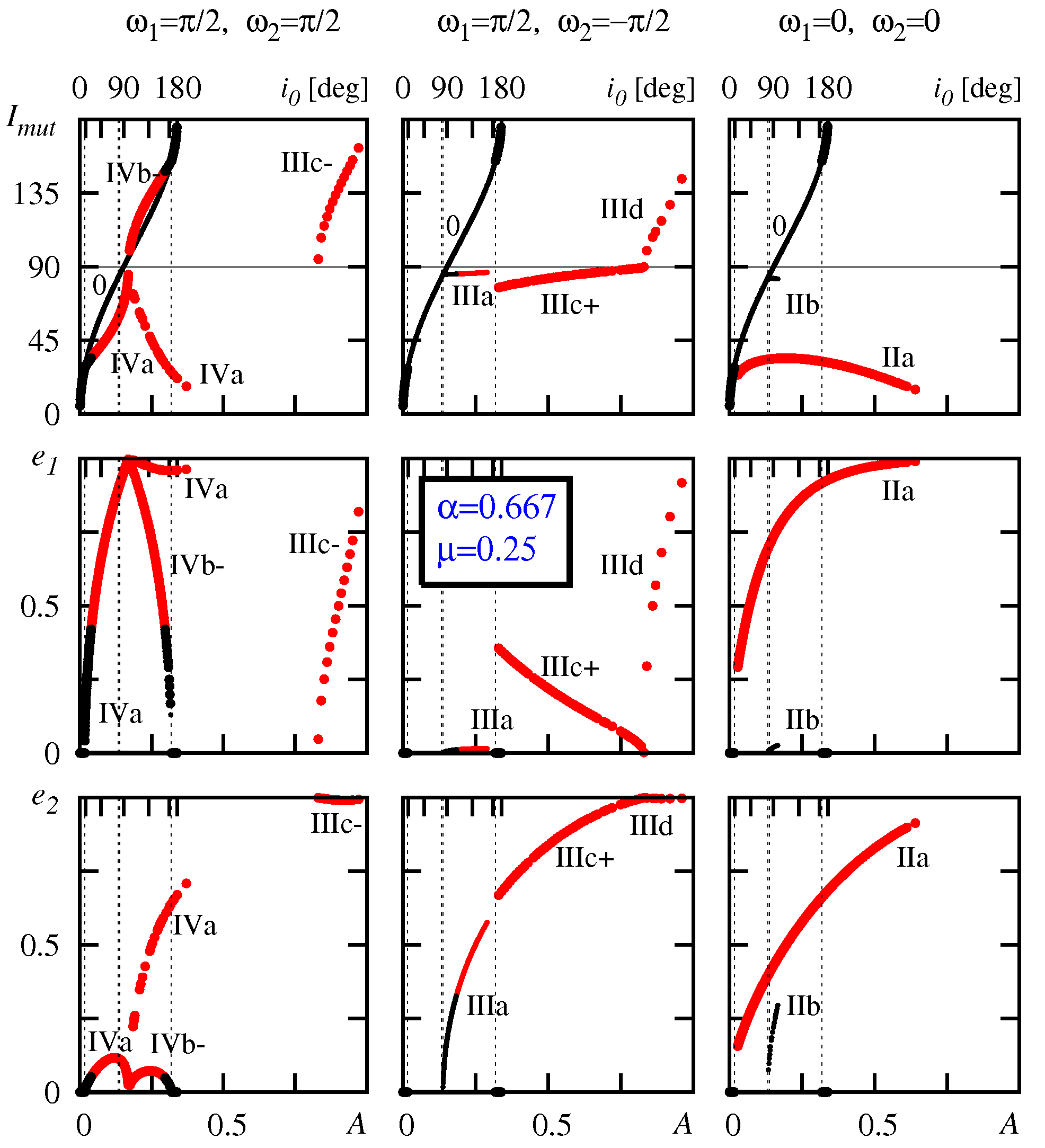}
    }
}
}
\caption{
Families of stationary solutions obtained for $\alpha=0.667$ and the  following
mass ratios: $\mu=2.0$ -- the top left-hand plots,  $\mu=1.0$ -- the top
right-hand plot, $\mu=0.5$ -- the bottom left-hand, and $\mu=0.25$ --  the
bottom right-hand plots. Large filled circle are for Lyapunov stable (or
linearly stable) equilibria, smaller filled circles are for unstable
equilibria,  red filled circles are for solutions found in regions, where the
power series of $\Hsec$ in $\alpha$ would diverge. The stationary solutions are
classified according with the quadrant of the \RP{}-plane, in which they
appear, hence the columns in each sub-group of diagrams [for fixed
$(\alpha,\mu$) written in the legend] are for the   following
$(\omega_1,\omega_2)$-pairs:  $(\pi/2,\pi/2)$ -- the left column, 
$(\pi/2,-\pi/2)$ -- the middle column, and  $(0,0)$ --  the right column. Each
sub-group of stability diagrams has panels for the mutual inclination (the top
row), and for the eccentricities  (the middle and the bottom rows,
respectively). The $x$-axis of each diagram is labeled by $\nAMD$ and $i_0$.
Particular families of solutions which are identified in this work are labeled
with Roman numbers and appropriate Latin letters. See the text for more
details. 
}
\label{fig:fig15}
\end{figure*}

%
The stationary solution at the origin $(e_1=0,e_2=0)$ (see an example in
Fig.~\ref{fig:fig6}) was investigated in detail by \cite{Libert2007b} for
$\alpha=0.1$, $\mu=0.25$, i.e., for relatively distant orbits (or typically
hierarchical configuration). In our classification, this family is marked with
''0'' in all stability diagrams of Figs.~\ref{fig:fig13}--\ref{fig:fig15}. 
Solutions of this family are also studied in detail by
\cite{Krasinsky1972,Krasinsky1974} in terms of the quadrupole approximation of
$\Hsec$, and we already did many references to these works and its results.
Here, we start to follow more closely the evolution  of family ''0'' for
$(\alpha,\mu)\equiv (0.333,0.5)$ with respect to $\nAMD$.   Similarly,
Fig.~\ref{fig:fig10} and Fig.~\ref{fig:fig11} reveal topology of the \RP{}-plane
and  the evolution of zero-eccentricity equilibria for a different pair of
parameters, $(\alpha,\mu)\equiv (0.2,2)$.

When the mutual inclination  remains relatively small (see
Figs.~\ref{fig:fig7}--\ref{fig:fig9}),  the zero-eccentricity equilibrium is
Lyapunov stable  because it corresponds to the maximum of $\Hsec$.  When $\nAMD$
increases, a bifurcation of this solution appears for $i_0 \sim 43^{\circ}$ (as
mentioned already, the L-K bifurcation). Inspecting the $\cP_S$-plane
(Fig.~\ref{fig:fig8}), we may notice that at the bifurcation point a new family
IVa appears, and it remains Lyapunov stable up to extremely large $e_1$. After
$e_1$ reaches the limit of~1 (simultaneously,  the mutual  inclination is close
to $\pi/2$),  the family bifurcates again: the family IVa remains on a branch
with large $e_1$ while a new branch of solutions (family IVb-) can be continued
to large mutual inclinations with simultaneous decrease of $e_1$. Family IVb-
can be regarded as the retrograde case of the L-K resonance with  $\Imut >
\pi/2$. Note that families IVa and IVb- are quasi-symmetric with respect to
$\Imut \sim \pi/2$, regarding the eccentricity and inclination of the inner
orbit. This symmetry is more ''exact'' for smaller $\alpha$ (hierarchical
configurations). We notice, that the quadrupole term approximation leads to
exact symmetry of the equilibria (see Fig.~\ref{fig:fig4} and the relevant
comments in Sect.~4.3).

We may also discover unstable family IIIb, which has the elliptic point located
in quadrant~III as well as a {\em saddle} of family IVb+ corresponding to
linearly stable solution.  In the later case, the neighboring trajectories are
characterized by librations of $\omega_1$ around $\pi/2$ and also (within a
limited vicinity of the libration center), by librations of $\omega_2$ around
$-\pi/2$. A further increase of $\nAMD$ leads to a shift of solutions IVa, IVb-,
and IIIb towards the border of permitted motions. Finally, for $\nAMD>0.4$, the
zero-eccentricity  equilibria vanish at all, and the energy plane is divided
onto two distinct islands.  By inspecting the $\cP_{C}$-plane (see the third
panel in top row in Figs.~\ref{fig:fig7},\ref{fig:fig9} and \ref{fig:fig11}) we
can also detect the second bifurcation of the zero-eccentricity equilibria which
is associated with the apparence of a saddle at the origin accompanied by two
{\em unstable} elliptic points close to $e_1 \sim 0$. These solutions can be
classified as members of family IIb because they emerge  in quadrant~II of the
\RP{}-plane.  The saddles visible in $\cP_S$ (Fig.~\ref{fig:fig8}) represent
members of family IIIa  [see also Fig.~\ref{fig:fig10} for the second pair of
$(\alpha,\mu)$].

Thanks to the semi-numerical algorithm, we  can follow the zero-eccentricity
family not only for larger $\alpha$, up to 0.667, but also in wide ranges of the
mass ratios, between $0.25$ and $2$.  Stability diagrams in
Figs.~\ref{fig:fig13}--\ref{fig:fig15} illustrate bifurcations  of equilibria
for different parameter pairs. Obviously, the zero-eccentricity solutions
appear for all combinations of the primary parameters, moreover, the evolution
of this family is very complex. Bifurcations of family ``0'' lead to new classes
of equilibria and qualitative changes of the $\Hsec$ topology  as seen at the
\RP{}-plane.

%
\subsection{Family Ia}
\label{sec:Ia}
Family Ia appears for a limited range of ($\alpha,\mu$) in quadrant~I of the
\RP{}-plane. We detected it for $\mu\ge 1$ (more massive inner planet). This
type of stationary solutions is characterized by small $e_1$ and a range of
$e_2$ between 0 and a value permitted by the  equation of the collision line. It
emerges from a bifurcation of the zero-eccentricity solution in the range of
large $\nAMD$ and is always unstable. 
%
\subsection{Families IVa, IVb-, IVb+, IIIb --- the L-K resonance}
%
We have already seen that the L-K equilibrium
appears in quadrant IV of the \RP{}-plane (family IVa) and is tightly related
to family ''0'' because it emerges from its ``first'' bifurcation on  the
$\nAMD$-axis. The second  L-K bifurcation leads to family IVb- associated with
saddles in quadrant~IV and to a pair of a saddle (IVb+) and an elliptic point
(family IIIb) in quadrant III. These structures are particularly well seen in 
the middle row of Fig.~\ref{fig:fig10}. Stability diagrams in
Figs.~\ref{fig:fig13}--\ref{fig:fig15} tell us that equilibria of families IVb- and
IVb+ are linearly stable while solutions of family IIIb are unstable.
Curiously,  equilibria of family IVb+ might be identified with non-restricted
case of the L-K resonance characterized by librations of  $\omega_1$ around
$\pm\pi/2$ with possible simultaneous librations of $\Delta\varpi$ around
$0$ (or, equivalently, $\omega_2$ around $\mp\pi/2$).  

\begin{figure*}
\centerline{
\vbox{
\hbox{
\hbox{\includegraphics[width=2.2in]{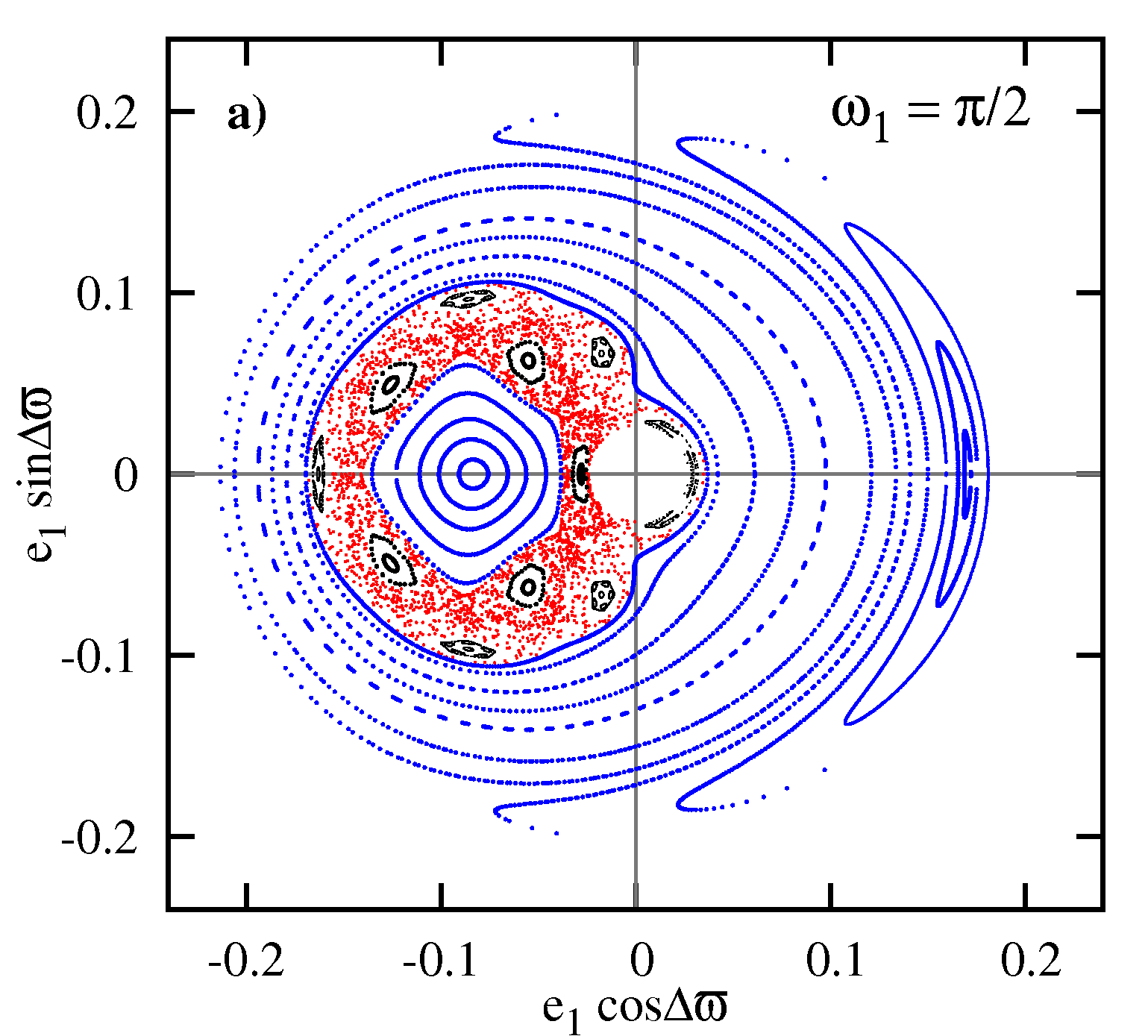}}
\hspace*{0.3cm}
\hbox{\includegraphics[width=2.2in]{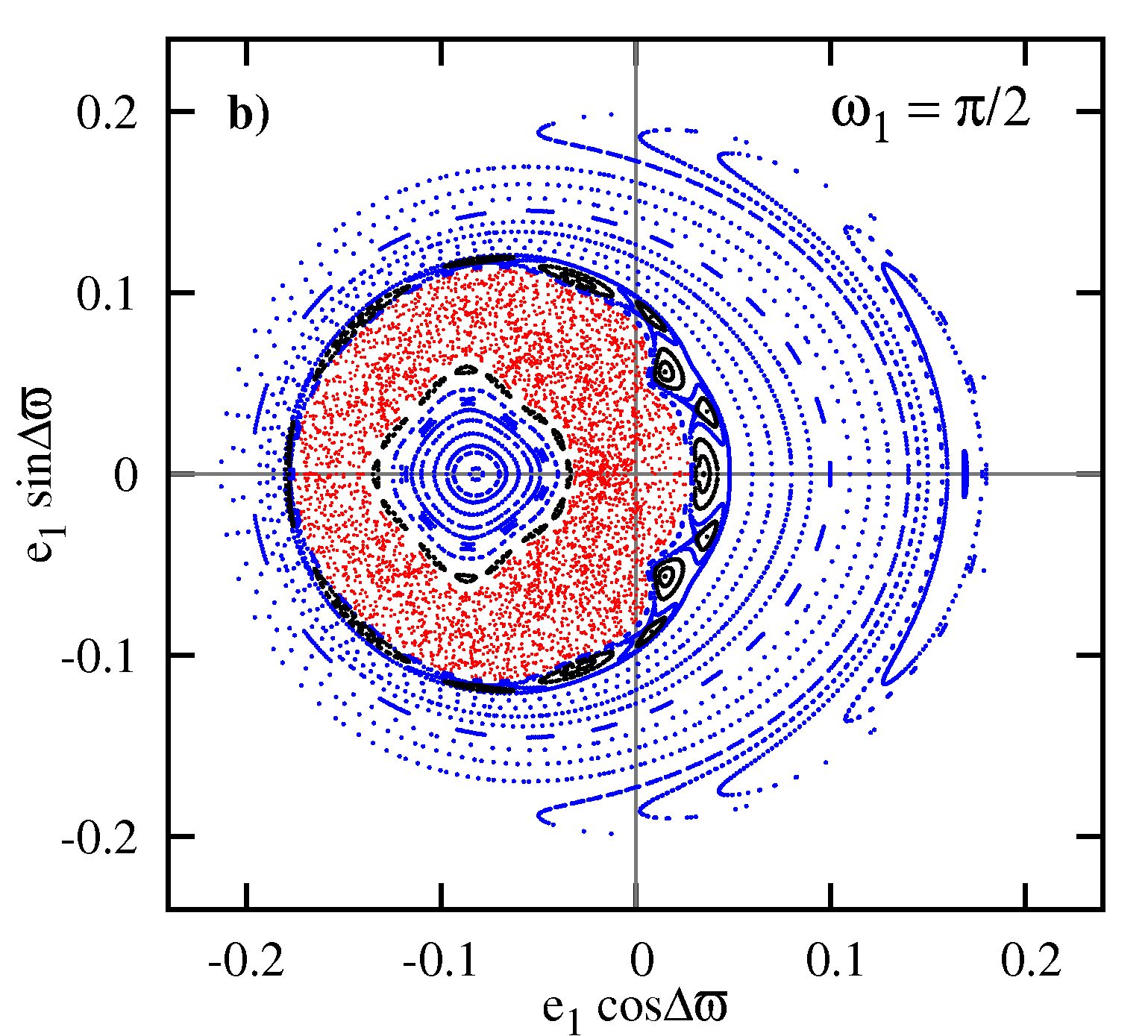}}
\hspace*{0.3cm}
\hbox{\includegraphics[width=2.2in]{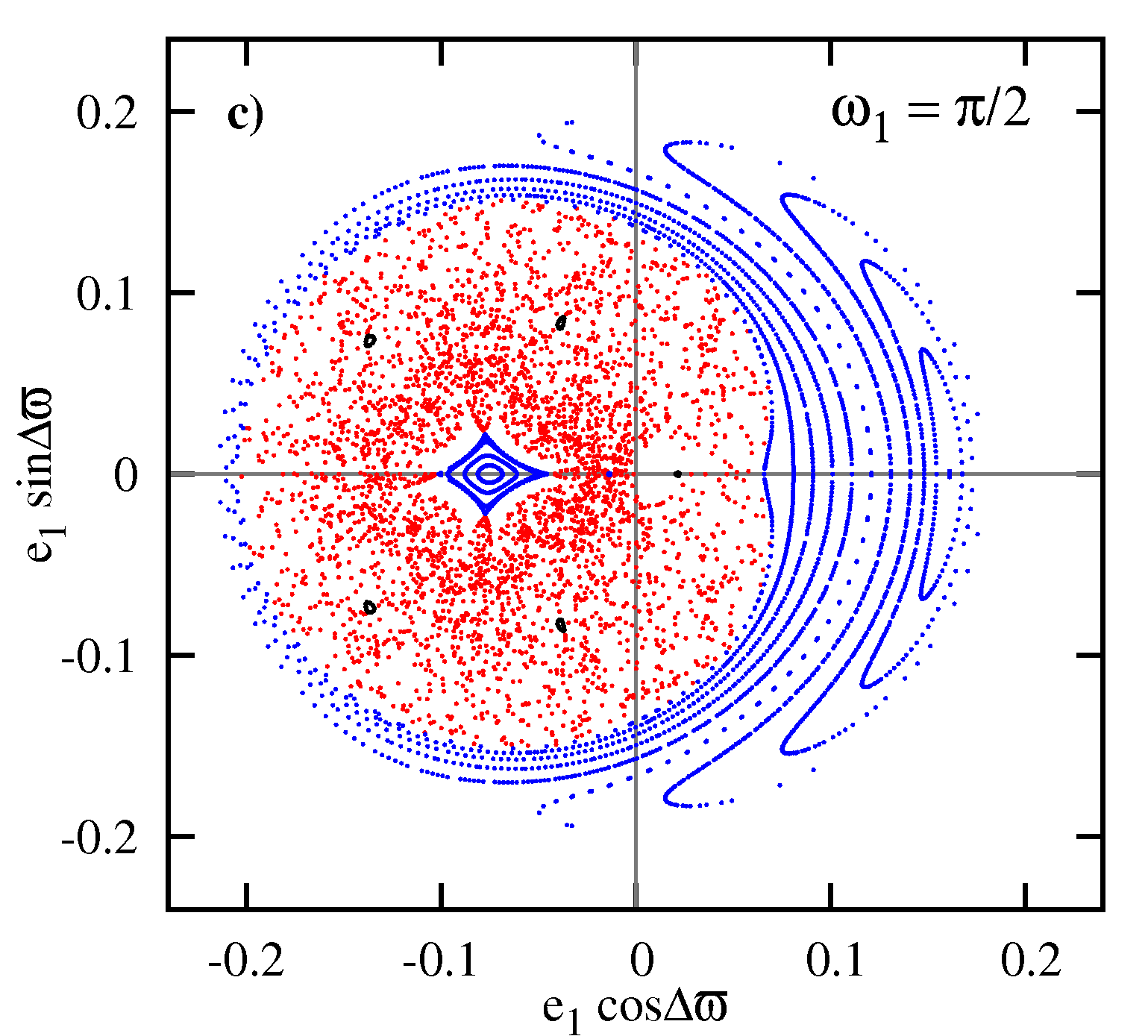}}
}
\hbox{
\hbox{\includegraphics[width=2.2in]{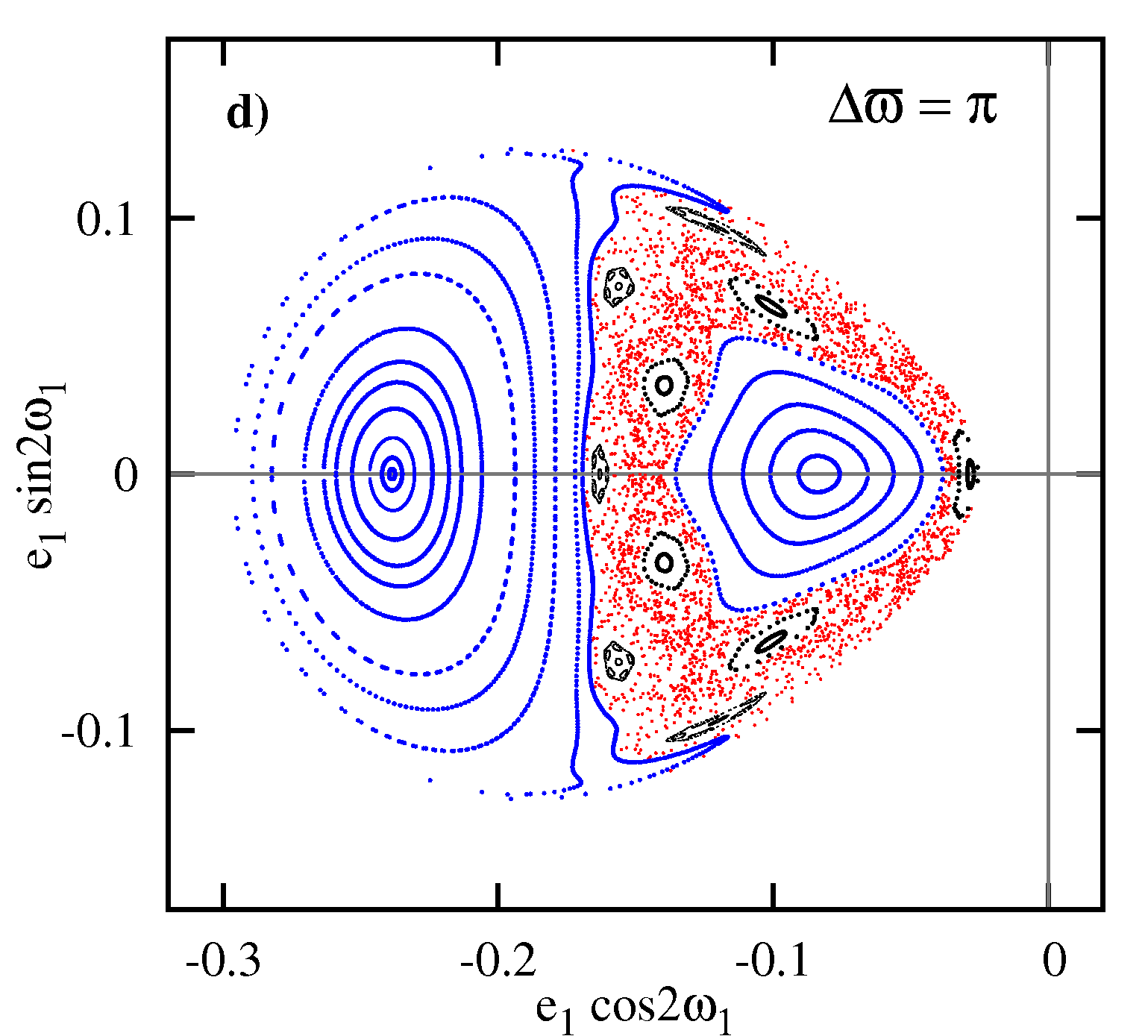}}
\hspace*{0.3cm}
\hbox{\includegraphics[width=2.2in]{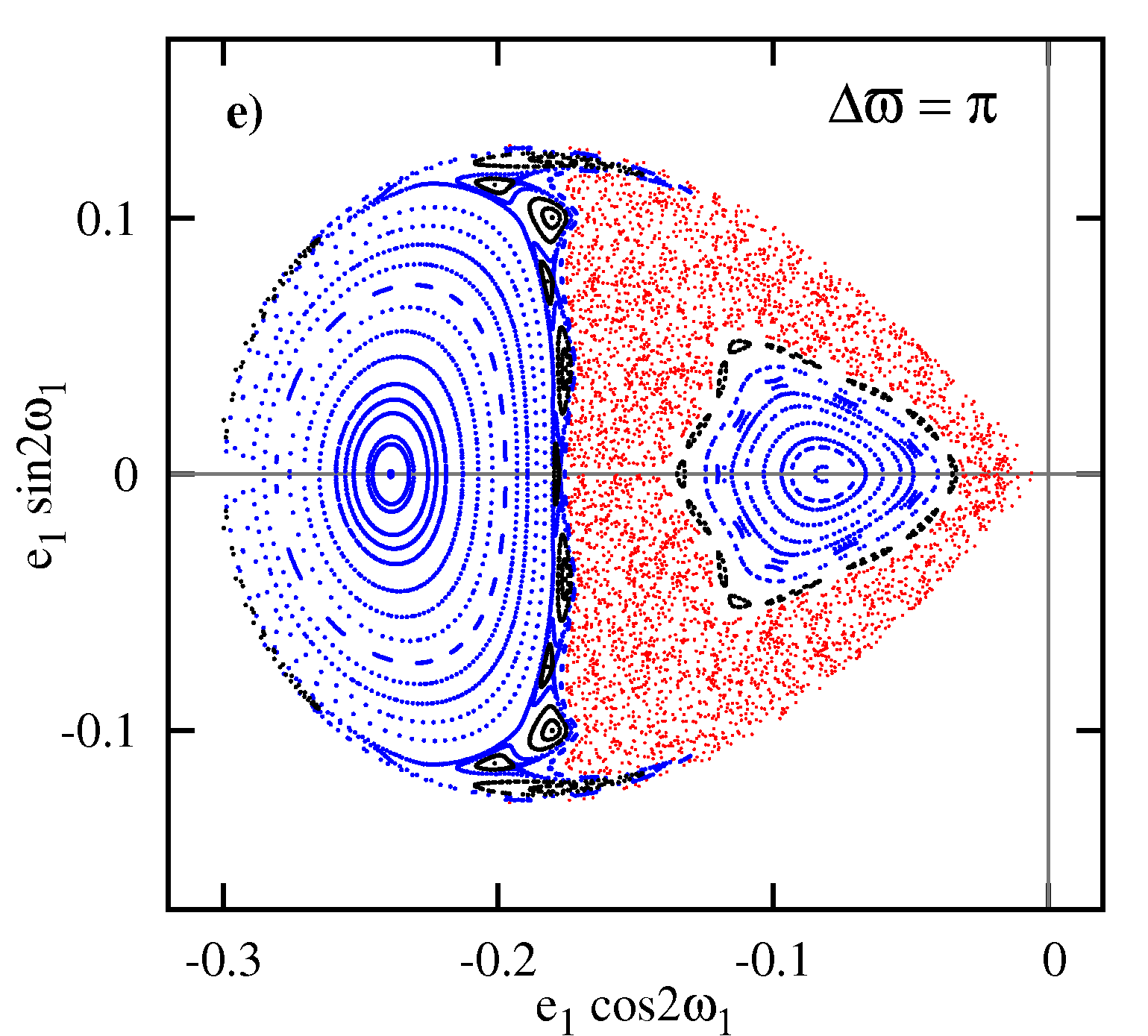}}
\hspace*{0.3cm}
\hbox{\includegraphics[width=2.2in]{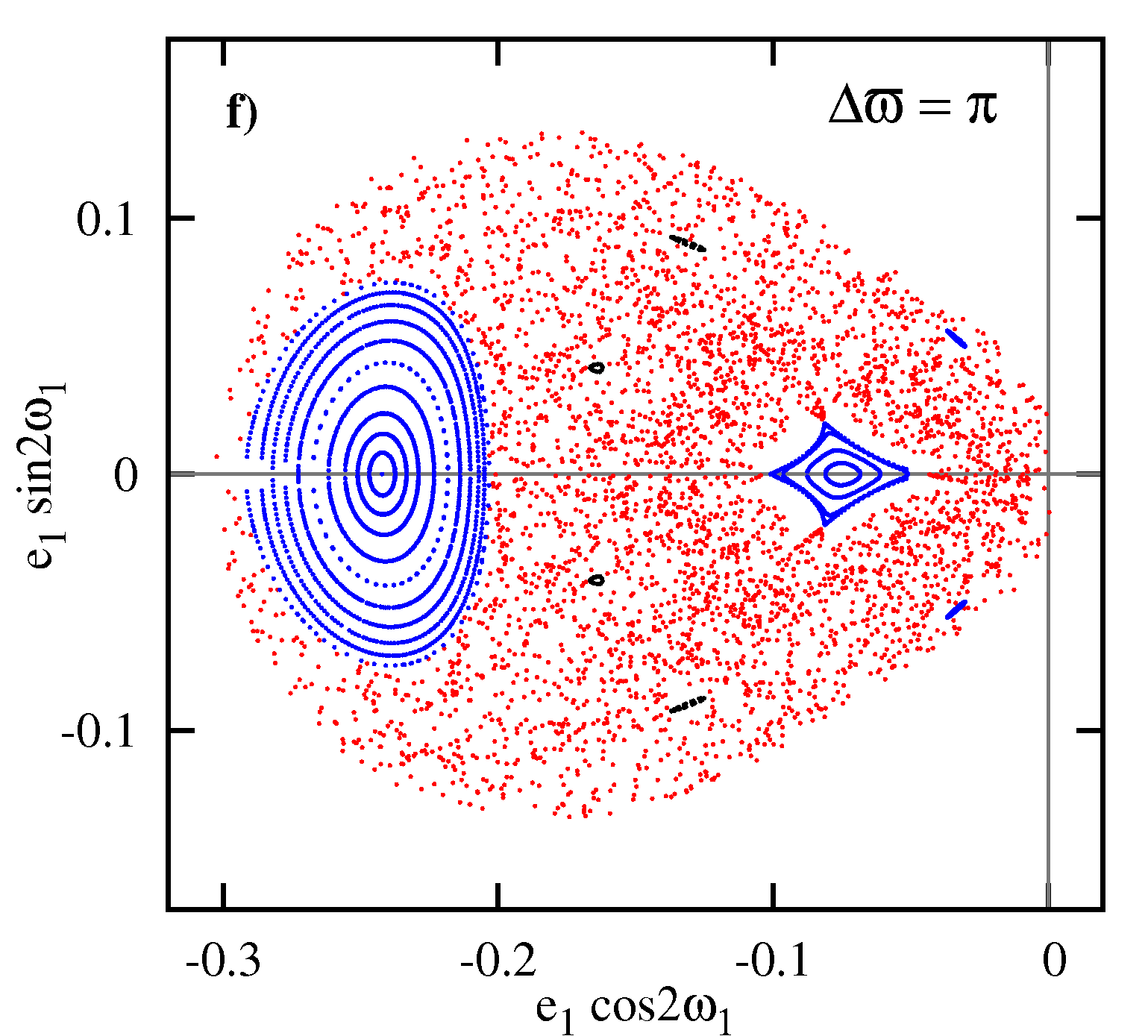}}
}
}
}
\caption{
Poincar\'e cross-sections  ($\omega_1=\pi/2$, $d\omega_1/dt<0$)  in the top row,
($\Delta\varpi=\pi$, $d\Delta\varpi/dt<0$) in the bottom row, computed for 3D
configuration of two-planet system with orbital parameters corresponding to the
best--fit parameters of the HD~12661 planetary system: 
$m_0=1.07~\mbox{M}_{\odot}$, 
$m_1=2.3~\mbox{m}_{\mbox{\idm{J}}}$, 
$m_2=1.57~\mbox{m}_{\mbox{\idm{J}}}$, 
$a_1=0.83~\mbox{au}$, 
$a_2=2.56~\mbox{au}$.
$\nAMD=0.085678$. 
The plots in columns from the left to the right are
for the following secular energies:   
$\mathcal{E}_{a,d}=-5.106073 \times 10^{-5}$, 
$\mathcal{E}_{b,e}=-5.106139 \times 10^{-5}$,                     
$\mathcal{E}_{c,f}=-5.106490 \times 10^{-5}$ (in canonical 
units of 1$M_{\sun}$, 1~au, 1~yr and $k=2\pi$).
These energies are chosen in the neighborhood of the Lidov-Kozai
resonance. Red dots are for chaotic motions, blue dots
mark quasi-periodic solutions. See the text for more details.
}
\label{fig:fig16}
\end{figure*}

To examine more closely the secular dynamics in the regime of the {\em classic}
L-K  resonance (family IVa), we computed the Poincar\'e cross sections for the
secular Hamiltonian having two degrees of freedom. These cross-sections are
obtained by integrating the equations of motion over a few Myr time-scale. The
parameters are selected as for the  HD~12661 system (see the caption to
Fig.~\ref{fig:fig16}). The cross-section planes are chosen as follows: 
\[ 
\cS_1 = \{e_1 \cos{\Delta{\varpi}} \times e_1 \sin{\Delta{\varpi}}\}, 
\quad \cS_2 = \{e_1 \cos{2 \omega_1} \times e_1 \sin{2 \omega_1}\}. 
\] 
The surface of section $\cS_1$ is defined by $\omega_1=\pi/2$
($d\omega_1/dt<0$), and the plane $\cS_2$ by $\Delta\varpi=\pi$
($d\Delta\varpi/dt<0$), respectively.  The top panels of Fig.~\ref{fig:fig16}
are for the $\cS_1$-plane, bottom panels of Fig.~\ref{fig:fig16} are for the
$\cS_2$-plane.   The cross sections are computed for energy curves in the
neighborhood of the L-K quasi-separatrix: panels in the left-hand column are for
the initial conditions lying on the energy level within the {\em
quasi-separatrix} curve encompassing the L-K resonance center, panels in the
middle column are for the energy level corresponding to the quasi-separatrix
curve, and the right-hand panels are for the energy level encompassing the
quasi-separatrix. In the cross-sections, we can detect a few high-order secular
resonances, invariant curves representing quasi-periodic orbits and relatively
large regions of chaotic motions. The appearance of chaotic dynamics in the 3-D
problem is the new feature as compared to the co-planar dynamics.  We recall
that in the later case, the averaging leads to  one-degree of freedom {\em
integrable} system.

Actually, the smooth invariant curves seen in the  Poincar\'e cross sections
assure us that quasi-analytic averaging makes it possible to derive very precise
numerical solutions of the secular equations of motion. The Poincar\'e
cross-sections, although obtained by complex algorithm relying on the numerical
integration of the mean Hamiltonian and the equations of motion derived by {\em
numerical differentiation}, make it possible to study the dynamics of the
secular system in detail, and in the whole permitted range of the orbital
parameters. 
\subsection{Families~IIa and IIb}
\begin{figure}
\centering
\hskip-7mm\includegraphics[width=2.8in]{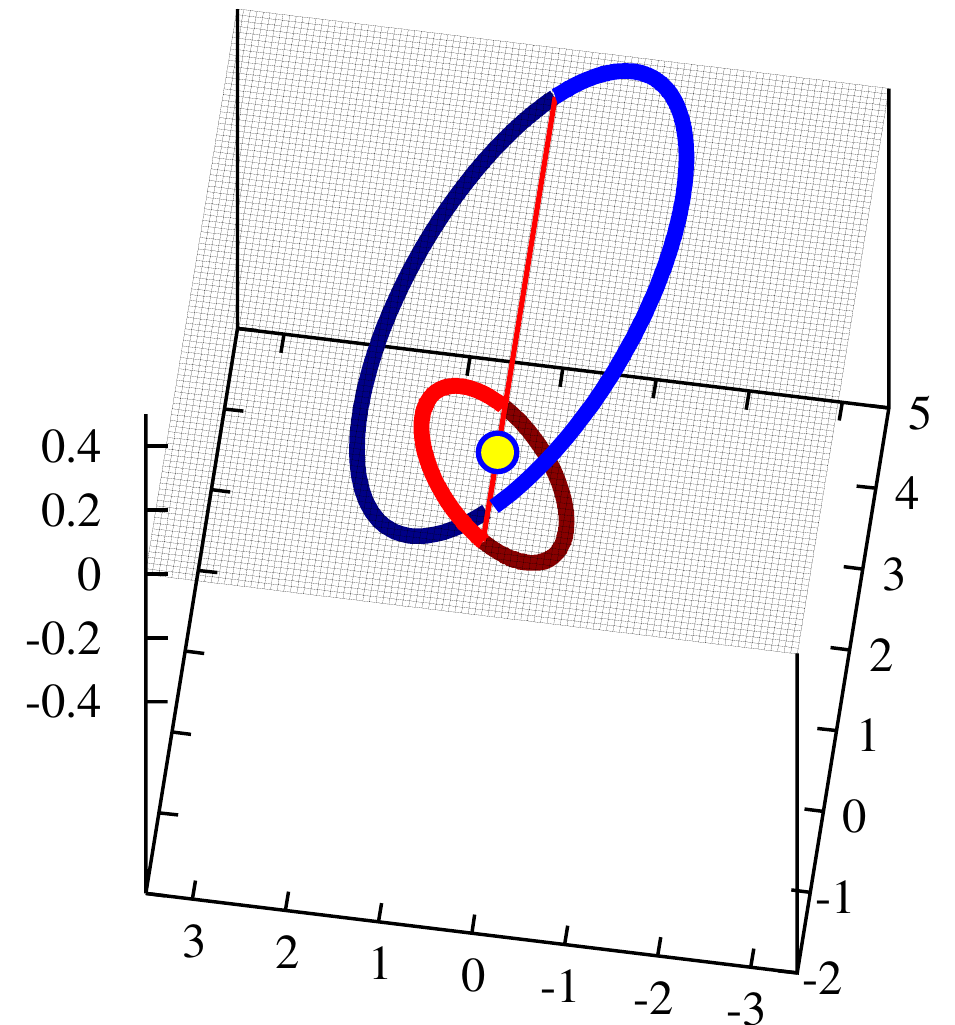}
\caption{
An example stationary configuration of family IIa computed for $a_1=0.83$~au,
$a_2=2.56$~au, $e_1=0.325, e_2=0.7204$.   The mutual inclination of orbits in
this case is $\Imut\sim 24^{\circ}$. 
}
\label{fig:fig17}
\end{figure}

Equilibria of family IIa are very special, because they are found over the
collision line of planetary orbits, hence beyond the formal limit of convergence of the expansion
of $\Hsec$ in $\alpha$. We call them the {\em chained} stationary
configurations because both secular orbits are connected like the links of a
chain (see Fig.~\ref{fig:fig17} for an illustration).  These solutions appear
at small mutual inclinations $\Imut$. In spite of large eccentricities, the
mean orbits cannot cross each other thanks 
 to a particular spatial orientation. These equilibria are generically stable because  they correspond to
the maxima of $\Hsec$. That can be seen in the stability diagrams
(Figs.~\ref{fig:fig13}--\ref{fig:fig15}). With increasing $\nAMD$, family IIa
emerges close to the collision line and then ``moves'' towards the border of
the \RP{}-plane. Curiously, the eccentricity 
$e_2$ of equilibria IIa spans moderate and large
values, and this family can be detected for all pairs of parameters analyzed in
this work. The most prominent example of family IIa is shown in stability
diagrams computed for $\alpha=0.667$ (see Fig.~\ref{fig:fig15}). We stress,
that these solutions are {\em non-resonant}. 
Family IIb is characterized by very {\em small} eccentricity of the inner planet
and large mutual inclination of orbits (let us recall that it it may emerge for
$\omega_1=\omega_2=0$). This family appears at bifurcational inclination
$I_1^{+}$ and can be revealed by the quadrupole theory  \citep{Krasinsky1974}.  
It is unstable --- although  it appears in the \RP{}-plane as an elliptic
point.  In fact, the secular Hamiltonian is not  sign definite function in its
neighborhood. The linear stability analysis reveals complex eigenvalues of the
linearized equations. 
%
\subsection{Families of quadrant III of the \RP{}-plane}
%
For small $\nAMD$, equilibria appearing in quadrant III of the \RP{}-plane  can
be  associated  with bifurcations of the zero-family solutions. These are 
equilibria of family IIIa at a saddle seen in the $\cP_S$-plane  (see two last
panels in the top row of Fig.~\ref{fig:fig8}) and appearing closely to the
$e_1=0$ axis,  and they are always unstable. Other solutions in quadrant III are
associated with a bifurcation of the L-K resonance in the regime of large $e_1$
(families IIIb, IVb+). We note, that in our ''taxonomy'', the name of IVb+ for
the saddle associated with elliptic point IIIb is justified by the fact that
this point can move between quadrant~IV and quadrant III (see a sequence of
panels in the {\em middle} row of Fig.~\ref{fig:fig10}). Equilibria IIIa appear
for {\em different} critical values of $\nAMD$ (or mutual inclination $i_0$) 
than solutions of family IIb. This can be particularly well seen  in
Fig.~\ref{fig:fig15} for large $\alpha$. Moreover, the quadrupole order theory
predict that these families appear for the same $i_0$. 

In the regime of large $\nAMD$, a plethora of quadrant~III solutions appears. We
classify them with symbols IIIc, IIIc-, IIIc+, IIId, IIIe, IIIf  (see, e.g., the
right-hand panel of Fig.~\ref{fig:fig3}) and they are associated with large
mutual inclinations of orbits and $e_2 \sim 1$, when the energy plane become
disconnected. Some of these families can be linearly stable (see the stability
diagrams in Figs.~\ref{fig:fig13}--\ref{fig:fig15}). In general, the numerical
continuation of these families is very difficult because they evolve close to
the boundary of permitted motions in the \RP{}-plane, and in the regime of large
mutual inclinations and eccentricities. Then the numerical procedure sometimes
fails, due to not precise enough determination of the second order derivatives,
and that may also explain some gaps in the family curves which are present in
the stability diagrams.  Also  problems with the continuation of these families
hinder precise and proper identification of some solutions.
%
\section{Conclusions} 
\label{sec:concl}
The semi-analytical averaging is a powerful technique helpful to reduce 
limitations of the analytical theories.  Although it relies on purely numerical
algorithms, its solid theoretical background is vital for the interpretation
and understanding of the results of numerical experiments. In this work, we
demonstrate that it makes it possible to study the secular dynamics of
two-planet system in wide ranges of  semi-major axes and masses ratio.  The
appropriate scaling of the problem parameters  helps us to represent the phase
space of the secular system globally. Assuming non-resonant configurations, and
that orbits are distant enough from collision zones, the averaged system may be
reduced to two degrees of freedom. Hence, to carry out the analysis,  we can
apply geometric tools, like the representative planes of initial conditions,
Poincar\'e cross sections,  continuations of stationary solutions with respect
to parameters, which are very helpful to  understand the structure of the phase
space and, in turn, the long-term behavior of the planetary system.

Equilibria are the basic class of solutions which can be investigated with
relatively simple tools. Our analysis reveals a number of families of
stationary solutions in the 3D secular problem of two planets. To the best of
our knowledge, some of them are yet  unknown in the literature and are related
to unusual orbital configurations. For instance, we found the so called chained
stationary configuration which are non-resonant,  can be found in the regime of
{\em small} mutual inclinations and large eccentricities, and  are located over
geometric collision line of orbits. In spite of such extreme dynamical
situation,  these secular solutions are Lyapunov stable and exist in wide
ranges of semi-major and masses ratio. Simultaneously, such orbital
configurations prohibit application of analytical methods  relying on 
power series
expansion of the perturbations.  The semi-analytic averaging helps to generalize
analytical results obtained for low-order expansions of the secular
Hamiltonian.

We obtained some interesting results regarding the Lidov-Kozai equilibrium in
the non-restricted problem. We found that this resonance may be associated with
librations of  $\Delta\varpi$ around $\pi$ in the neighboring trajectories 
(not only with librations of the inner pericenter around $\pm\pi/2$). These
librations are possible for relatively large ratios of  semi-major axes and
planetary masses. We found that the L-K resonance may also  appear in the
regime of large eccentricity of the inner orbit, and then it would be
associated with librations of $\omega_1$ around $\pm\pi/2$ with simultaneous
librations of $\Delta\varpi$ around $0$. The parametric evolution of the L-K
resonance is related to the stability of the zero-eccentricity equilibria.
There is a link between bifurcations of this family (and changes of its
stability) with an appearance of new families of stationary solutions in other
parts of the phase space (hence, with changes of its global topology). 

Our work illustrates qualitatively different view of the 3D dynamics as
compared to the coplanar configurations. It is already known 
\citep{Michtchenko2004} that the  non-resonant, coplanar systems of two
point-mass planets fulfilling  the averaging theorem and interacting through
Newtonian forces are integrable. The secular dynamics of such systems are
basically trivial and may be reduced to one degree of freedom. Under the same
assumptions, the spatial configurations may exhibit strong chaos and extremely
complex secular phenomena. 

In the approximation of Newtonian, point-mass interactions, the dynamics depend
on masses and semi-major axes ratios, hence our results are valid both for
planetary systems with small planets, as well as for systems comprising of
brown dwarfs or even sub-stellar companions. However, the dynamics of real
systems may strongly depend on the magnitude of the mutual interactions.
Moreover,  many stable equilibria are found for large values of $\nAMD$.
According to the notion of $\AMD$ \citep{Laskar2000}, in such cases the real
configurations are unlikely long-term stable even close to secularly stable
equilibria.  In that sense, the results of stability  analysis may be too
optimistic. We skip a study of such effects in this work, because it would make
the paper necessarily very lengthly. We should keep in mind that introduction
of relativistic, tidal, and stellar quadrupole-moment perturbations,  may
affect the secular dynamics dramatically
\citep{Migaszewski2008c,Migaszewski2008d}. 

The quasi-global technique applied in this paper has been proved an efficient
and effective tool for the analysis of the secular 3-D model. Nevertheless, we
learned from the work that the quasi-analytical approach is not a perfect tool.
Due to limitations of the numerical algorithms, the continuation of  families
of equilibria and analysis of their stability is particularly difficult when we
reach limits of permitted motion or $\Hsec$ is a weakly varying function. We
can also overlook some solutions. Unfortunately, that leave us sometimes with
open questions.  
\section*{Acknowledgments}
We thank Makiko Nagasawa for a review that improved the manuscript. This work
is supported by the Polish Ministry of Sciences and Education, Grant No.
1P03D-021-29. C.M. is also supported by Nicolaus Copernicus University Grant
No.~408A.
\bibliographystyle{mn2e}
\bibliography{ms}

\begin{thebibliography}{}

\bibitem[\protect\citeauthoryear{{Adams} \& {Laughlin}}{{Adams} \&
  {Laughlin}}{2003}]{Adams2003}
{Adams} F.~C.,  {Laughlin} G.,  2003, Icarus, 163, 290

\bibitem[\protect\citeauthoryear{{Brouwer} \& {Clemence}}{{Brouwer} \&
  {Clemence}}{1961}]{Brouwer1961}
{Brouwer} D.,  {Clemence} G.~M.,  1961, {Methods of celestial mechanics}.
New York: Academic Press, 1961

\bibitem[\protect\citeauthoryear{{Brumberg}}{{Brumberg}}{1995}]{Brumberg1995}
{Brumberg} V.~A.,  1995,
Analytical Techniques of Celestial Mechanics,
 Springer-Verlag Berlin Heidelberg New York

\bibitem[\protect\citeauthoryear{{Butler}, {Wright}, {Marcy}, {Fischer},
  {Vogt}, {Tinney}, {Jones}, {Carter}, {Johnson}, {McCarthy} \&
  {Penny}}{{Butler} et~al.}{2006}]{Butler2006}
{Butler} R.~P., et. al, 2006, ApJ, 646, 505

\bibitem[\protect\citeauthoryear{{Fabrycky} \& {Tremaine}}{{Fabrycky} \&
  {Tremaine}}{2007}]{Fabrycky2007}
{Fabrycky} D.,  {Tremaine} S.,  2007, ApJ, 669, 1298

\bibitem[\protect\citeauthoryear{{Ferraz-Mello}, {Michtchenko}, {Beaug{\'e}} \&
  {Callegari} Jr.}{{Ferraz-Mello} et~al.}{2005}]{FerrazMello2005}
{Ferraz-Mello} S.,  {Michtchenko} T.~A.,  {Beaug{\'e}} C.,    {Callegari} Jr.
  N.,  2005, in {Dvorak} R.,  {Freistetter} F.,   {Kurths} J.,  eds.,
Vol.~683 of Lecture Notes in Physics, Berlin
  Springer Verlag, {Extrasolar Planetary Systems}.
pp 219--+

\bibitem[\protect\citeauthoryear{{Ferrer} \& {Osacar}}{{Ferrer} \&
  {Osacar}}{1994}]{Ferrer1994}
{Ferrer} S.,  {Osacar} C.,  1994, Celestial Mechanics and Dynamical Astronomy,
  58, 245

\bibitem[\protect\citeauthoryear{{Fischer}, {Marcy}, {Butler}, {Vogt}, {Frink}
  \& {Apps}}{{Fischer} et~al.}{2001}]{Fischer2001}
{Fischer} D.~A.,  et. al, 2001, ApJ, 551, 1107

\bibitem[\protect\citeauthoryear{{Fischer}, {Marcy}, {Butler}, {Vogt}, {Henry},
  {Pourbaix}, {Walp}, {Misch} \& {Wright}}{{Fischer}
  et~al.}{2003}]{Fischer2003}
{Fischer} D.~A.,  {Marcy} G.~W.,  {Butler} R.~P.,  {Vogt} S.~S.,  {Henry}
  G.~W.,  {Pourbaix} D.,  {Walp} B.,  {Misch} A.~A.,    {Wright} J.~T.,  2003,
  ApJ, 586, 1394

\bibitem[\protect\citeauthoryear{{Ford}, {Kozinsky} \& {Rasio}}{{Ford}
  et~al.}{2000}]{Ford2000}
{Ford} E.~B.,  {Kozinsky} B.,    {Rasio} F.~A.,  2000, ApJ, 535, 385

\bibitem[\protect\citeauthoryear{{Go{\'z}dziewski}}{{Go{\'z}dziewski}}{2003a}]%
{Gozdziewski2003a}
{Go{\'z}dziewski} K.,  2003a, A\&A, 398, 1151

\bibitem[\protect\citeauthoryear{{Go{\'z}dziewski}}{{Go{\'z}dziewski}}{2003b}]%
{Gozdziewski2003c}
{Go{\'z}dziewski} K.,  2003b, Celestial Mechanics and Dynamical Astronomy, 85,
  79

\bibitem[\protect\citeauthoryear{{Go{\'z}dziewski} \&
  {Konacki}}{{Go{\'z}dziewski} \& {Konacki}}{2004}]{Gozdziewski2004}
{Go{\'z}dziewski} K.,  {Konacki} M.,  2004, ApJ, 610, 1093

\bibitem[\protect\citeauthoryear{{Go{\'z}dziewski} \&
  {Migaszewski}}{{Go{\'z}dziewski} \& {Migaszewski}}{2006}]{Gozdziewski2006}
{Go{\'z}dziewski} K.,  {Migaszewski} C.,  2006, A\&A, 449, 1219

\bibitem[\protect\citeauthoryear{{Innanen}, {Zheng}, {Mikkola} \&
  {Valtonen}}{{Innanen} et~al.}{1997}]{Innanen1997}
{Innanen} K.~A.,  {Zheng} J.~Q.,  {Mikkola} S.,    {Valtonen} M.~J.,  1997, AJ,
  113, 1915

\bibitem[\protect\citeauthoryear{{Ji}, {Liu}, {Kinoshita}, {Zhou}, {Nakai} \&
  {Li}}{{Ji} et~al.}{2003}]{Ji2003b}
{Ji} J.,  {Liu} L.,  {Kinoshita} H.,  {Zhou} J.,  {Nakai} H.,    {Li} G.,
  2003, ApJL, 591, L57

\bibitem[\protect\citeauthoryear{{Khalil}}{{Khalil}}{2001}]{Khalil2001}
{Khalil} H.~K.,  2001, Nonlinear Systems.
Prenitce Hall

\bibitem[\protect\citeauthoryear{{Kinoshita} \& {Nakai}}{{Kinoshita} \&
  {Nakai}}{2007}]{Kinoshita2007}
{Kinoshita} H.,  {Nakai} H.,  2007, Celestial Mechanics and Dynamical
  Astronomy, 98, 67

\bibitem[\protect\citeauthoryear{{Kozai}}{{Kozai}}{1962}]{Kozai1962}
{Kozai} Y.,  1962, AJ, 67, 579

\bibitem[\protect\citeauthoryear{{Krasinsky}}{{Krasinsky}}{1972}]{Krasinsky197%
2}
{Krasinsky} G.~A.,  1972, Celestial Mechanics, 6, 60

\bibitem[\protect\citeauthoryear{{Krasinsky}}{{Krasinsky}}{1974}]{Krasinsky197%
4}
{Krasinsky} G.~A.,  1974, in {Kozai} Y.,  ed., 
Vol.~62 of IAU Symposium,
pp 95--116

\bibitem[\protect\citeauthoryear{{Laskar}}{{Laskar}}{2000}]{Laskar2000}
{Laskar} J.,  2000, Physical Review Letters, 84, 3240

\bibitem[\protect\citeauthoryear{{Laskar} \& {Robutel}}{{Laskar} \&
  {Robutel}}{1995}]{Laskar1995}
{Laskar} J.,  {Robutel} P.,  1995, Celestial Mechanics and Dynamical Astronomy,
  62, 193

\bibitem[\protect\citeauthoryear{{Lee} \& {Peale}}{{Lee} \&
  {Peale}}{2003}]{Lee2003}
{Lee} M.~H.,  {Peale} S.~J.,  2003, ApJ, 592, 1201

\bibitem[\protect\citeauthoryear{{Libert} \& {Henrard}}{{Libert} \&
  {Henrard}}{2006}]{Libert2006}
{Libert} A.-S.,  {Henrard} J.,  2006, Icarus, 183, 186

\bibitem[\protect\citeauthoryear{{Libert} \& {Henrard}}{{Libert} \&
  {Henrard}}{2007a}]{Libert2007a}
{Libert} A.-S.,  {Henrard} J.,  2007a, A\&A, 461, 759

\bibitem[\protect\citeauthoryear{{Libert} \& {Henrard}}{{Libert} \&
  {Henrard}}{2007b}]{Libert2007b}
{Libert} A.-S.,  {Henrard} J.,  2007b, Icarus, 191, 469

\bibitem[\protect\citeauthoryear{{Lidov} \& {Ziglin}}{{Lidov} \&
  {Ziglin}}{1976}]{Lidov1976}
{Lidov} M.~L.,  {Ziglin} S.~L.,  1976, Celestial Mechanics, 13, 471

\bibitem[\protect\citeauthoryear{{Lidov}}{{Lidov}}{1961}]{Lidov1961}
{Lidov} R.,  1961, Izd. Akad. Nauk SSSR (1963), 1, 119

\bibitem[\protect\citeauthoryear{{Markeev}}{{Markeev}}{1978}]{Markeev1978}
{Markeev} A.~P.,  1978, {Libration points in Celestial Mechanics and
  Astrodynamics}.
Moscow, Nauka

\bibitem[\protect\citeauthoryear{{Meyer} \& {Schmidt}}{{Meyer} \&
  {Schmidt}}{1986}]{Meyer1986}
{Meyer} K.~R.,  {Schmidt} D.~S.,  1986, J. Diff. Eq., 62, 222

\bibitem[\protect\citeauthoryear{{Michtchenko} \& {Ferraz-Mello}}{{Michtchenko}
  \& {Ferraz-Mello}}{2001}]{Michtchenko2001}
{Michtchenko} T.~A.,  {Ferraz-Mello} S.,  2001, AJ, 122, 474

\bibitem[\protect\citeauthoryear{{Michtchenko}, {Ferraz-Mello} \&
  {Beaug{\'e}}}{{Michtchenko} et~al.}{2006}]{Michtchenko2006}
{Michtchenko} T.~A.,  {Ferraz-Mello} S.,    {Beaug{\'e}} C.,  2006, Icarus,
  181, 555

\bibitem[\protect\citeauthoryear{{Michtchenko} \& {Malhotra}}{{Michtchenko} \&
  {Malhotra}}{2004}]{Michtchenko2004}
{Michtchenko} T.~A.,  {Malhotra} R.,  2004, Icarus, 168, 237

\bibitem[\protect\citeauthoryear{{Migaszewski} \&
  {Go{\'z}dziewski}}{{Migaszewski} \&
  {Go{\'z}dziewski}}{2008}]{Migaszewski2008a}
{Migaszewski} C.,  {Go{\'z}dziewski} K.,  2008, MNRAS, 388, 789

\bibitem[\protect\citeauthoryear{{Migaszewski} \&
  {Go{\'z}dziewski}}{{Migaszewski} \&
  {Go{\'z}dziewski}}{2009a}]{Migaszewski2008d}
{Migaszewski} C.,  {Go{\'z}dziewski} K.,  2009a, arXiv:0901.0102

\bibitem[\protect\citeauthoryear{{Migaszewski} \&
  {Go{\'z}dziewski}}{{Migaszewski} \&
  {Go{\'z}dziewski}}{2009b}]{Migaszewski2008c}
{Migaszewski} C.,  {Go{\'z}dziewski} K.,  2009b, MNRAS, 392, 2

\bibitem[\protect\citeauthoryear{{Miller} \& {Hamilton}}{{Miller} \&
  {Hamilton}}{2002}]{Miller2002}
{Miller} M.~C.,  {Hamilton} D.~P.,  2002, ApJ, 576, 894

\bibitem[\protect\citeauthoryear{{Morbidelli}}{{Morbidelli}}{2002}]{Morbidelli%
2002}
{Morbidelli} A.,  2002, {Modern celestial mechanics : aspects of {S}olar
  {S}ystem dynamics}.
Taylor {\&} Francis, London

\bibitem[\protect\citeauthoryear{{Murray} \& {Dermott}}{{Murray} \&
  {Dermott}}{2000}]{Murray2000}
{Murray} C.~D.,  {Dermott} S.~F.,  2000, {Solar System Dynamics}.
Cambridge University Press

\bibitem[\protect\citeauthoryear{{Press}, {Teukolsky}, Vetterling \&
  {Flannery}}{{Press} et~al.}{1992}]{Press1992}
{Press} W.~H.,  {Teukolsky} S.~A.,  Vetterling W.~T.,    {Flannery} B.~P.,
  1992, Numerical Recipes in C. The Art of Scientific Computing.
Cambridge Univ. Press

\bibitem[\protect\citeauthoryear{{Rodr{\'{\i}}guez} \&
  {Gallardo}}{{Rodr{\'{\i}}guez} \& {Gallardo}}{2005}]{Rodriguez2005}
{Rodr{\'{\i}}guez} A.,  {Gallardo} T.,  2005, ApJ, 628, 1006

\bibitem[\protect\citeauthoryear{{Sokolskii}}{{Sokolskii}}{1975}]{Sokolskii197%
5}
{Sokolskii} A.~G.,  1975, Prikladnaia Matematika i Mekhanika, 39, 366

\bibitem[\protect\citeauthoryear{{Thomas} \& {Morbidelli}}{{Thomas} \&
  {Morbidelli}}{1996}]{Michel1996}
{Thomas} F.,  {Morbidelli} A.,  1996, Celestial Mechanics and Dynamical
  Astronomy, 64, 209

\bibitem[\protect\citeauthoryear{{Thommes} \& {Lissauer}}{{Thommes} \&
  {Lissauer}}{2003}]{Thommes2003}
{Thommes} E.~W.,  {Lissauer} J.~J.,  2003, ApJ, 597, 566

\bibitem[\protect\citeauthoryear{{Veras} \& {Armitage}}{{Veras} \&
  {Armitage}}{2007}]{Veras2007}
{Veras} D.,  {Armitage} P.~J.,  2007, ApJ, 661, 1311

\bibitem[\protect\citeauthoryear{{Veras} \& {Ford}}{{Veras} \&
  {Ford}}{2008}]{Veras2008}
{Veras} D.,  {Ford} E.~B.,  2008, ApJ, (arXiv:0811.0001)

\bibitem[\protect\citeauthoryear{{Wright}, {Upadhyay}, {Marcy}, {Fischer},
  {Ford} \& {Johnson}}{{Wright} et~al.}{2008}]{Wright2008}
{Wright} J.~T.,  {Upadhyay} S.,  {Marcy} G.~W.,  {Fischer} D.~A.,  {Ford}
  E.~B.,    {Johnson} J.~A.,  2008, ApJ, (arXiv:0812.1582)

\end{thebibliography}

\end{document}